\documentclass[lettersize,journal]{IEEEtran}
\usepackage{amsmath,amsfonts}
\usepackage{algorithmic}
\usepackage{array}
\usepackage[caption=false,font=normalsize,labelfont=sf,textfont=sf]{subfig}
\usepackage{booktabs}
\usepackage{textcomp}
\usepackage{stfloats}
\usepackage{multirow}
\usepackage{url}
\usepackage[ruled,linesnumbered]{algorithm2e}
\usepackage{verbatim}
\usepackage{graphicx}
\usepackage{subfig} 
\usepackage{subfloat}
\usepackage{float}
\usepackage{bm}
\def\BibTeX{{\rm B\kern-.05em{\sc i\kern-.025em b}\kern-.08em
		T\kern-.1667em\lower.7ex\hbox{E}\kern-.125emX}}
\usepackage{balance}

\usepackage[T1]{fontenc}
\begin{document}
	\title{Adaptive Marginalized Semantic Hashing for Unpaired Cross-Modal Retrieval}
	\author{Kaiyi Luo,
		Chao Zhang, 
		Huaxiong Li,
		Xiuyi Jia,
		Chunlin Chen
		\IEEEcompsocitemizethanks{
			\IEEEcompsocthanksitem K. Luo, C. Zhang, H. Li and C. Chen are with the Department of Control Science and Intelligence Engineering, Nanjing University, Nanjing 210093, China. E-mail:  ky\_luo@smail.nju.edu.cn, chzhang@smail.nju.edu.cn, huaxiongli@nju.edu.cn, clchen@nju.edu.cn.
			\IEEEcompsocthanksitem X. Jia is with the School of Computer
			Science and Engineering, Nanjing University of Science and Technology, Nanjing 210014, China. E-mail: jiaxy@njust.edu.cn.
		}
	}
	
	\markboth{}%
	{Shell \MakeLowercase{\textit{et al.}}: Bare Demo of IEEEtran.cls for Computer Society Journals}
	
	\maketitle
	
	\begin{abstract}
		In recent years, Cross-Modal Hashing (CMH) has aroused much attention due to its fast query speed and efficient storage. Previous literatures have achieved promising results for Cross-Modal Retrieval (CMR) by discovering discriminative hash codes and modality-specific hash functions. Nonetheless, most existing CMR works are subjected to some restrictions: 1) It is assumed that data of different modalities are fully paired, which is impractical in real applications due to sample missing and false data alignment, and 2) binary regression targets including the label matrix and binary codes are too rigid to effectively learn semantic-preserving hash codes and hash functions. To address these problems, this paper proposes an Adaptive Marginalized Semantic Hashing (AMSH) method which not only enhances the discrimination of latent representations and hash codes by adaptive margins, but also can be used for both paired and unpaired CMR. As a two-step method, in the first step, AMSH generates semantic-aware modality-specific latent representations with adaptively marginalized labels, which enlarges the distances between different classes, and exploits the labels to preserve the inter-modal and intra-modal semantic similarities into latent representations and hash codes. In the second step, adaptive margin matrices are embedded into the hash codes, and enlarge the gaps between positive and negative bits, which improves the discrimination and robustness of hash functions. On this basis, AMSH generates similarity-preserving hash codes and robust hash functions without strict one-to-one data correspondence requirement. Experiments are conducted on several benchmark datasets to demonstrate the superiority and flexibility of AMSH over some state-of-the-art CMR methods.
	\end{abstract}
	

	\section{Introduction}
	\IEEEPARstart{T}{he} recent years has witnessed the explosive growth of multimedia such as texts and images. Cross-Modal Retrieval (CMR) has attracted much interest to search relevant items of different modalities \cite{SDH}, such as searching desired images with text descriptions. Unlike uni-modal retrieval that searches within one modality, the biggest challenge of CMR is to bridge the heterogeneous gaps among modalities, which is caused by the different feature distributions and dimensionality across modalities \cite{SCRATCH, ALECH,WASH}. Some CMR methods \cite{CCA,JFS,Corr-Cross-AE,overview} were proposed to eliminate the heterogeneity by learning common real-valued representations for all modalities. Despite the effectiveness of these methods, query and training efficiency is compromised, and huge memory is required for data storage. Thus, how to efficiently and accurately search semantically similar items across modalities remains a challenge \cite{SMFH,CALM}.
	
	Hashing technique has been widely applied in CMR due to its fast query speed and efficient storage. Many Cross-Modal Hashing (CMH) methods have been proposed in recent years \cite{deep1,deep2,deep3,deep4,deep5,FUH,RFDH}. The basic idea of hashing methods is to map data of different modalities into a common Hamming space, where each sample can be represented by binary codes. In this way, distances between data points can be measured by XOR operation, which requires low computational costs. CMH methods can be roughly categorized into supervised methods \cite{SMFH,CSDH,DCH,SCRATCH,SRLCH,MTFH} and unsupervised ones \cite{CMFH,CCQ,FSH,CRE}. Unsupervised CMH methods seek to generate robust hash codes and hash functions without label information by exploiting data structures. Typical works include Collective Matrix Factorization Hashing (CMFH) \cite{CMFH}, Fusion Similarity Hashing (FSH) \cite{FSH} and Collective Reconstructive Embeddings (CRE) \cite{CRE}. Supervised CMH methods utilize labels to guide the hash learning, with inter-modal and intra-modal similarities well preserved. Due to the heterogeneous gaps among different modalities, a common strategy of supervised CMH methods is to seek a shared latent representation, with data across modalities fully paired. To preserve semantic similarities, some methods \cite{SCRATCH,ALECH} train a linear classifier between labels and hash codes for semantic embedding. Typical works include Supervised Matrix Factorization Hashing (SMFH) \cite{SMFH}, Discrete Cross-modal Hashing (DCH) \cite{DCH}, Scalable disCRete mATrix faCtorization Hashing (SCRATCH) \cite{SCRATCH} and Subspace Relation Learning for Cross-modal Hashing (SRLCH) \cite{SRLCH}. Generally, supervised methods outperform unsupervised ones, since labels are high-level semantic information, which can better represent data similarities and correlations \cite{ALECH}.
	
	Although most supervised CMH methods have achieved promising retrieval results, there are still some restrictions to be solved: (1) One crucial problem is the strict inter-modal data alignment assumption. Most supervised CMH methods assumed that data of all modalities are fully paired, which facilitates the elimination of heterogeneous gaps by projecting multi-modal data into a shared latent representation. However, in real practice, collected data may suffer from data missing and misalignment problem, making the one-to-one correspondence assumption invalid. This scenario is referred to as Unpaired Cross-Modal Retrieval (UCMR). Recently, some works have been proposed to address Semi-paired Cross-Modal Retrieval \cite{SPDH,UAPMH,CALM,flexible}, where only partial data correspondence is given. However, semi-paired hashing methods are not able to solve the UCMR problem due to their reliance on paired data. To the best of our knowledge, there are few literatures dedicated to the UCMR problem \cite{GSPH,UCMH,RUCMH,FMH}. (2) Some supervised  CMH methods train a classifier to preserve high-level semantic information based on the original logical label matrix, which ignores the distance between different classes. In the multi-class classification, it is desired that different classes are as far away from each other as possible to enhance discrimination. (3) Most previous methods directly adopt least squares regression between multi-modal feature and binary codes, which compromises the robustness of hash functions, because the binary codes are too rigid for robust and discriminative regression. 
	\begin{figure*}[t]
		\centering
		\includegraphics[scale=1.05]{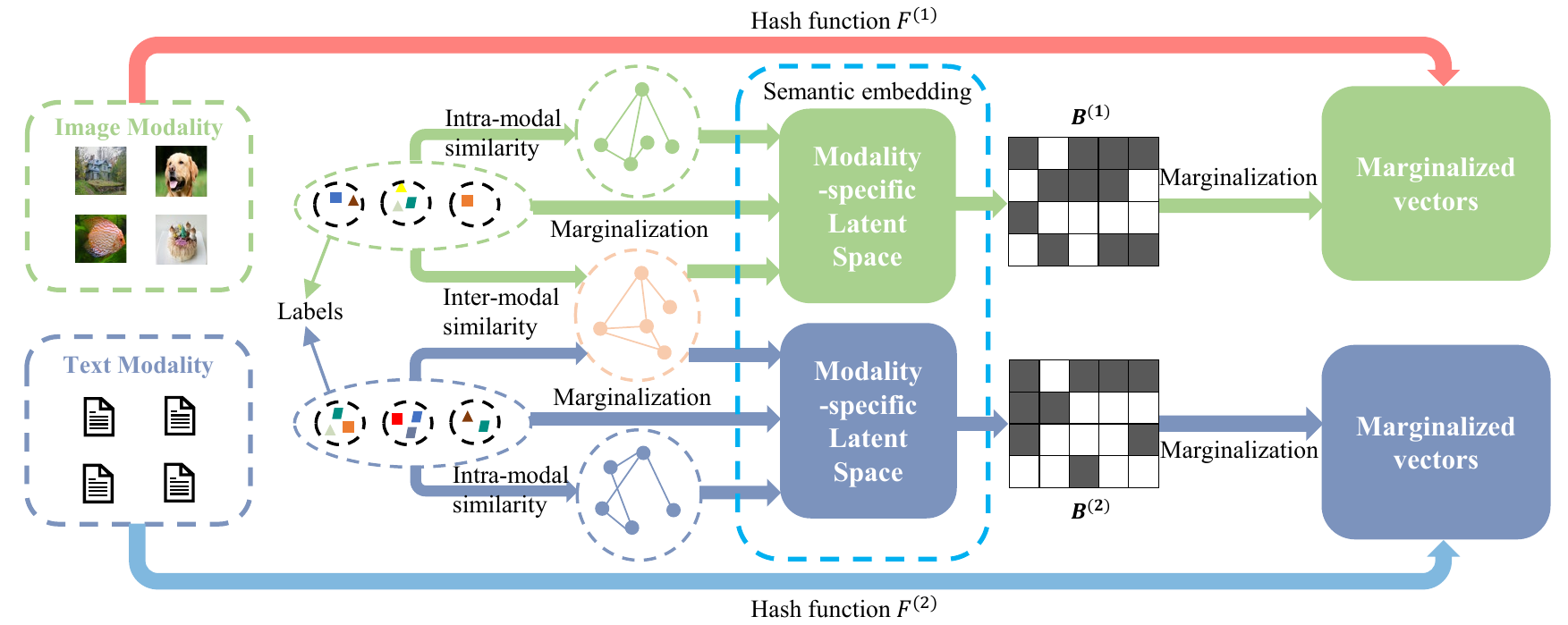}
		\caption{The overall workflow of AMSH.}
		\label{workflow}
	\end{figure*} 
	
	To tackle these limitations, we propose an Adaptive Marginalized Semantic Hashing (AMSH) method for CMR with unpaired data. The overall framework of AMSH is illustrated in Fig. \ref{workflow}. AMSH is a two-step method. In the first step, to tackle the UCMR problem, AMSH generates modality-specific latent representations, which are embedded with semantic information by preserving inter-modal and intra-modal similarities. Besides, AMSH introduces the adaptive margin matrices into the linear classifier between the label matrix and the latent representation to alleviate the rigid zero-one linear regression, which enhances the discrimination of latent representations and hash codes. In the second step, to address the rigid binary linear regression problem, AMSH incorporates the adaptive margin matrices to adaptively enlarge the distances between positive and negative bits. Thus discriminative semantic-aware hash codes and robust hash functions can be obtained. The main contributions are summarized as follows:
	\begin{itemize}
		\item We present a novel supervised method, i.e., AMSH, which can not only learn discriminative hash codes and hash functions, but also solve unpaired CMR problems.
		\item We introduce the adaptive margin matrices which are utilized to enlarge the distances between different classes, and the distances between positive and negative hash code bits, to enhance the discrimination of hash codes and hash functions.
		\item Compared with some state-of-the-art CMH methods, extensive experiments are conducted on several benchmark datasets, demonstrating the effectiveness of the proposed AMSH.
	\end{itemize} 
	
	The remainder of this paper is organized as follows. Section II briefly reviews some existing related works. Section III presents the formulation of AMSH and the optimization scheme. Experiments are conducted in Section IV. Section V concludes the paper.
	\section{RELATED WORK}
	\subsection{Supervised Cross-Modal Hashing}
	Supervised CMH methods bridge the semantic gaps among different modalities by embedding label information into hash codes, which generally outperform unsupervised ones. Semantic Correlation Maximization (SCM) hashing \cite{SCM} constructs the similarity matrix with label information, and further applies spectral relaxation to solve the binary NP-hard problem. Semantics-Preserving Hashing (SePH) method \cite{SePH} constructs a probability distribution with semantic affinities, and approximates the distribution to the target binary codes. These methods relax the binary constraints into continuous ones. However, relaxation strategies may cause quantization error. Some other discrete hashing methods have been proposed without relaxation. DCH \cite{DCH} views the binary codes as features to train classifiers, which generates discriminative binary codes and avoids quantization error. SMFH \cite{SMFH} learns unified hash codes via collective matrix factorization and measures the pairwise similarity with labels. SRLCH \cite{SRLCH} and SCRATCH \cite{SCRATCH} exploit relation information and matrix factorization to improve hash learning. scalaBle Asymmetric Discrete Cross-modal Hashing (BATCH) \cite{BATCH} exploits labels to preserve the pairwise similarity. Some methods also explore data structures for better hash learning. Luo et al. proposed a Supervised Discrete Manifold-Embedded Cross-Modal Hashing (SDMCH) \cite{SDMCH} which explores manifold correlations among data of different modalities. Adaptive Label correlation based asymmEtric Cross-modal Hashing (ALECH) \cite{ALECH} and Two-stEp Cross-modal Hashing (TECH) \cite{TECH} both consider label correlation for better semantic discrimination. Fast Cross-Modal Hashing (FCMH) \cite{FCMH} preserves not only global similarities but also the local similarities. Fast Discriminative Discrete Hashing (FDDH) \cite{FDDH} adaptively enlarges semantic distance of different classes to increase label discrimination. However, all these methods cannot address the UCMR problem, for they are based on the assumption that data across modalities are fully paired, which is impractical in real applications. 
	
	Recently, with the development of deep learning, hashing methods have been greatly boosted by Deep Neural Networks (DNNs). Many DNNs-based CMR methods have been proposed \cite{DCMH,PRDH,AADAH,MCCN}. The performance of these methods generally outperforms shallow models due to the representation capacity of DNNs. Typical works include Deep Cross-Modal Hashing (DCMH) \cite{DCMH}, Pairwise Relationship Guided Deep Hashing (PRDH) \cite{PRDH} and Attention-Aware Deep Adversarial Hashing (AADAH) \cite{AADAH}. Nevertheless, DNNs-based methods suffer from high computational costs, and it is challenging to determine the network structures for different modalities. Besides, most DNNs-based methods cannot be applied to the UCMR problem. 
	
	\subsection{Unpaired Cross-Modal Hashing}
	Many CMH methods are based on the assumption that data of different modalities have one-to-one correspondence. However, in real applications, data misalignment and missing may occur, making fully paired data unobtainable. One realistic challenge is to address the UCMR problem, which is less explored by researchers. To our knowledge, there are several UCMR-oriented works. Generalized Semantic Preserving Hashing (GSPH) \cite{GSPH} seeks to factorize the similarity matrix into the dot product of two binary matrices, thus embedding semantic information into the to-be-learned binary codes, and applies kernel logistic regression to obtain modality-specific hash functions. Unpaired Cross-Modal Hashing (UCMH) \cite{UCMH} utilizes matrix factorization to generate modality-specific latent representations. To preserve semantic similarities, UCMH constructs the Laplacian matrices with labels, which is time-consuming. Robust Unsupervised Cross-Modal Hashing (RUCMH) \cite{RUCMH} considers the data reconstruction relations in the subspace to bridge the heterogeneous gaps between modalities, and adopts $\ell_{2,1}$ norm to learn hash functions, which can alleviate the model sensitivity to data noise. Nonetheless, RUCMH is an unsupervised method, which ignores embedding the label information into the hash functions. Enhanced Discrete Multi-modal Hashing (EDMH) \cite{EDMH} and Flexible Multi-modal Hashing (FMH) \cite{FMH} both adopt an asymmetric strategy to keep similarities. However, they do not consider label discrimination, with the distances between different classes the same. Besides, most of these methods directly adopt linear regression to obtain modality-specific hash functions without considering the rigid binary problem, which impairs the robustness of hash functions.
	\section{PROPOSED METHOD}
	\subsection{Notations}
	In this paper, matrices are denoted by boldface uppercase letters, and vectors boldface lowercase letters. For a matrix $\mathbf{M}$, $M_{ij}$ is the entry of the $i$th row and $j$th column. Given $m$ modalities, $\mathbf{X}^{(i)} = [\boldsymbol{x}^{(i)}_1, \boldsymbol{x}^{(i)}_2,\ldots,\boldsymbol{x}^{(i)}_{n_i}]\in\mathbb{R}^{d_{i}\times n_i}$, is the training data from the $i$th modality, where $d_{i}$ is the feature dimensionality and $n_i$ is the size of data. We assume that the samples are zero-centered in each modality, i.e., $\sum_{j=1}^{n_{i}} \boldsymbol{x}_{j}^{(i)}=\mathbf{0}, i=1, \ldots, m$. $\mathbf{L}^{(i)}=[\boldsymbol{l}^{(i)}_{1}, \boldsymbol{l}^{(i)}_{2}, \ldots, \boldsymbol{l}^{(i)}_{n_i}] \in\{0,1\}^{c \times n_i}$ represents the label matrix of the $i$th modality, where $c$ is the number of classes. $L^{(i)}_{jk} = 1$ if the $k$th sample from the $i$th modality is associated with the $j$th category and $0$ otherwise. Hash code matrices are denoted by $\mathbf{B}^{(i)}\in\{-1,1\}^{r\times n_i}$, where $r$ is the code length. $\left\{\mathbf{F}^{(i)}\right\}_{i=1}^{m}$ are the modality-specific hash functions that connect the original data space to the Hamming space. $\operatorname{tr}(\cdot)$ is the trace operation. $\operatorname{sgn}(\cdot)$ is the sign function that returns 1 if the input is positive and -1 otherwise. $\mathbf{I}_n$ is an $n\times n$ identity matrix, and $\mathbf{0}_n$ is an $n$-dimensional column vector with all items being 0.
	\subsection{Hash Code Learning}
	\subsubsection{Adaptive Marginalized Regression}
\begin{figure}[t]
		\center
		\includegraphics[width=8cm,height=4cm]{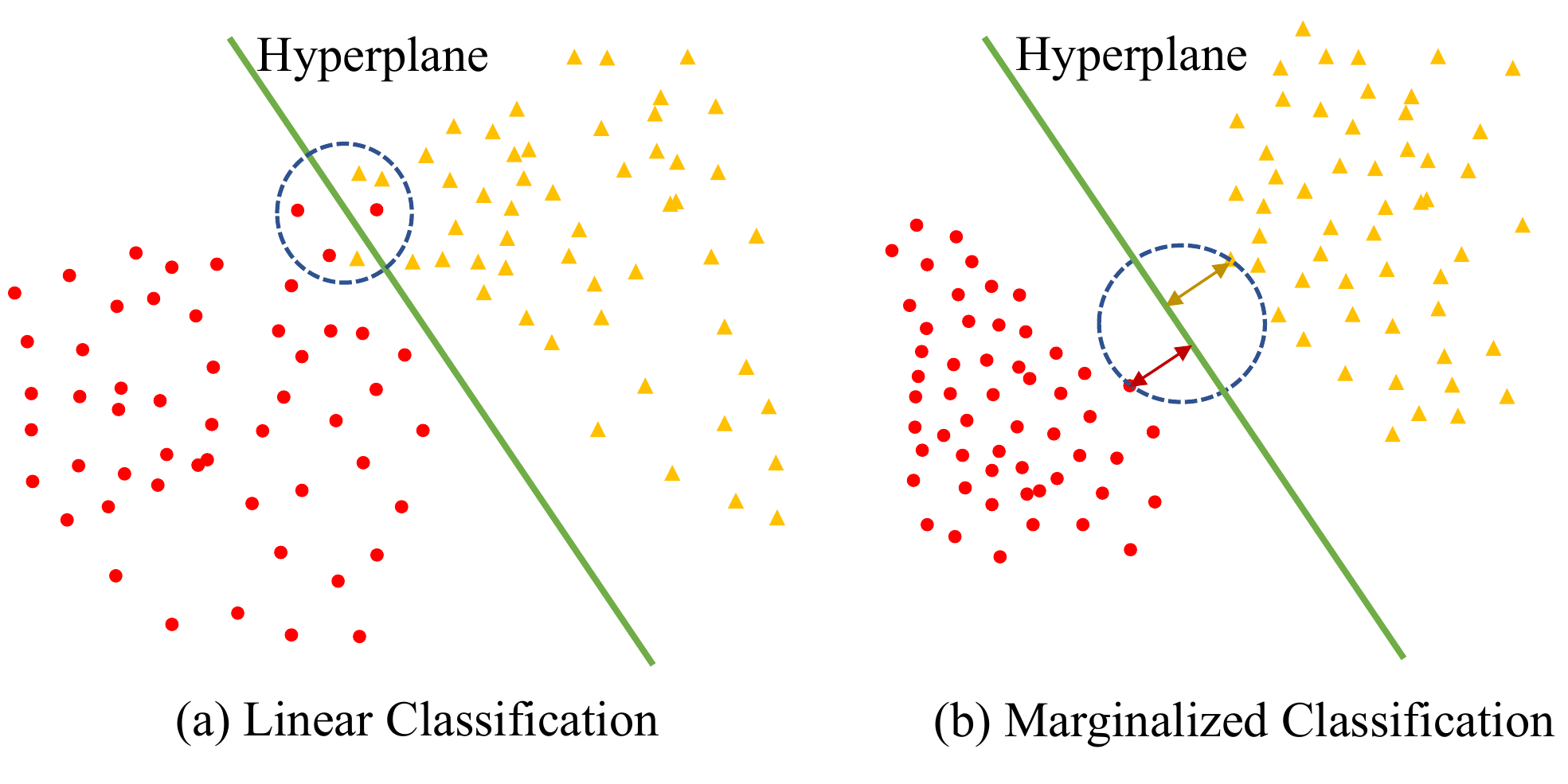}
		
		\caption{Illustration of marginalized features. Margin factors pull data of different categories away from each other so that they are easy to be classified.}
		\label{margin}
	\end{figure}

	To eliminate the heterogeneous gap across modalities and leverage the semantic information, most previous methods learn a common latent representation, and construct a linear regression between the latent feature space and the label space, which can be described as follows:  
	\begin{equation}
		\min _{\mathbf{P}, \mathbf{V}}\|\mathbf{L} - \mathbf{P V}\|_{F}^{2},
		\label{11}
	\end{equation}
	where $\mathbf{P} \in \mathbb{R}^{c \times r}$ is a projection matrix, and $\mathbf{V}$ is the shared latent representation. Eq. (\ref{11}) is a traditional regression model, and it is expected to generate a shared latent representation that can be correctly classified by a simple linear regression model. However, in the unpaired setting, the shared latent representation cannot be obtained. A straightforward solution is to construct multiple regression models by learning modality-specific representations, which can be written as:
	\begin{equation}
		\min _{\mathbf{P}^{(i)}\mathbf{V}^{(i)}}\sum_{i=1}^{m}\|\mathbf{L}^{(i)} - \mathbf{P}^{(i)}\mathbf{V}^{(i)}\|_{F}^{2}
		\label{111},
	\end{equation} 
	where $m$ is the number of modalities. The above objective function aims to generate semantic-preserving latent representations for each modality by establishing linear transform relationships between the representation and the strict zero-one label matrix. However, the zero-one label matrix is too rigid to learn an effective projection matrix for fitting class labels, which may degrade the representation quality and limit the flexibility. Besides, from the perspective of robust classification, it is expected that the distances between data points in different classes are as large as possible after transformation, as shown in Fig. \ref{margin}. Inspired by \cite{DLSR, PRDR}, to alleviate this problem, we introduce adaptive margin factors to refine the regression targets  so that the margins of different classes are enlarged.
	
	For its purpose, we introduce an index matrix $\mathbf{R}^{(i)}\in\mathbb{R}^{c\times n_i}$ for each modality that satisfies:
	\begin{equation}
		R_{j k}^{(i)}= \begin{cases}+1, & \text { if } \mathbf{L}^{(i)}_{j k}=1 \\ -1, & \text { otherwise }\end{cases},
	\end{equation}
	where `+1' denotes the positive direction and `-1' the negative direction. The margin matrix $\mathbf{E}^{(i)} = \left\{\varepsilon_{j k} \geq 0\right\} \in \mathbb{R}^{c \times n_i}$ is used to perform the marginalization operation on every element of $\mathbf{L}^{(i)}$. For example, $\boldsymbol{l}^{(i)} = [0,1,0,1]$ is marginalized into $\boldsymbol{l}^{(i)} = [-\varepsilon_1,1+\varepsilon_2,-\varepsilon_3,1+\varepsilon_4]$, where $\varepsilon_1, \varepsilon_2, \varepsilon_3$ and $\varepsilon_4$ are adaptive margin factors. Eq. (\ref{111}) can be rewritten as:
	\begin{equation}
		\begin{aligned}
			&\min _{\mathbf{P}^{(i)}, \mathbf{V}^{(i)}, \mathbf{E}^{(i)}}\sum_{i=1}^{m}\|\mathbf{L}^{(i)} + \mathbf{R}^{(i)}\odot\mathbf{E}^{(i)} - \mathbf{P}^{(i)}\mathbf{V}^{(i)}\|_{F}^{2} \\
			&\text { s.t. }  \mathbf{E}^{(i)} \geq \mathbf{0},
		\end{aligned}
		\label{first}
	\end{equation}
	where $\odot$ is the Hadamard product operator. In this way, the latent representations can be regressed with adaptive marginalized targets, which enlarges the distances between positive and negative classes and enhances the discrimination of features. 
	\subsubsection{Semantic Similarity Embedding}
	With the supervision of refined regression targets, Eq. (\ref{first}) can embed the semantic information into latent representations to some extent. However, it mainly focuses on the inter-class separability for classification, while the intra-class compactness is not fully explored. For effective representation learning, it is desired that the semantically similar data points should also be close in the latent subspace. In this paper, we use the pairwise semantic similarities to improve the intra-class compactness. The semantic similarity $S_{uk}^{(ij)}$ between two samples, $\boldsymbol{x}_u^{(i)}$ and $\boldsymbol{x}_k^{(j)}$, is defined as follows:
	\begin{equation}
		S_{uk}^{(ij)} = \frac{\mathbf{L}_{* u}^{(i)T} \mathbf{L}_{* k}^{(j)}}{\left\|\mathbf{L}^{(i)}_{* u}\right\|_{2} \cdot\left\|\mathbf{L}^{(j)}_{* k}\right\|_{2}},
	\end{equation}
	where the similarity is measured with the cosine distance. To have a compact form, we define the column-normalized label matrix as:
	\begin{equation}
		\widetilde{\mathbf{L}}^{(i)} = \left[\frac{\mathbf{L}_{* 1}^{(i)}}{\left\|\mathbf{L}_{* 1}^{(i)}\right\|_{2}}, \frac{\mathbf{L}_{* 2}^{(i)}}{\left\|\mathbf{L}_{* 2}^{(i)}\right\|_{2}}, \ldots, \frac{\mathbf{L}_{* n_i}^{(i)}}{\left\|\mathbf{L}_{* n_i}^{(i)}\right\|_{2}} \right].
	\end{equation}
	Thus, the pairwise similarity matrix between two modalities can be obtained by $\mathbf{S}^{(ij)} = \widetilde{\mathbf{L}}^{(i)T}\widetilde{\mathbf{L}}^{(j)}$. $\mathbf{S}^{(ij)}$ is not explicitly computed during the optimization to reduce computational costs. It is noted that $S^{(ij)}_{uk} \in [0, 1]$ characterizes the fine-grained semantic similarities among samples, especially for multi-label data. Two samples that have more shared labels will have larger $S^{(ij)}_{uk}$. To improve the semantic information of modality-specific representations, we preserve the inter-modal similarities into it by solving the following problem:
	\begin{equation}
		\min _{\mathbf{V}^{(i)}}\sum_{i\neq j}\|\mathbf{V}^{(j)T}\mathbf{V}^{(i)} - r\mathbf{S}^{(ji)}\|_{F}^{2}
		\label{second}.
	\end{equation}
	Apart from inter-modal semantic similarity preservation, the intra-modal similarities should also be maintained:
	\begin{equation}
		\min _{\mathbf{V}^{(i)}}\sum_{i = 1}^{m}\|\mathbf{V}^{(i)T}\mathbf{V}^{(i)} - r\mathbf{S}^{(ii)}\|_{F}^{2}
		\label{3}.
	\end{equation}
	However, the symmetric dot product in Eq. (\ref{3}) makes the optimization difficult. To reduce the quantization error and accelerate the optimization, we adopt an asymmetric strategy:
	\begin{equation}
		\min _{\mathbf{V}^{(i)},\mathbf{B}^{(i)}}\sum_{i = 1}^{m}(\|\mathbf{B}^{(i)T}\mathbf{V}^{(i)} - r\mathbf{S}^{(ii)}\|_{F}^{2} + \|\mathbf{B}^{(i)} - \mathbf{V}^{(i)}\|_{F}^{2})
		\label{third}.
	\end{equation}
	In this way, the hash codes $\mathbf{B}$ can be obtained by approximating $\mathbf{V}$, and the semantic information in $\mathbf{V}$ can be transferred into $\mathbf{B}$.
	\subsubsection{Overall Objective Function}
	Combine Eq. (\ref{first}), Eq. (\ref{second}) and Eq. (\ref{third}), the overall objective function is
	\begin{equation}
		\begin{aligned}
			&\min \sum_{i=1}^{m}(\|\mathbf{L}^{(i)} + \mathbf{R}^{(i)}\odot\mathbf{E}^{(i)} - \mathbf{P}^{(i)}\mathbf{V}^{(i)}\|_{F}^{2}+\eta\|\mathbf{B}^{(i)} - \mathbf{V}^{(i)}\|_{F}^{2}\\
			& + \lambda\|\mathbf{B}^{(i)T}\mathbf{V}^{(i)} - r\mathbf{S}^{(ii)}\|_{F}^{2})  + \beta\sum_{i\neq j}\|\mathbf{V}^{(j)T}\mathbf{V}^{(i)} - r\mathbf{S}^{(ji)}\|_{F}^{2} \\
			&\text { s.t. }\left\{\begin{array}{l}
				\mathbf{E}^{(i)} \geq \mathbf{0}, \mathbf{B}^{(i)}\in\{-1,1\}^{r\times n_i} ; \\
				\mathbf{V}^{(i)}\mathbf{V}^{(i)T}=n_i\mathbf{I}_{r}, \mathbf{V}^{(i)} \mathbf{1}_{n_i}=\mathbf{0}_{r}.
			\end{array}\right.
		\end{aligned}
		\label{overall}
	\end{equation}
	In Eq. (\ref{overall}), two widely used constraints,  $\mathbf{V}^{(i)}\mathbf{V}^{(i)T}=n_i\mathbf{I}_{r}, \mathbf{V}^{(i)} \mathbf{1}_{n_i}=\mathbf{0}_{r}$, are imposed for bit decorrelation and balance. By optimizing the above problem, $\mathbf{V}^{(i)}$ containing semantic information can be obtained by utilizing labels,  thus generating discriminative hash codes for all modalities.
	\subsection{Optimization}
	To address Eq. (\ref{overall}), we develop an alternate discrete optimization method by updating one variable with others fixed in each step.
	
	\textbf{Update $\mathbf{P}^{(i)}$.} Remove the irrelevant terms, and the subproblem w.r.t $\mathbf{P}^{(i)}$ is
	\begin{equation}
		\min_{\mathbf{P}^{(i)}} \|\mathbf{L}^{(i)} + \mathbf{R}^{(i)}\odot\mathbf{E}^{(i)} - \mathbf{P}^{(i)}\mathbf{V}^{(i)}\|_{F}^{2}.
	\end{equation}
	With the constraint $\mathbf{V}^{(i)}\mathbf{V}^{(i)T}=n_i\mathbf{I}_{r}$, the closed-form solution for $\mathbf{P}^{(i)}$ can be obtained by setting the derivative w.r.t $\mathbf{P}^{(i)}$ to zero:
	\begin{equation}
		\mathbf{P}^{(i)} = (\mathbf{L}^{(i)} + \mathbf{R}^{(i)}\odot\mathbf{E}^{(i)})\mathbf{V}^{(i)T}/n_i
		\label{P}
	\end{equation}
	
	\textbf{Update $\mathbf{V}^{(i)}$.} Constrained by $\mathbf{V}^{(i)}\mathbf{V}^{(i)T}=n_i\mathbf{I}_{r}$, Eq. (\ref{overall}) can be transformed into the following problem:
	\begin{equation}
		\begin{aligned}
			&\max _{\mathbf{V}^{(i)}} \operatorname{Tr}\left(\mathbf{Z} \mathbf{V}^{(i)T}\right) \\
			&\text { s.t. } \mathbf{V}^{(i)} \mathbf{V}^{(i)T}=n_i \mathbf{I}_{r}, \mathbf{V}^{(i)} \mathbf{1}_{n_i}=\mathbf{0}_{r},
		\end{aligned}
	\end{equation}
	where $\mathbf{Z} = \mathbf{P}^{(i)T}(\mathbf{L}^{(i)} + \mathbf{R}^{(i)}\odot\mathbf{E}^{(i)}) + \eta\mathbf{B}^{(i)} + \lambda r\mathbf{B}^{(i)}\widetilde{\mathbf{L}}^{(i)T}\widetilde{\mathbf{L}}^{(i)} + \beta r\sum_{i\neq j}\mathbf{V}^{(j)}\widetilde{\mathbf{L}}^{(j)T}\widetilde{\mathbf{L}}^{(i)}$. It can be optimized with the following theorem.
	\newtheorem{thm}{\bf Theorem}[section]
	\begin{thm}\label{V}
		\cite{theorem1} \textit{Given the optimization problem
			\begin{equation}
				\max _{\mathbf{V}} \operatorname{Tr}\left(\mathbf{Z} \mathbf{V}^{T}\right) \text { s.t. } \mathbf{V} \mathbf{V}^{T}=n \mathbf{I}_{r}, \mathbf{V} \mathbf{1}_{n}=\mathbf{0}_{r},
			\end{equation}
			the eigendecomposition is performed on $\mathbf{Z}\mathbf{J}\mathbf{Z}^T$ with $\mathbf{J}$ defined as $\mathbf{J}=\mathbf{I}_{n}-\frac{1}{n} \mathbf{1}_{n} \mathbf{1}_{n}^{T}$: 
			\begin{equation}
				\mathbf{Z J} \mathbf{Z}^{T}=[\mathbf{N}, \widetilde{\mathbf{N}}]\left[\begin{array}{cc}
					\boldsymbol{\nabla} & \mathbf{0} \\
					\mathbf{0} & \mathbf{0}
				\end{array}\right][\mathbf{N}, \widetilde{\mathbf{N}}]^{T},
			\end{equation}
			where $\boldsymbol{\nabla} \in \mathbb{R}^{r^{\prime} \times r^{\prime}}$ is the diagonal eigenvalue matrix and $\mathbf{N} \in \mathbb{R}^{r \times r^{\prime}}$ is comprised of the corresponding eigenvectors, with $r^{\prime}$ the rank of $\mathbf{Z J} \mathbf{Z}^{T}$. Then Gram-Schmidt process is applied on $\widetilde{\mathbf{N}}$ to obtain the orthogonal matrix $\overline{\mathbf{N}} \in \mathbb{R}^{r \times\left(r-r^{\prime}\right)}$. Define $\mathbf{K}=\mathbf{J} \mathbf{Z}^{T} \mathbf{N} \boldsymbol{\nabla}^{-1 / 2}$ and a random orthogonal matrix is denoted by $\overline{\mathbf{K}} \in \mathbb{R}^{n \times (r - r^{\prime})}$. The optimal solution for $\mathbf{V}$ is given by $\widehat{\mathbf{V}}=\sqrt{n}[\mathbf{N}, \overline{\mathbf{N}}][\mathbf{K}, \overline{\mathbf{K}}]^{T}$. if $r=r^{\prime}$, $\overline{\mathbf{N}}, \widetilde{\mathbf{N}}$ and $\mathbf{K}$ are empty, which means $\widehat{\mathbf{V}}$ is unique.
		}
		
	\end{thm}
	
	\textbf{Update $\mathbf{B}^{(i)}$.} With other variables fixed, the subproblem for $\mathbf{B}^{(i)}$ is 
	\begin{equation}
		\begin{aligned}
			&\min_{\mathbf{B}^{(i)}}\eta\|\mathbf{B}^{(i)} - \mathbf{V}^{(i)}\|_{F}^{2}+ \lambda\|\mathbf{B}^{(i)T}\mathbf{V}^{(i)} - r\mathbf{S}^{(ii)}\|_{F}^{2}\\
			&\text { s.t. } \mathbf{B}^{(i)}\in\{-1,1\}^{r\times n_i}.
		\end{aligned}
		\label{q3}
	\end{equation}
	With some algebraic manipulation, (\ref{q3}) can be transformed into:
	\begin{equation}
		\begin{aligned}
			&\max_{\mathbf{B}^{(i)}}\operatorname{Tr}((\eta\mathbf{V}^{(i)} + \lambda r\mathbf{V}^{(i)}\widetilde{\mathbf{L}}^{(i)T}\widetilde{\mathbf{L}}^{(i)})\mathbf{B}^{(i)T})\\
			&\text { s.t. } \mathbf{B}^{(i)}\in\{-1,1\}^{r\times n_i}.
		\end{aligned}
	\end{equation}
	Apparently, the optimal solution for $\mathbf{B}^{(i)}$ is given by
	\begin{equation}
		\mathbf{B}^{(i)} = \operatorname{sgn}(\eta\mathbf{V}^{(i)} + \lambda r\mathbf{V}^{(i)}\widetilde{\mathbf{L}}^{(i)T}\widetilde{\mathbf{L}}^{(i)}).
		\label{B}
	\end{equation}
	
	\textbf{Update $\mathbf{E}^{(i)}$.} Remove the terms without $\mathbf{E}^{(i)}$, and the subproblem is 
	\begin{equation}
		\begin{aligned}
			&\min_{\mathbf{E}^{(i)}}\|\mathbf{H} - \mathbf{R}^{(i)}\odot\mathbf{E}^{(i)} \|_{F}^{2} \\
			&\text { s.t. } 	\mathbf{E}^{(i)} \geq \mathbf{0},
		\end{aligned}
		\label{q4}
	\end{equation}
	where $\mathbf{H} = \mathbf{P}^{(i)}\mathbf{V}^{(i)} - \mathbf{L}^{(i)}.$ According to the property of Frobenius norm, Eq. (\ref{q4}) can be decoupled into $c\times n_i$ subproblems. For an element of $\mathbf{H}$, the subproblem is 
	\begin{equation}
		\min _{E_{u k}}\left(H_{u k}-R_{u k}^{(i)} E_{u k}^{(i)}\right)^{2}, \quad \text { s.t. } \quad E_{u k}^{(i)} \geq 0.
		\label{q41}
	\end{equation}
	The optimal solution of Eq. (\ref{q41}) is given by $E_{u k}^{(i)}=\max \left(R_{u k}^{(i)} H_{u k}^{(i)}, 0\right)$. Similarly, the optimal solution of Eq. (\ref{q4}) can be obtained by
	\begin{equation}
		\mathbf{E}^{(i)}=\max \left(\mathbf{R}^{(i)}\odot\mathbf{H}^{(i)}, \mathbf{0}\right).
		\label{E}
	\end{equation}
	
	By optimizing Eq. (\ref{overall}) with the above steps w.r.t $\mathbf{P}^{(i)}$, $\mathbf{V}^{(i)}$, $\mathbf{B}^{(i)}$ and $\mathbf{E}^{(i)}$, Eq. (\ref{overall}) can gradually converge to a stable value or reach the maximum iteration. 
	\subsection{Hash Functions Learning}
	With the optimal hash codes generated from the first step, AMSH then learns the modality-specific hash functions to perform CMR with new queries. Previous methods usually adopt the traditional least squares regression for hash functions learning, which can be described as:
	\begin{equation}
		\min _{\mathbf{F}^{(i)}}\sum_{i = 1}^{m}\|\mathbf{B}^{(i)}-\mathbf{F}^{(i)} \mathbf{X}^{(i)}\|_{F}^{2},
		\label{step2_ori}
	\end{equation}
	where $\mathbf{F}^{(i)}\in\mathbb{R}^{r\times d_i}$ is the hash function. Nonetheless, the binary codes in Eq. (\ref{step2_ori}) are also too strict for effective regression, and makes the hash functions sensitive to noise. Thus, we introduce adaptive margin matrices, and the hash functions can be obtained by minimizing the following problem:
	\begin{equation}
		\begin{aligned}
			&\min_{\mathbf{F}^{(i)},\mathbf{M}^{(i)}} \sum_{i = 1}^{m}\|\mathbf{B}^{(i)} + \mathbf{B}^{(i)}\odot\mathbf{M}^{(i)} - \mathbf{F}^{(i)}\mathbf{X}^{(i)}\|_F^{2}\\
			&\text { s.t. } \mathbf{M}^{(i)} \geq \mathbf{0}.
		\end{aligned}
	\end{equation}
	Intuitively, the margin matrix makes the regression target more flexible and robust by enlarging the distances between positive and negative bits, which makes AMSH more robust to noise and eases the learning of hash functions. 
	
	The kernel method is widely adopted to capture the nonlinear data structure. We map the original data from the $i$th modality to the kernel space with the following function:
	\begin{equation}
		\phi(\boldsymbol{x})=\left[\exp \left(\frac{-\left\|\boldsymbol{x}-\boldsymbol{x}_{1}\right\|}{2 \delta^{2}}\right), \ldots, \exp \left(\frac{-\left\|\boldsymbol{x}-\boldsymbol{x}_{k}\right\|}{2 \delta^{2}}\right)\right]^{T},
	\end{equation}
	where $\boldsymbol{x}_{1},\boldsymbol{x}_{2}...\boldsymbol{x}_{k}$ are $k$ randomly selected samples and $\delta$ is the bandwidth defined as the mean distance between other training samples and the $k$ samples. The kernel function is applied to all modalities, and the objective function is
	\begin{equation}
		\begin{aligned}
			&\min_{\mathbf{F}^{(i)},\mathbf{M}^{(i)}} \sum_{i = 1}^{m}\|\mathbf{B}^{(i)} + \mathbf{B}^{(i)}\odot\mathbf{M}^{(i)} - \mathbf{F}^{(i)}\phi(\mathbf{X}^{(i)})\|_F^{2}\\
			&\text { s.t. } \mathbf{M}^{(i)} \geq \mathbf{0}.
		\end{aligned}
		\label{step2}
	\end{equation}
	The optimal $\mathbf{F}^{(i)}$ and $\mathbf{M}^{(i)}$ can be efficiently obtained by performing the alternate optimization scheme, similar to the hash code learning. 
	
	\textbf{Update $\mathbf{F}^{(i)}$.} Set the derivative of (\ref{step2}) w.r.t $\mathbf{F}^{(i)}$ to zero, and the optimal solution can be computed by
	\begin{equation}
		\mathbf{F}^{(i)} = (\mathbf{B}^{(i)} + \mathbf{B}^{(i)}\odot\mathbf{M}^{(i)})\phi(\mathbf{X}^{(i)})^{T}(\phi(\mathbf{X}^{(i)})\phi(\mathbf{X}^{(i)})^{T})^{-1}.
		\label{F}
	\end{equation} 
	
	\textbf{Update $\mathbf{M}^{(i)}$.} Similar to optimizing $\mathbf{E}^{(i)}$ in the hash code learning, the optimal solution can be obtained by
	\begin{equation}
		\mathbf{M}^{(i)}=\max \left(\mathbf{B}^{(i)}\odot\mathbf{C}^{(i)}, \mathbf{0}\right),
		\label{M}
	\end{equation}
	where $\mathbf{C}^{(i)} = \mathbf{F}^{(i)}\phi(\mathbf{X}^{(i)}) - \mathbf{B}^{(i)}$. 
	\begin{algorithm}[t]
		\caption{Optimization for AMSH}\label{optimization}
		\KwIn{Training data $\{\mathbf{X}^{(i)}\}_{i = 1}^{m}$, code length $r$, label matrices $\{\mathbf{L}^{(i)}\}_{i = 1}^{m}$, parameters  $\eta$, $\lambda$, $\beta$.}
		\KwOut{Hash functions $\{\mathbf{F}^{(i)}\}_{i = 1}^{m}$ and hash codes $\{\mathbf{B}^{(i)}\}_{i = 1}^{m}$.}
		Initialize $\mathbf{V}^{(i)}, \mathbf{E}^{(i)}, \mathbf{P}^{(i)}$, $\mathbf{B}^{(i)}, \mathbf{F}^{(i)}, \mathbf{M}^{(i)}$ randomly\;
		\While{not converged}
		{Update $\mathbf{P}^{(i)}$ with Eq. (\ref{P})\;
			Update $\mathbf{V}^{(i)}$ according to theorem \ref{V}\;
			Update $\mathbf{B}^{(i)}$ with Eq. (\ref{B})\;
			Update $\mathbf{E}^{(i)}$ with Eq. (\ref{E})\;
		}
		Map the training data $\{\mathbf{X}^{(i)}\}_{i = 1}^{m}$ into the kernel space\;
		\While{not converged}
		{Update $\mathbf{F}^{(i)}$ with Eq. (\ref{F})\;
			Update $\mathbf{M}^{(i)}$ with Eq. (\ref{M})\;
		}
	\end{algorithm}
	
	After obtaining the modality-specific hash functions, for a new query from the $i$th modality, $\boldsymbol{x}^{(i)}$, its hash code can be determined by $\boldsymbol{b}^{(i)} = \operatorname{sgn}(\mathbf{F}^{(i)} (\phi(\boldsymbol{x}^{(i)}))).$ The training procedure is summarized in Algorithm \ref{optimization}.
	\subsection{Complexity Analysis}
	For simplicity, we assume that the numbers of samples of all modality is the same, i.e., $n_1 = n_2 = \ldots = n_m = n$. As can be observed from Algorithm \ref{optimization}, the optimization is designed in a two-step fashion. Concretely, for hash code learning, the time complexity includes $O\left(\tau\left(c n+c n r\right)\right)$ for updating $\mathbf{P}^{(i)}$, $O\left(\tau\left(r^3+cn+mrcn\right)\right)$ for updating $\mathbf{V}^{(i)}$, $O\left(\tau\left(rcn+rn\right)\right)$ for updating $\mathbf{B}^{(i)}$ and $O\left(\tau\left(rcn+cn\right)\right)$ for updating $\mathbf{E}^{(i)}$, where $\tau$ is the number of iteration. To obtain hash functions, updating $\mathbf{M}^{(i)}$ requires $O\left(\tau\left((rd_i+r)n\right)\right)$, and obtaining $\mathbf{F}^{(i)}$ requires $O\left(\tau\left((r+rd_i+d_i^2)n+d_i^3+d_i^2r\right)\right)$, where $d_i$ is the kernelized feature length of the $i$th modality. With $r, c, d_i, m \ll n$, the complexity required for optimizing AMSH is linear to the size of datasets. Thus, AMSH is an efficient algorithm and is scalable to large-scale datasets.
	\subsection{Convergence Analysis}
	In this subsection, the convergence property of AMSH is discussed. For hash code learning, we denote the objective function value of the $t$th iteration as $\mathcal{J}\left(\mathbf{P}^{t}, \mathbf{V}^{t}, \mathbf{B}^{t}, \mathbf{E}^{t}\right)$. For each variable to be minimized, there is a closed-form solution. Thus, for the $(t+1)$th iteration, it can be derived that $\mathcal{J}\left(\mathbf{P}^{t}, \mathbf{V}^{t}, \mathbf{B}^{t}, \mathbf{E}^{t}\right)\geq\mathcal{J}\left(\mathbf{P}^{t+1}, \mathbf{V}^{t}, \mathbf{B}^{t}, \mathbf{E}^{t}\right)\geq\mathcal{J}\left(\mathbf{P}^{t+1}, \mathbf{V}^{t+1}, \mathbf{B}^{t}, \mathbf{E}^{t}\right)\geq \mathcal{J}\left(\mathbf{P}^{t+1}, \mathbf{V}^{t+1}, \mathbf{B}^{t+1}, \mathbf{E}^{t}\right)\geq\mathcal{J}\left(\mathbf{P}^{t+1}, \mathbf{V}^{t+1}, \mathbf{B}^{t+1}, \mathbf{E}^{t+1}\right)$, which means the objective function is monotonously decreasing with each iteration. Besides, the property of norms determines the non-negativity of the objective function. According to the bounded monotone convergence theorem \cite{convergence}, the objective function is guaranteed to converge to a local optimal solution. For hash function learning, similar analysis can be applied to demonstrate the convergence. To experimentally explain the convergence property of AMSH, we provide the changing curve of the objective function during training in Section IV-G. 
	\section{EXPERIMENT}
	In this section, experiments are conducted on several benchmark datasets to demonstrate the superiority of AMSH.
	\subsection{Datasets}
	Three datasets are used to conduct experiments, including MIRFlickr-25K \cite{FLI}, NUS-WIDE \cite{NUS} and IAPR TC-12 \cite{IRC}, which are widely adopted for CMR evaluation. TABLE \ref{statistic} presents the basic information of the three datasets.
	\begin{table}[t]
		\caption{Dataset statistics.}
		\centering
		\setlength{\tabcolsep}{4mm}{
			\begin{tabular}{cccc}
				\toprule  
				Statistics& MIRFlickr-25K& NUS-WIDE& IAPR TC-12\\
				\hline  
				Total & 16,738& 186,577& 20,000\\
				Query & 836& 2,000& 2,000\\
				Training& 15,902&184,577& 18,000\\
				Image feature& 150&500&512\\
				Text feature& 500&1000&2912\\
				Category &24&10&255\\
				\toprule 
			\end{tabular}
		}
		
		\label{statistic}
	\end{table}
	
	\textbf{MIRFlickr-25K} originally consists of 25,000 image-text pairs, collected from Flickr website, with each one annotated with at least one of the 24 semantic labels. Each image is represented with a 150-D edge histogram feature vector, and each text a 500-D vector. We select textual tags that appear at least 20 times, and removes the samples without textual tags. Finally, the dataset contains 16,738 samples. Following \cite{DCH}, we randomly select 5\% as the query set, and the rest 95\% as the training set.
	
	\textbf{NUS-WIDE} is a large database with 269,684 instances associated with 81 semantic concepts. Following \cite{SRLCH} and \cite{SCRATCH}, we select 186,577 instances with 10 most frequent labels. The images and texts are represented as 500-D SIFT feature vectors and 1000-D binary tagging vectors, respectively. 2,000 samples are randomly selected as the query set, and the rest form the training set.
	
	\textbf{IAPR TC-12} is comprised of 20,000 worldwide images with related textual descriptions. Each instance is classified as at least one of the 255 semantic labels. An image and a text are denoted by a 512-D GIST feature vector and a 2,912-dimensional BOW vector, respectively. 2,000 image-text pairs are randomly selected as the query set, and the rest as the training set.
	\subsection{Baselines and Implementation}
	To demonstrate the effectiveness of AMSH, we evaluate it in two different settings, i.e., the fully paired scenario and the unpaired scenario. For the fully paired scenario, We compare our AMSH with several state-of-the-art cross-modal hashing methods, including CMFH \cite{CMFH}, FSH \cite{FSH}, SMFH \cite{SMFH}, CRE \cite{CRE}, DCH \cite{DCH}, SCRATCH \cite{SCRATCH}, SRLCH \cite{SRLCH}, BATCH \cite{BATCH}. In these methods, CMFH, FSH and CRE are unsupervised ones and the rest are supervised ones. These baselines are implemented with the codes provided by the authors, and the parameters of these baselines are carefully tuned according to the original paper to reach their best performance. For the unpaired scenario, SAPMH \cite{UAPMH}, GSPH \cite{GSPH}, UCMH \cite{UCMH}, RUCMH \cite{RUCMH}, EDMH \cite{EDMH} and FMH \cite{FMH} are adopted for comparison. Among these baselines, RUCMH is an unsupervised method and the others are supervised ones. For AMSH, the retrieval performance is relatively insensitive to the model parameters in appropriate ranges, which will be shown in Section IV-F, and we set $\eta = 1$, $\lambda = 1e-3$ and $\beta = 1e-3$ on all datasets. The number of instance anchors for kernel feature learning is set to 1500 for both image modality and textual modality. The maximum iteration number $\tau$ is set to 15. All the experiments are conducted on a personal computer with MATLAB R2021b, Win10 system, Intel i7-10875H CPU @ 2.3GHz, 16GB RAM.
	\begin{table*}[t]
		\centering
		\caption{The MAP results of AMSH and baselines on MIRFlickr-25K, NUS-WIDE and IAPR TC-12 with paired data.}
		\renewcommand{\arraystretch}{1.1}
		\begin{tabular}{p{1.5cm}<{\centering}p{1.8cm} p{0.6cm}<{\centering} p{0.6cm}<{\centering} p{0.6cm}<{\centering} p{0.7cm}<{\centering} p{0.1cm}<{\centering} p{0.6cm}<{\centering} p{0.6cm}<{\centering} p{0.6cm}<{\centering} p{0.7cm}<{\centering}p{0.1cm}<{\centering}
				p{0.7cm}<{\centering} p{0.7cm}<{\centering} p{0.7cm}<{\centering}
				p{0.7cm}<{\centering}}
			\toprule
			\multicolumn{1}{c}{\multirow{1}[4]{*}{Task}} & \multicolumn{1}{l}{\multirow{1}[4]{*}{Method}} & \multicolumn{4}{c}{MIRFlickr-25K}  &       & \multicolumn{4}{c}{NUS-WIDE} &       & \multicolumn{4}{c}{IAPR TC-12} \\
			\cmidrule{3-6}\cmidrule{8-11}\cmidrule{13-16}
			&   &  16 b   & 32 b   & 64 b   & 128 b   &      &  16 b    & 32 b    & 64 b    & 128 b & &  16 b    & 32 b   & 64 b    & 128 b \\
			\midrule
			\multirow{1}[26]{*}{$I\rightarrow T$}
			& CMFH \cite{CMFH} & 0.5838& 0.5789& 0.5825& 0.5858&&  0.3478&	0.3450&	0.3447&	0.3516&&0.2966&	0.2963&	0.2965&	0.2977\\
			& FSH \cite{FSH} &0.6185&	0.6254&	0.6288&	0.6280&&0.4735	&0.4771&	0.4802&	0.4951&&0.3887	&0.3969	&0.4087&	0.4149\\
			&CRE \cite{CRE} &0.6029&	0.6122&	0.6244&	0.6213&&0.4606	&0.4774	&0.4923&	0.4833&&0.3930	&0.4141	&0.4234	&0.4154\\
			& DCH \cite{DCH}&0.6976&	0.7171&	0.7226&	0.7318 &&0.5977	&0.6110&	0.6388	&0.6537&&0.4369	&0.4550	&0.4592	&0.4803\\
			& SMFH \cite{SMFH}&0.5641&	0.5556&	0.5593&	0.5628&&0.4178	&0.4248&	0.4270&	0.4313&&0.3341	&0.3365	&0.3398	&0.3440\\
			&SCRATCH \cite{SCRATCH}& 0.7076&	0.7088&	0.7207&	0.7251&&0.6202&	0.6193&	0.6356&	0.6287&&0.4490&	0.4613&	0.4714&	0.4853\\
			& SRLCH \cite{SRLCH} & 0.6381&	0.6474&	0.6874&	0.6809&&0.5918&	0.6185&	0.6330	&0.6470&&0.3560&	0.3520&	0.3823&	0.4065\\
			&BATCH \cite{BATCH} &0.7406&	0.7468&	0.7504&	0.7519&&0.6464&	0.6490&	0.6593&	0.6659&&0.4807&	0.5084&	0.5281&	0.5388\\
			&AMSH&\textbf{0.7455}&\textbf{0.7516}&\textbf{0.7572}&\textbf{0.7591}&& \textbf{0.6544}&\textbf{0.6589}&\textbf{0.6642}&\textbf{0.6690}&&\textbf{0.4888}&\textbf{0.5163}&\textbf{0.5375}&\textbf{0.5507}\\
			\midrule
			\multirow{1}[26]{*}{$T\rightarrow I$}
			& CMFH \cite{CMFH} & 0.5902&	0.5899&	0.5949&	0.5985 &&0.3362&	0.3363&	0.3367&	0.3365&&0.3487&	0.3500&	0.3476&	0.3453 \\
			& FSH \cite{FSH} &0.6103&	0.6162&	0.6183&	0.6185&&0.4991&	0.5076&	0.5105&	0.5192&&0.3987&	0.4083&	0.4245&	0.4338\\
			&CRE \cite{CRE} &0.6124&	0.6245&	0.6416&	0.6398&&0.4712&	0.4878&	0.5076&	0.5034&&0.4090&	0.4405&	0.4474&	0.4402\\
			& DCH \cite{DCH}&0.7689&	0.7900	&0.8072&0.8233 &&0.7147&	0.7408&	0.7728&	0.7821&&0.5128&	0.5513&	0.5702&	0.6084 \\
			& SMFH \cite{SMFH}&0.5678&	0.5554&	0.5598&	0.5633&&0.3974	&0.4031	&0.4079&	0.4138&&0.3327&	0.3357&	0.3394	&0.3426\\
			&SCRATCH \cite{SCRATCH}& 0.7808&	0.7773&	0.7999&	0.8050&&0.7611&	0.7821&	0.7908&	0.7826&&0.5325&	0.5767&	0.5989	&0.6285\\
			& SRLCH \cite{SRLCH} &0.6884&	0.7028&	0.7525&	0.7491&&0.7413&	0.7656&	0.7797&	0.7953&&0.4014&	0.4098&	0.4455	&0.4777\\
			&BATCH \cite{BATCH} &0.8134&0.8326	&0.8344	&0.8405&&0.7695&	0.7906&	0.8001&	0.7994&&0.5720	&0.6183	&0.6531&	0.6655\\
			&AMSH&\textbf{0.8217}&\textbf{0.8359}&\textbf{0.8407}&\textbf{0.8477}&&\textbf{0.7784}&\textbf{0.7975}&\textbf{0.8040}&\textbf{0.8092}&& \textbf{0.5914}&\textbf{0.6309}&\textbf{0.6632}&\textbf{0.6787} \\
			\bottomrule
		\end{tabular}%
		\label{pairedmap}%
	\end{table*}%
	\subsection{Evaluation Metrics}
	For evaluation, two basic CMR tasks are tested, i.e., Image-to-Text (i.e., ${I \rightarrow T}$) using images as query to search texts, and Text-to-Image (i.e., ${I \rightarrow T}$) using texts as query to search images. Mean Average Precision (MAP) and the Precision-Recall (PR) curve are adopted for evaluation. Average Precision (AP) is defined as:
	\begin{equation}
		A P(\mathbf{x})=\frac{1}{N_{g}} \sum_{k=1}^{K} P_{\mathbf{x}}(k) \varphi_{\mathbf{x}}(k),
	\end{equation}
	where $N_{g}$ is the number of true neighbors, and $P_{\mathbf{x}}(k)$ denotes the precision of the top $k$ retrieved instances. $\varphi_{\mathbf{x}}(k) = 1$ if the $k$th retrieved instance is the true neighbor of the query and $0$ otherwise. Ground-truth neighbors of a query are defined as samples with at least one shared semantic label with the query. $K$ is the size of the retrieval set \cite{BATCH,SCRATCH}. MAP is defined as:
	\begin{equation}
		M A P = \frac{1}{N_{q}} \sum_{i=1}^{N_{q}} A P(i),
	\end{equation}
	where $N_{q}$ is the size of the query set. higher MAP values indicate better performance.
	
	\subsection{Evaluation on Fully Paired Data}
\begin{figure*}[t]
		\center
		\includegraphics[width=4.2cm,height=3.3cm]{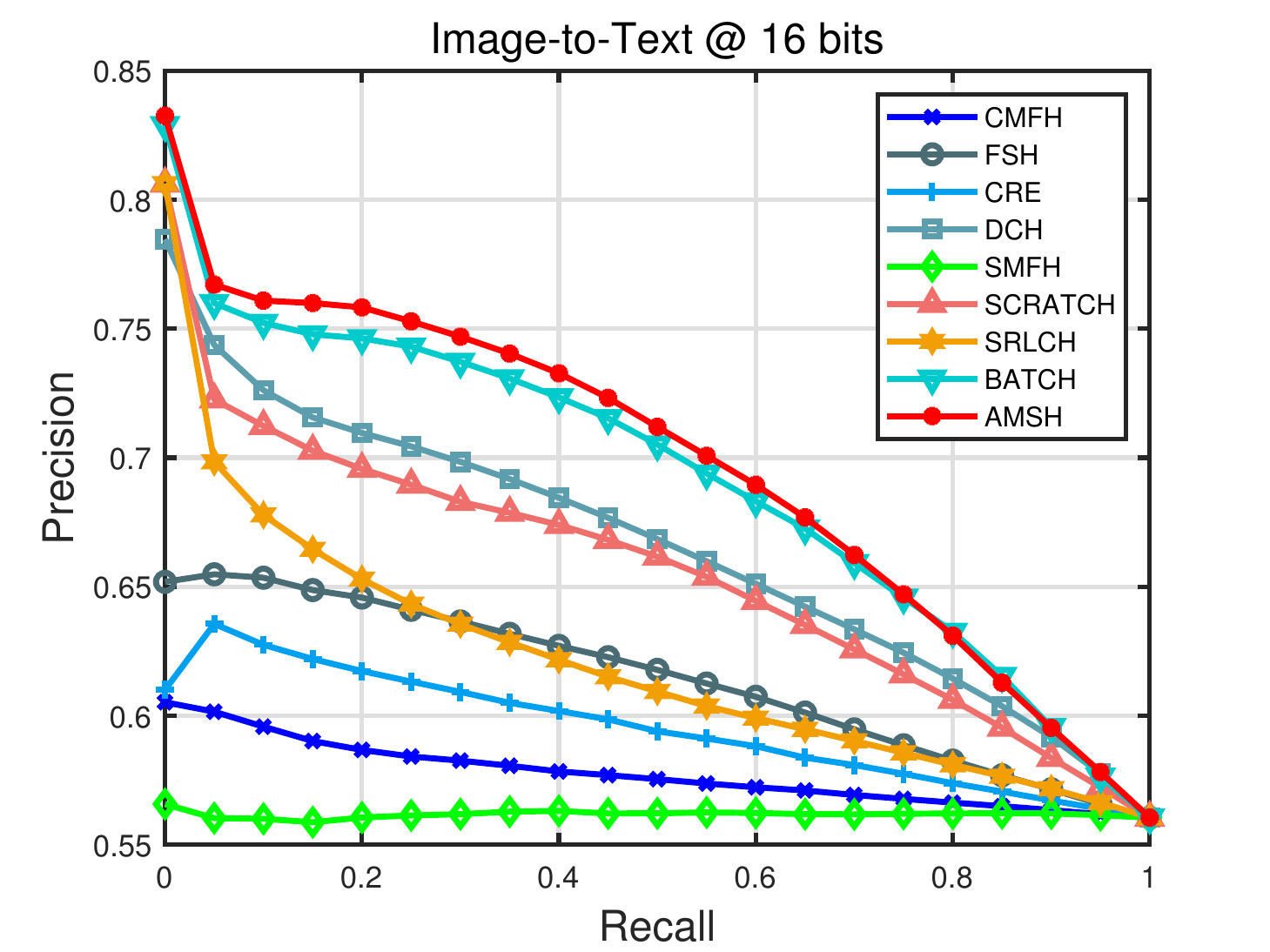}
		\includegraphics[width=4.2cm,height=3.3cm]{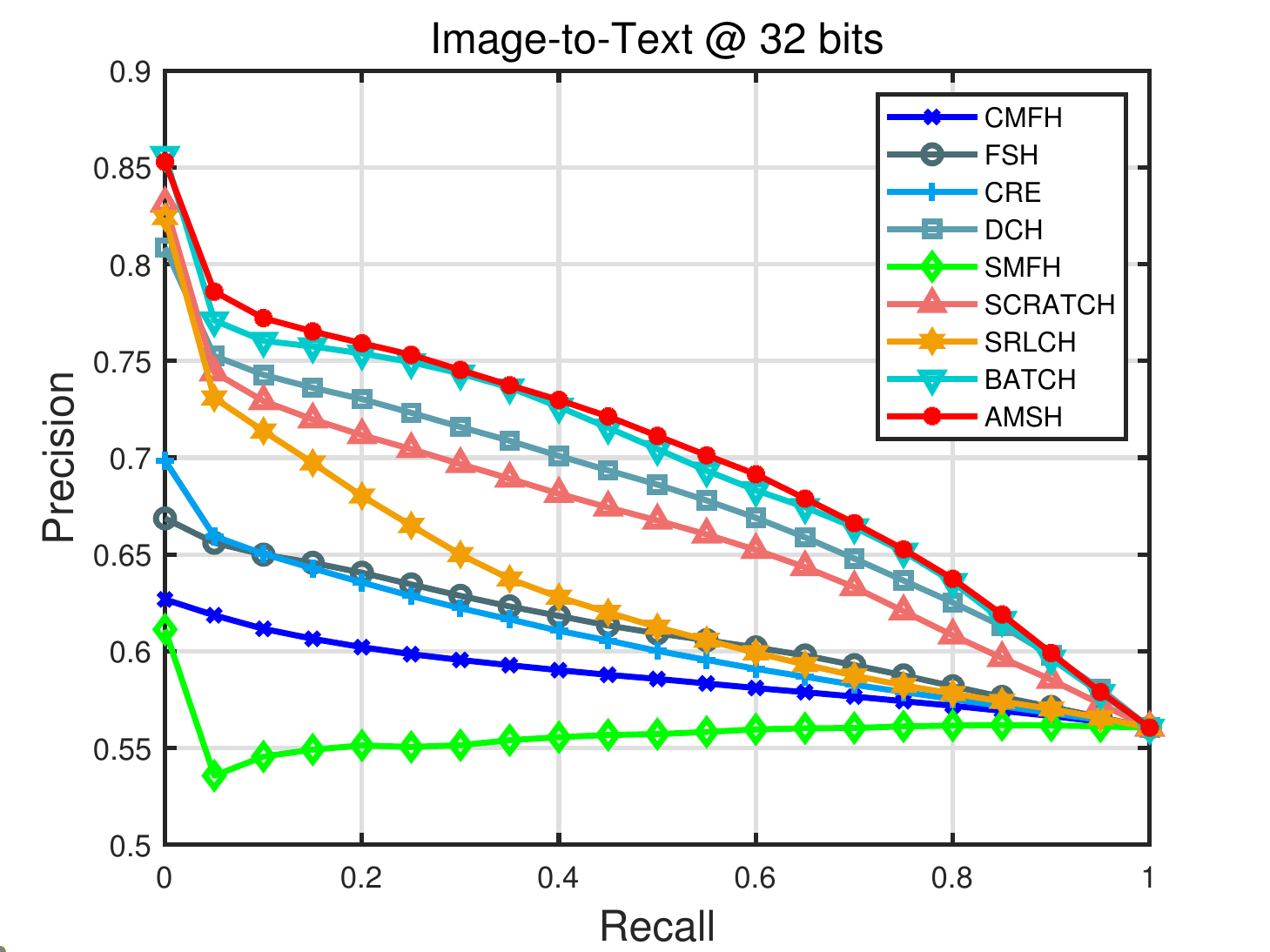}
		\includegraphics[width=4.2cm,height=3.3cm]{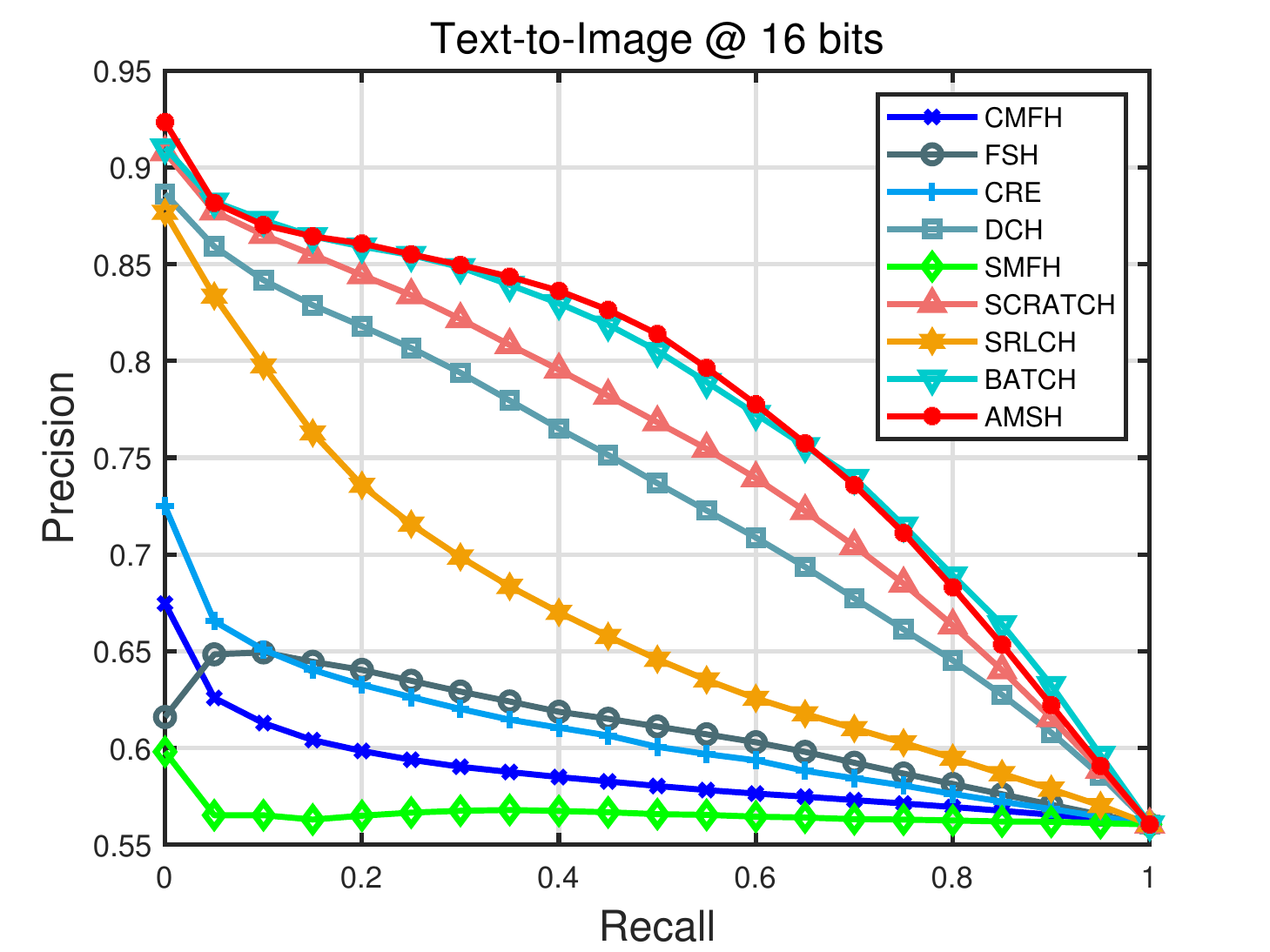}
		\includegraphics[width=4.2cm,height=3.3cm]{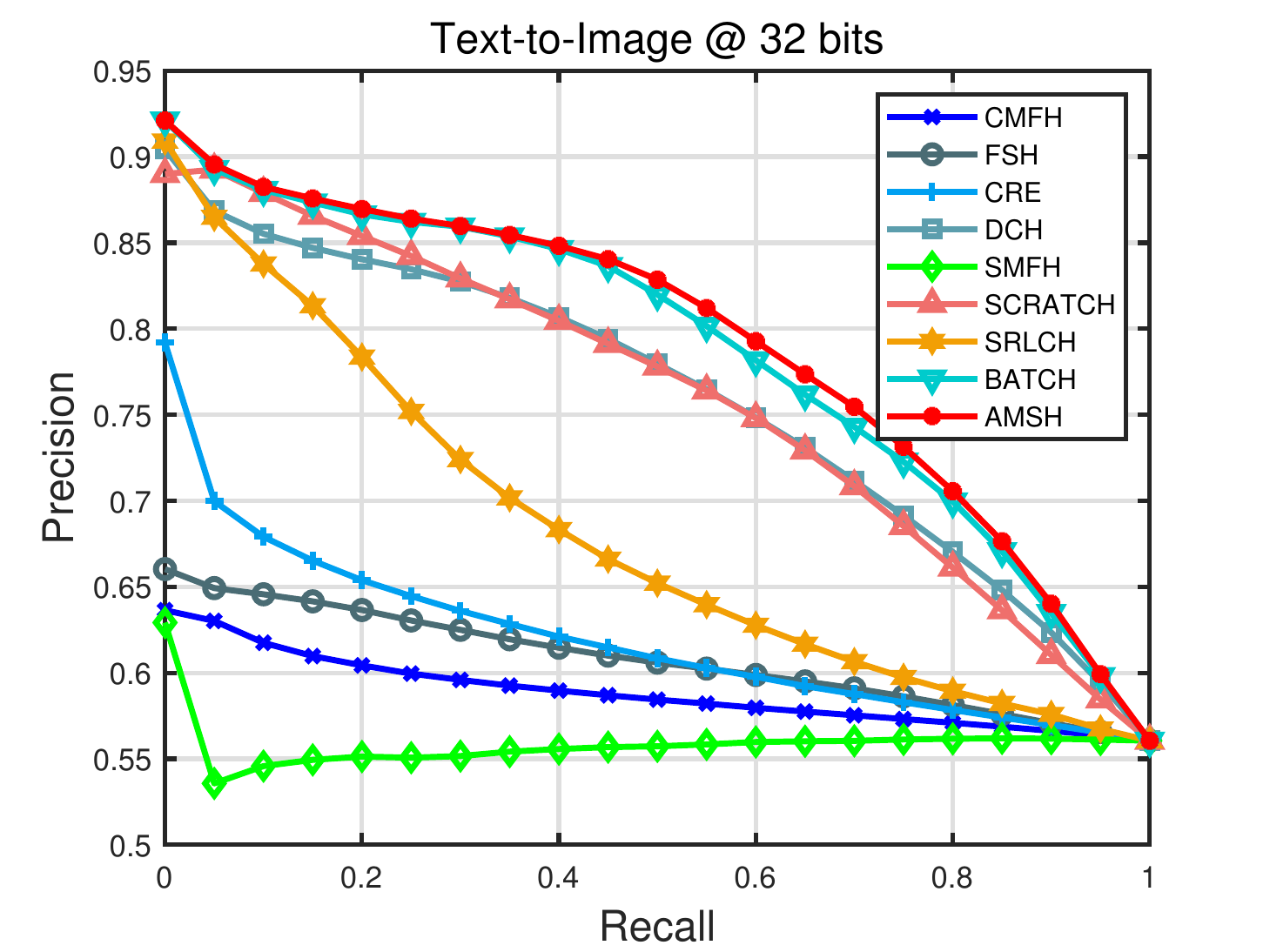}
		\includegraphics[width=4.2cm,height=3.3cm]{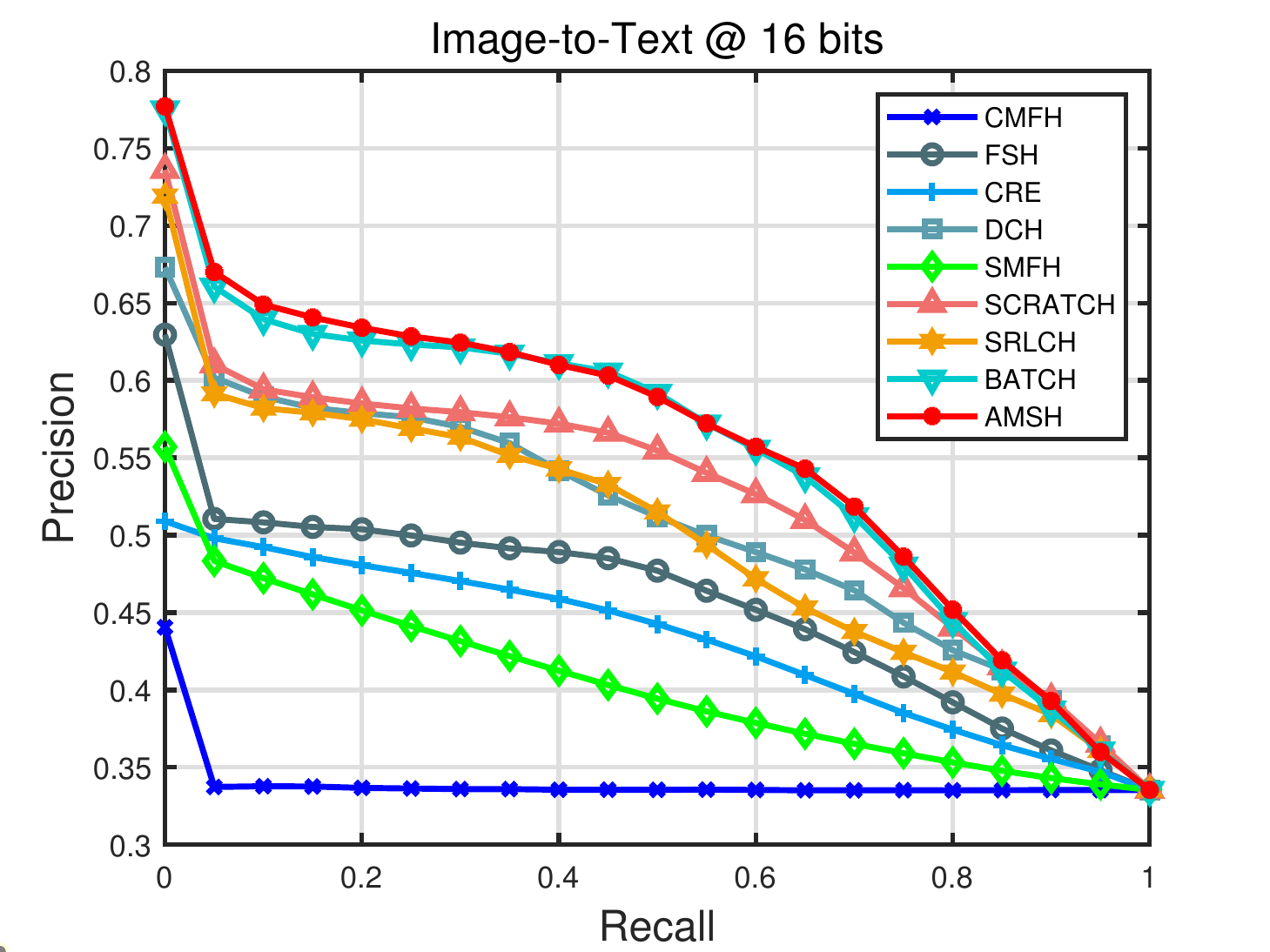}
		\includegraphics[width=4.2cm,height=3.3cm]{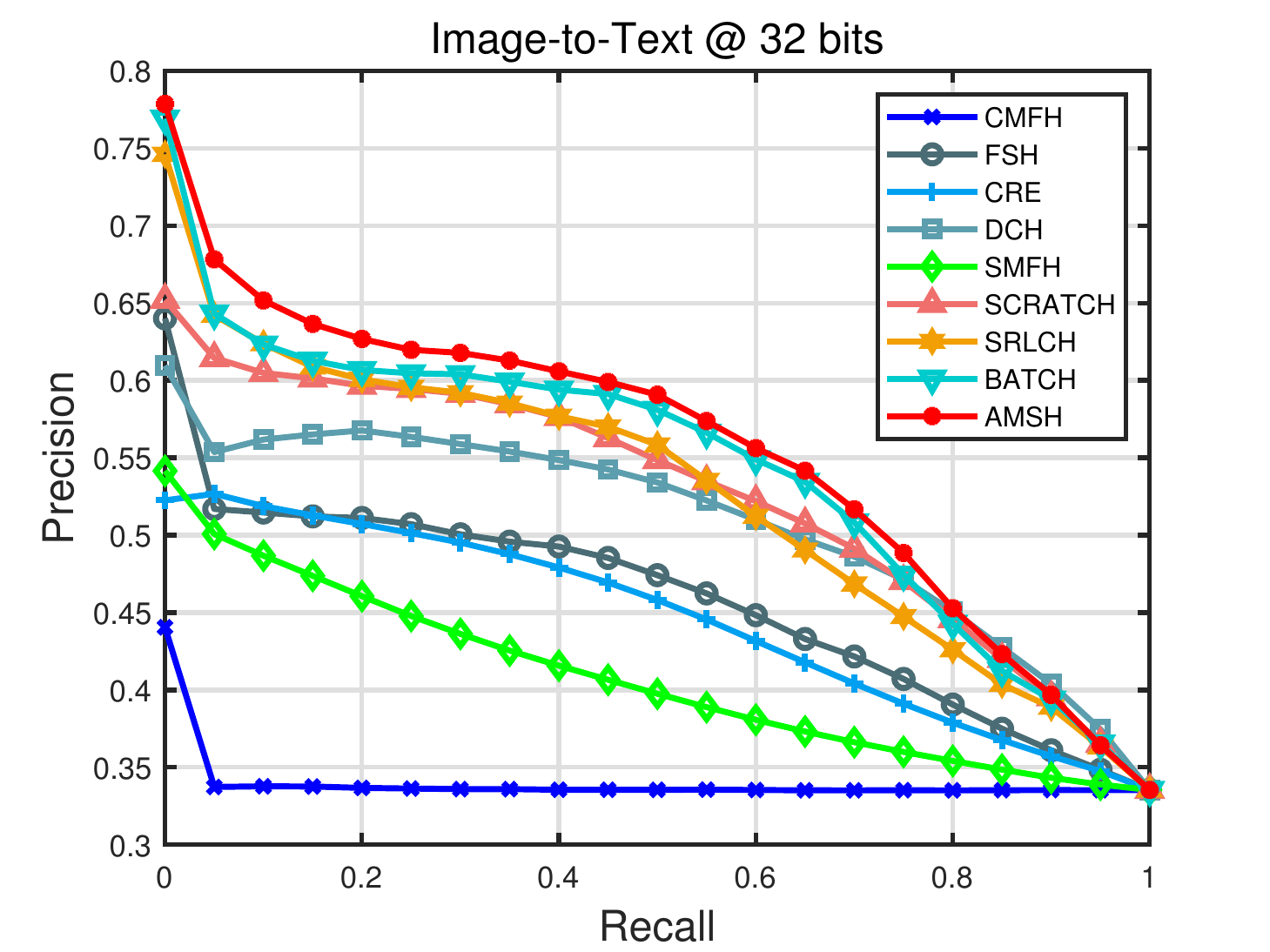}
		\includegraphics[width=4.2cm,height=3.3cm]{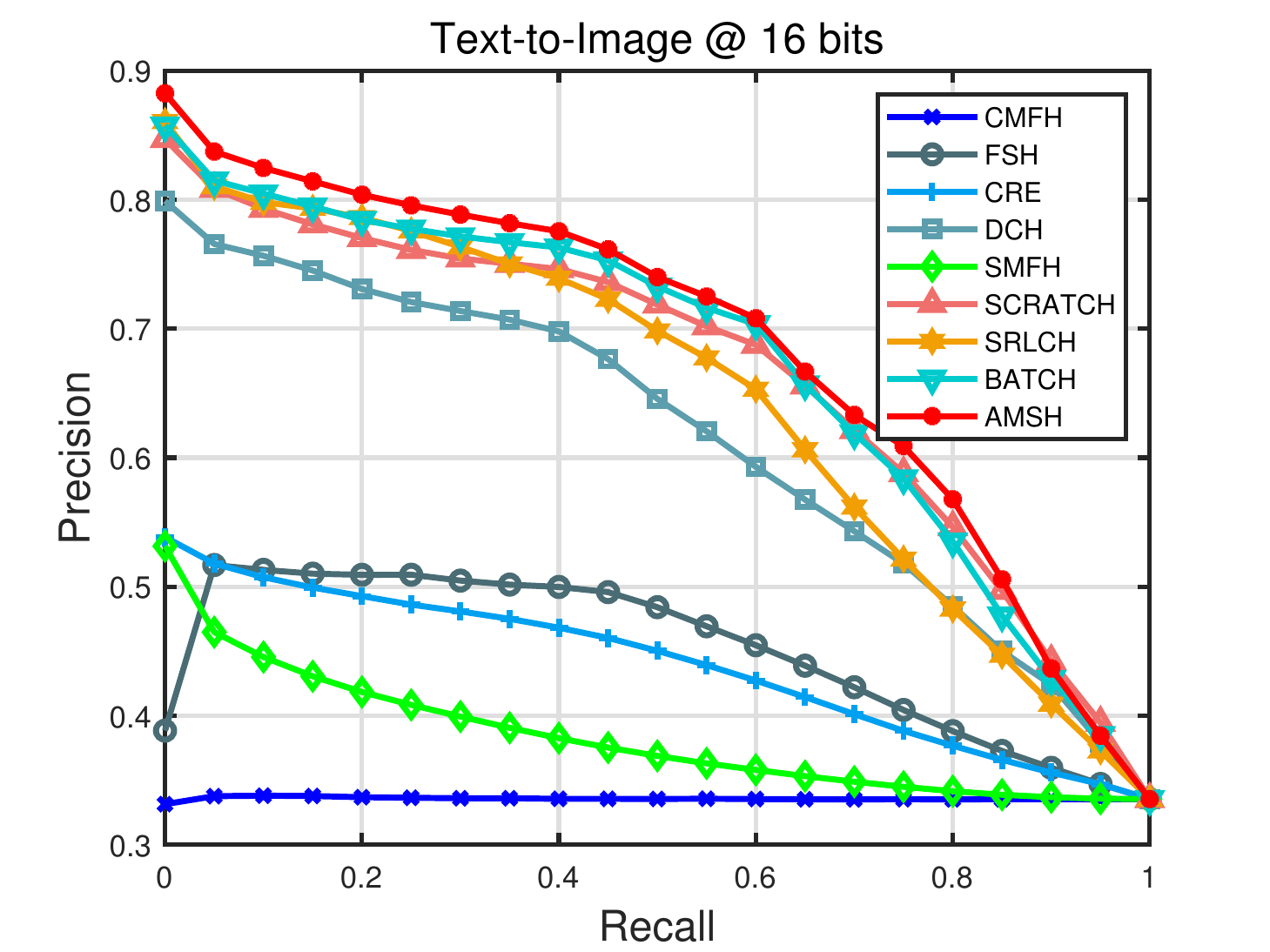}
		\includegraphics[width=4.2cm,height=3.3cm]{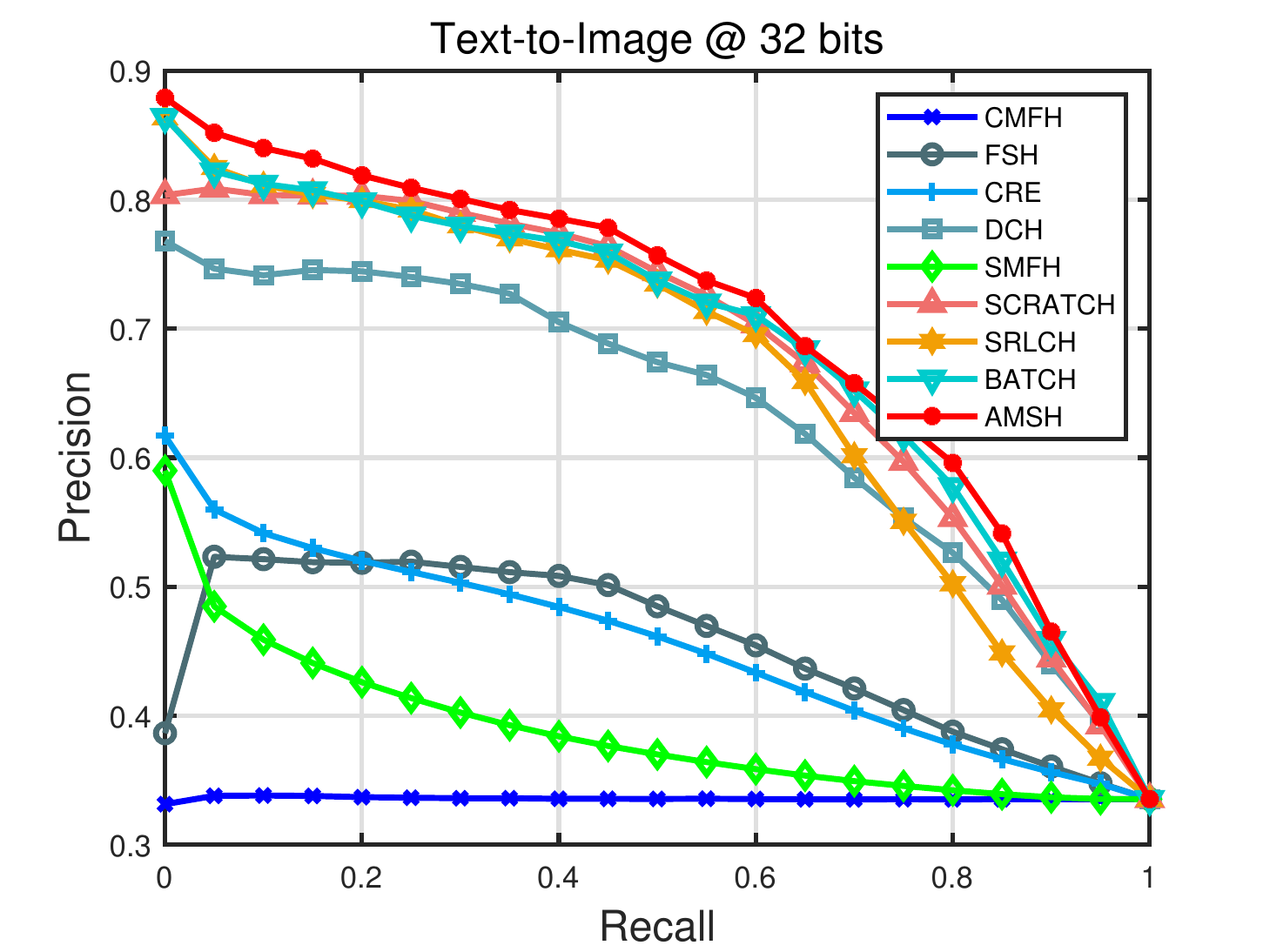}
		\includegraphics[width=4.2cm,height=3.3cm]{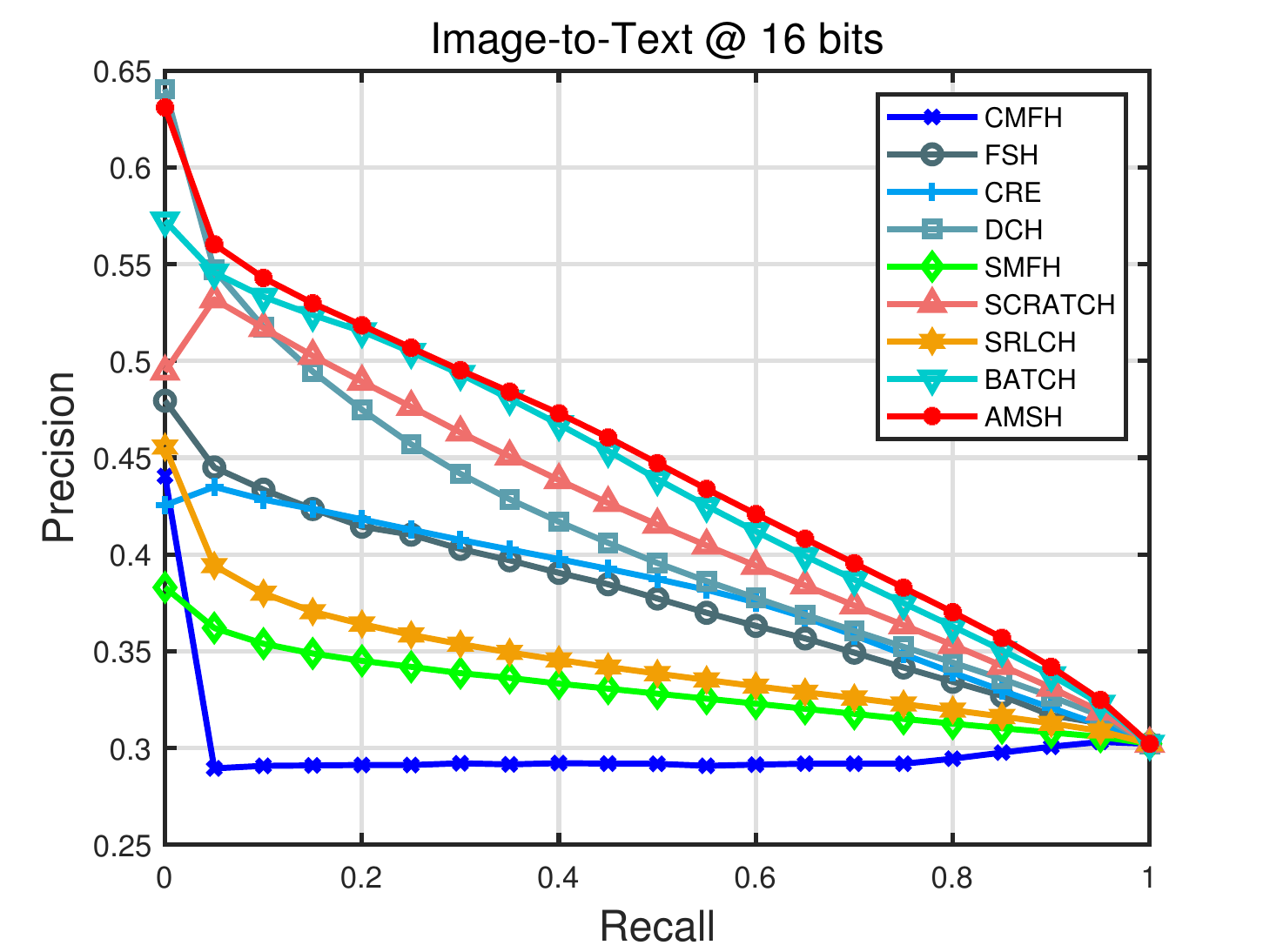}
		\includegraphics[width=4.2cm,height=3.3cm]{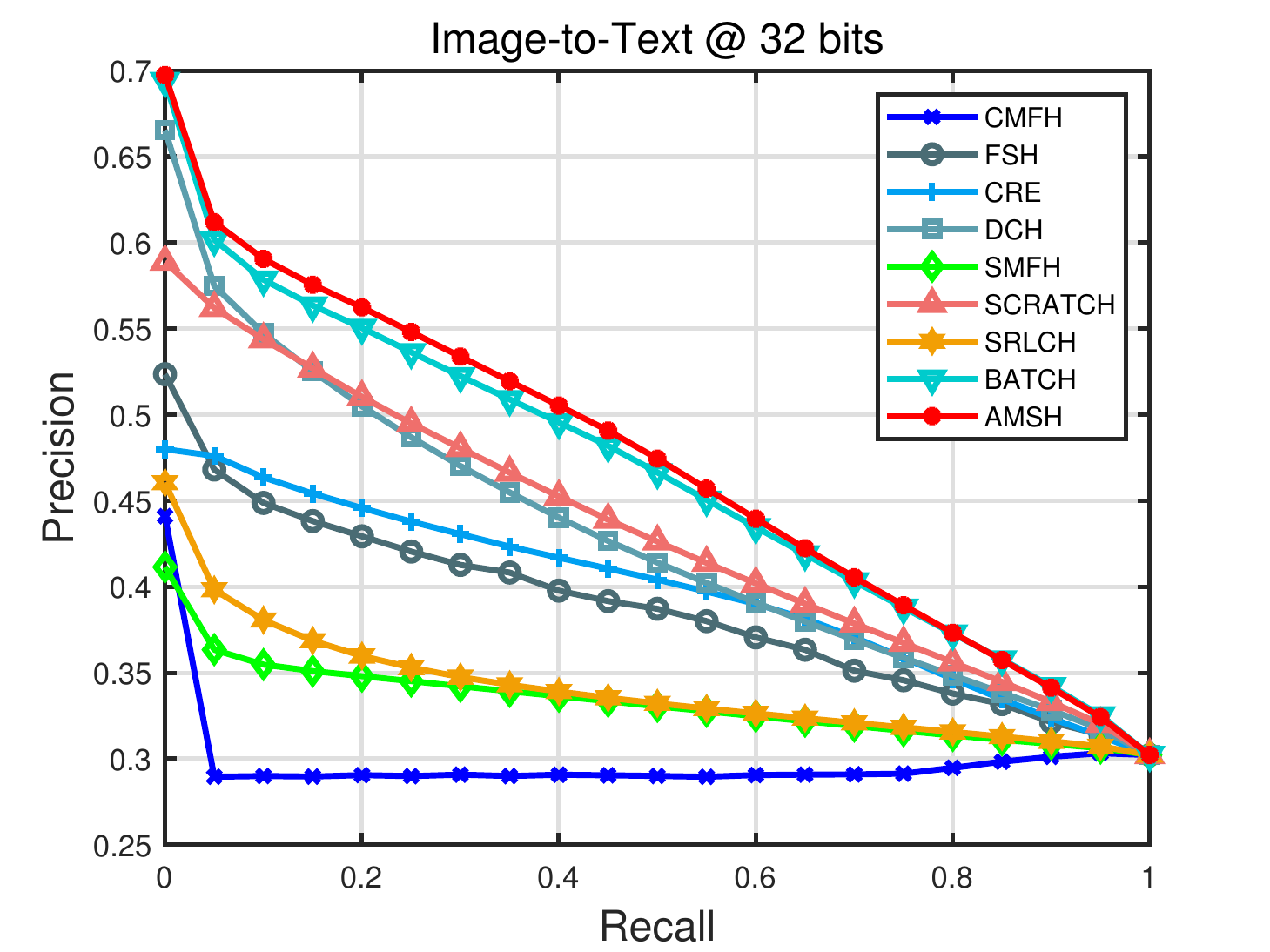}
		\includegraphics[width=4.2cm,height=3.3cm]{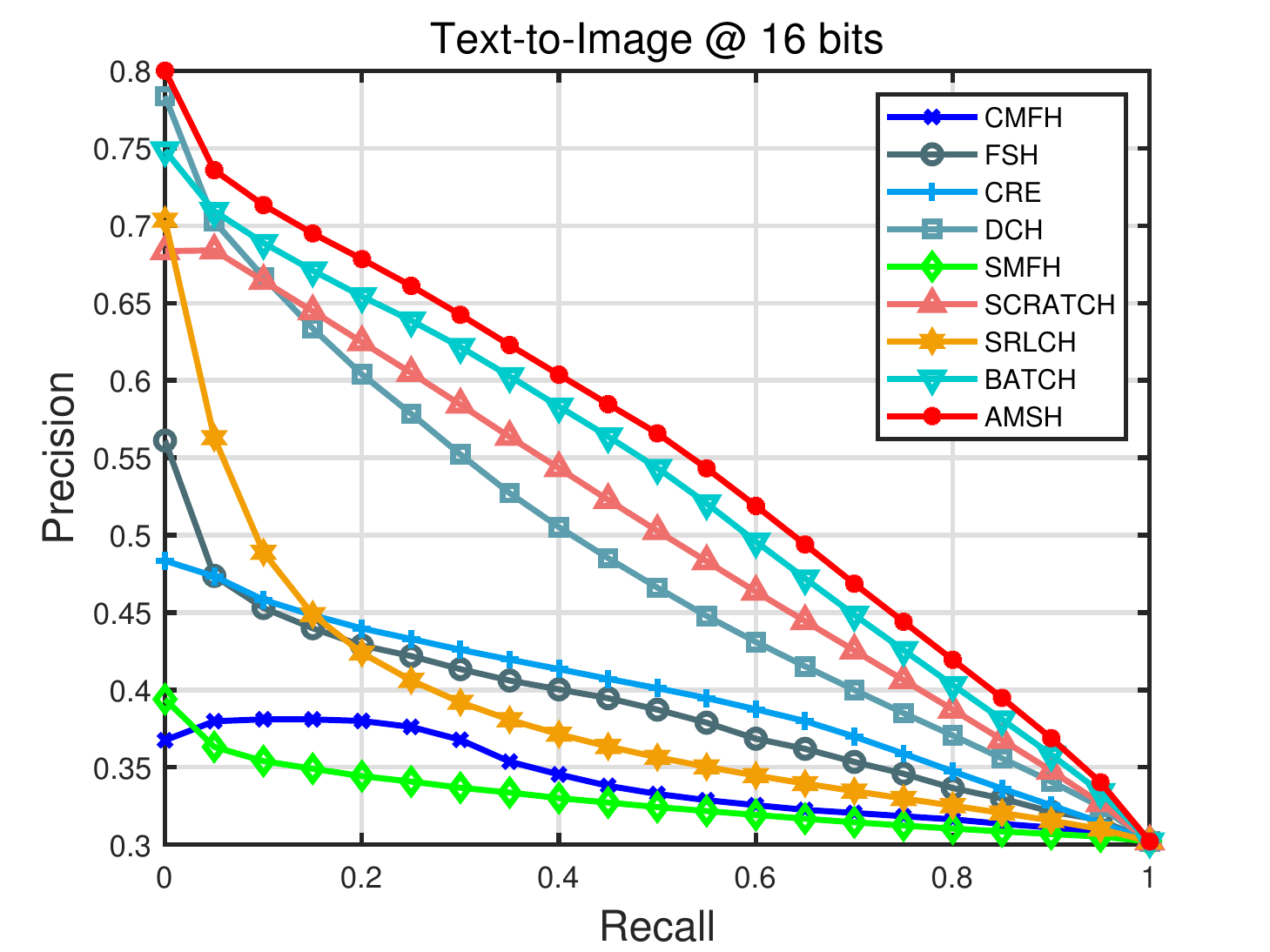}
		\includegraphics[width=4.2cm,height=3.3cm]{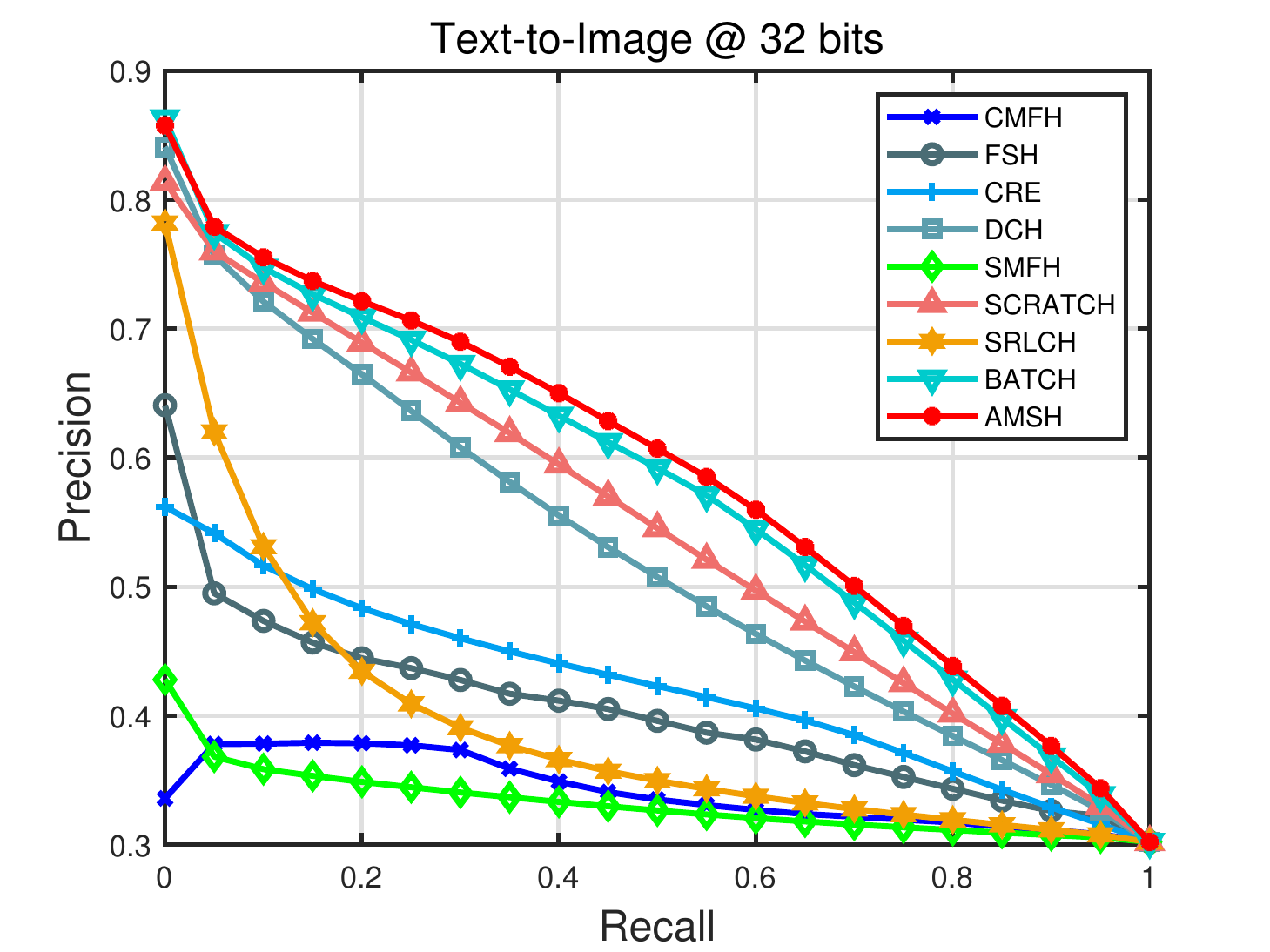}
		\caption{The PR curves of AMSH and baselines on MIRFlickr-25K (top), NUS-WIDE (middle) and IAPR TC-12 (bottom) with paired data.}
		\label{pairedpr}
	\end{figure*}

	We first evaluate AMSH in the fully paired scenario and vary the code length from 16 to 128 bits (i.e., 16, 32, 64 and 128). TABLE \ref{pairedmap} reports the MAP results of AMSH and the baselines on  MIRFlickr-25K, NUS-WIDE and IAPR TC-12, and Fig. \ref{pairedpr} illustrates the corresponding PR curves. From the experimental results, the following observations can be summarized.
	\begin{itemize}
		\item AMSH obtains higher MAP results than all the baselines on Image-to-Text and Text-to-Image tasks on three datasets with different code lengths, and the PR curves of AMSH are generally above these baselines. AMSH considers enlarging the distances between different classes to enhance label discrimination, contributing to more discriminative hash codes. For hash functions learning, adaptive marginalization eases the rigid binary linear regression to generate more robust hash functions. The experimental results demonstrate the effectiveness and superiority of AMSH over these methods.
		\item Generally speaking, supervised methods outperform unsupervised ones, which shows the effectiveness of using semantic information for hash learning. AMSH and BATCH embed high-level semantic information into hash codes by constructing the similarity matrix with labels. SCRATCH, SRLCH and DCH construct a linear classifier between hash codes and labels, which generates semantic-aware hash codes. Among the unsupervised methods, CRE and FSH achieve promising results, for CRE adopts a similar strategy to BATCH except replacing the labels with raw data, and FSH preserves fusion similarity for more discriminative hash codes.
		\item For most baselines, the performance gets better as the code length increases, since more semantic information can be stored with longer binary codes.
		\item The MAP results of these methods drop drastically on NUS-WIDE, compared to MIRFlickr-25k. One possible reason is that the size of NUS-WIDE is much larger than MIRFlickr-25k with large intra-class variations, making it hard to learn discriminative hash codes. Similar results can be observed on IAPR TC-12, with bigger drop in MAP results. 
		\item For most methods, MAP results on Text-to-Image task are generally higher than those on Image-to-Text task. one possible factor is that textual modality is more semantically similar to label information than image modality.
	\end{itemize}

	\subsection{Evaluation on Unpaired Data}
	In this subsection, we further explore the performance of AMSH in the unpaired setting. Specifically, we adopt the same benchmark datasets used in the last subsection, with all data randomly shuffled. Thus, data have no one-to-one correspondence. Six baselines are adopted for comparison, i.e., UCMH \cite{UCMH} , GSPH \cite{GSPH}, RUCMH \cite{RUCMH}, SAMPH \cite{UAPMH}, EDMH \cite{EDMH} and FMH \cite{FMH}, which can be used for unpaired data.
	\begin{table*}[t]
		\centering
		\caption{The MAP results of AMSH and baselines on MIRFlickr-25K, NUS-WIDE and IAPR TC-12 with unpaired data.}
		\renewcommand{\arraystretch}{1.1}
		\begin{tabular}{p{1.5cm}<{\centering}p{1.8cm} p{0.6cm}<{\centering} p{0.6cm}<{\centering} p{0.6cm}<{\centering} p{0.7cm}<{\centering} p{0.1cm}<{\centering} p{0.6cm}<{\centering} p{0.6cm}<{\centering} p{0.6cm}<{\centering} p{0.7cm}<{\centering}p{0.1cm}<{\centering}
				p{0.7cm}<{\centering} p{0.7cm}<{\centering} p{0.7cm}<{\centering}
				p{0.7cm}<{\centering}}
			\toprule
			\multicolumn{1}{c}{\multirow{1}[4]{*}{Task}} & \multicolumn{1}{l}{\multirow{1}[4]{*}{Method}} & \multicolumn{4}{c}{MIRFlickr-25K}  &       & \multicolumn{4}{c}{NUS-WIDE} &       & \multicolumn{4}{c}{IAPR TC-12} \\
			\cmidrule{3-6}\cmidrule{8-11}\cmidrule{13-16}
			&   &  16 b   & 32 b   & 64 b   & 128 b   &      &  16 b    & 32 b    & 64 b    & 128 b & &  16 b    & 32 b   & 64 b    & 128 b \\
			\midrule
			\multirow{1}[20]{*}{$I\rightarrow T$}
			&SAPMH \cite{UAPMH}&0.5643&	0.5636&	0.5673&	0.5684&&0.3384&	0.3386&	0.3380&	0.3384&&0.3041&	0.3044&	0.3042&	0.3042\\
			&UCMH \cite{UCMH}&0.5981&0.5597&0.5912&0.5927&&0.4091&	0.4002&	0.4061	&0.3937&&0.3070&	0.3070	&0.3070&0.3071\\
			&RUCMH \cite{RUCMH}&0.5620&	0.5607&	0.5626&	0.5614&&0.3399&	0.3487&	0.3467&	0.3396&&0.3252	&0.3018&	0.3040&	0.3031\\
			&GSPH \cite{GSPH}&0.6541&	0.6600&	0.6695&	0.6705&&0.5062&	0.5199&	0.5349&	0.5409	&&0.4014&	0.4222&	0.4459&	0.4572\\
			&EDMH \cite{EDMH}&0.6738&	0.6728&	0.6876&	0.6979&&0.4594&	0.5209&	0.5030&	0.5502&&0.3775&	0.4089&	0.4255&	0.4314\\
			&FMH \cite{FMH}&0.7130&	0.7338&	0.7354&	0.7337&&0.6473&	0.6480&	0.6297&	0.6409&&0.3791&	0.4192&	0.4331&	0.4904\\
			&AMSH&\textbf{0.7456}&\textbf{0.7515}&\textbf{0.7582}&\textbf{0.7575}&&\textbf{0.6580}&\textbf{0.6544}&\textbf{0.6565}&\textbf{0.6641}&&\textbf{0.4908}&\textbf{0.5182}&\textbf{0.5370}&\textbf{0.5482}\\
			\midrule
			\multirow{1}[20]{*}{$T\rightarrow I$}
			&SAPMH \cite{UAPMH}&0.5625&	0.5629&	0.5667&	0.5672&&0.3378&	0.3378&	0.3378&	0.3381&&0.3042&	0.3044&	0.3049&	0.3030\\
			
			&UCMH \cite{UCMH}&0.5852&	0.5556&	0.5804&	0.5838&&0.3982	&0.3915&	0.3978&	0.3895&&0.3064&	0.3065&	0.3066&	0.3066
			\\
			&RUCMH \cite{RUCMH}&0.5642&	0.5613&	0.5621&	0.5616&&0.3293	&0.3407&	0.3401&	0.3383&&0.3143&	0.3029&	0.3042&	0.3046
			\\
			&GSPH \cite{GSPH}&0.6420	&0.6529&	0.6607&	0.6616&&0.4812&	0.4916&	0.5010&	0.5090&&0.4021&	0.4183&	0.4382&	0.4503
			\\
			&EDMH \cite{EDMH}&0.6622&	0.6498&	0.6828&	0.6898&&0.5362&	0.5809&	0.5990&	0.6507&&0.3879&	0.3906&	0.4502&	0.4565
			\\
			&FMH \cite{FMH}&0.7914&	0.8109&	0.8169&	0.8246&&0.7344&	0.7289&	0.7244&	0.7493&&0.3821&	0.4259&	0.4433&	0.5080\\
			
			&AMSH&\textbf{0.8310}&\textbf{0.8309}&\textbf{0.8400}&\textbf{0.8436}&&\textbf{0.7915}&\textbf{0.7860}&\textbf{0.8048}&\textbf{0.8060}&&\textbf{0.5830}&\textbf{0.6328}&\textbf{0.6598}&\textbf{0.6771}\\
			\bottomrule
		\end{tabular}%
		\label{unpairedmap}%
	\end{table*}%
	\begin{figure*}[t]
		\center
		\includegraphics[width=4.2cm,height=3.3cm]{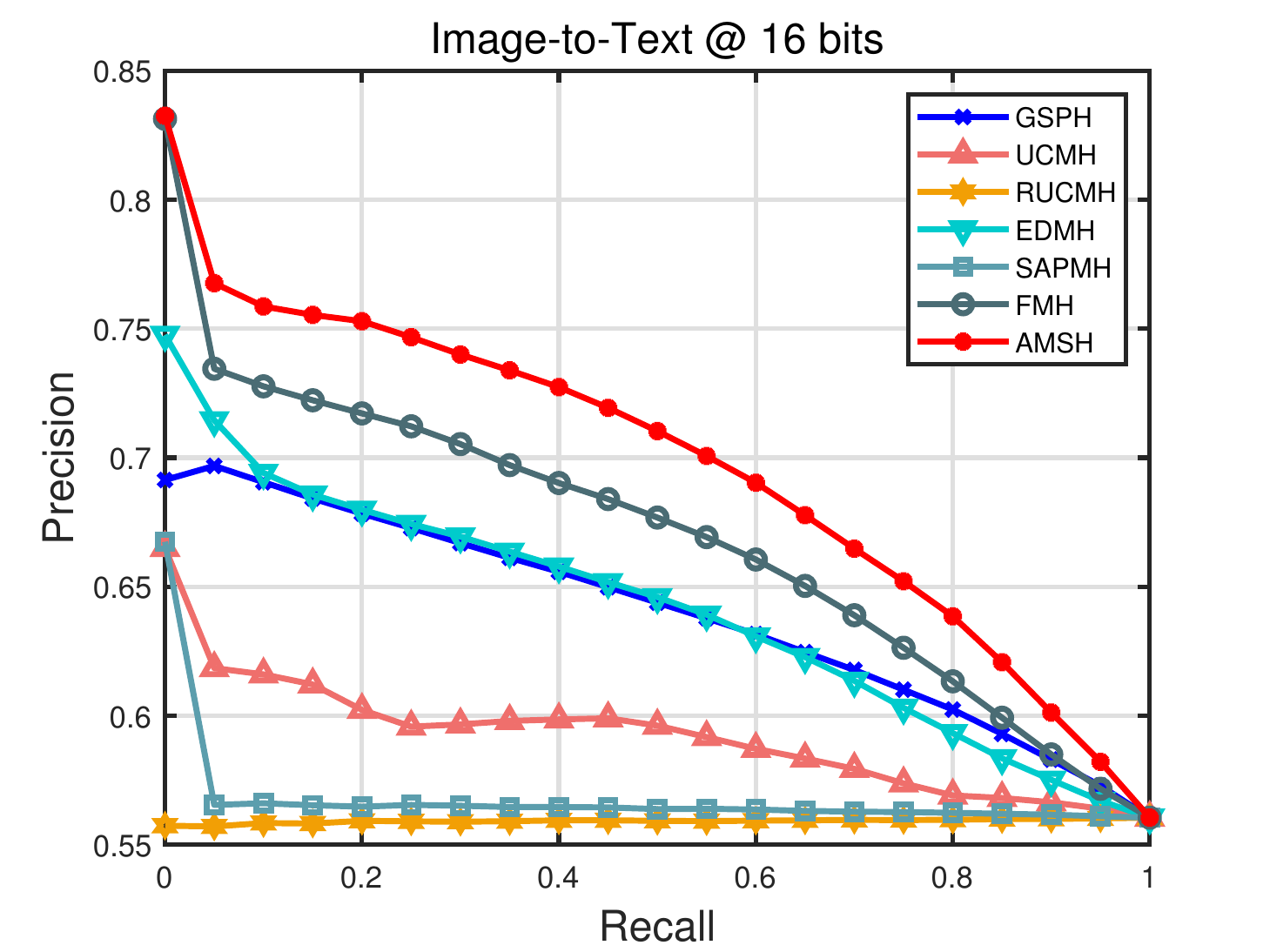}
		\includegraphics[width=4.2cm,height=3.3cm]{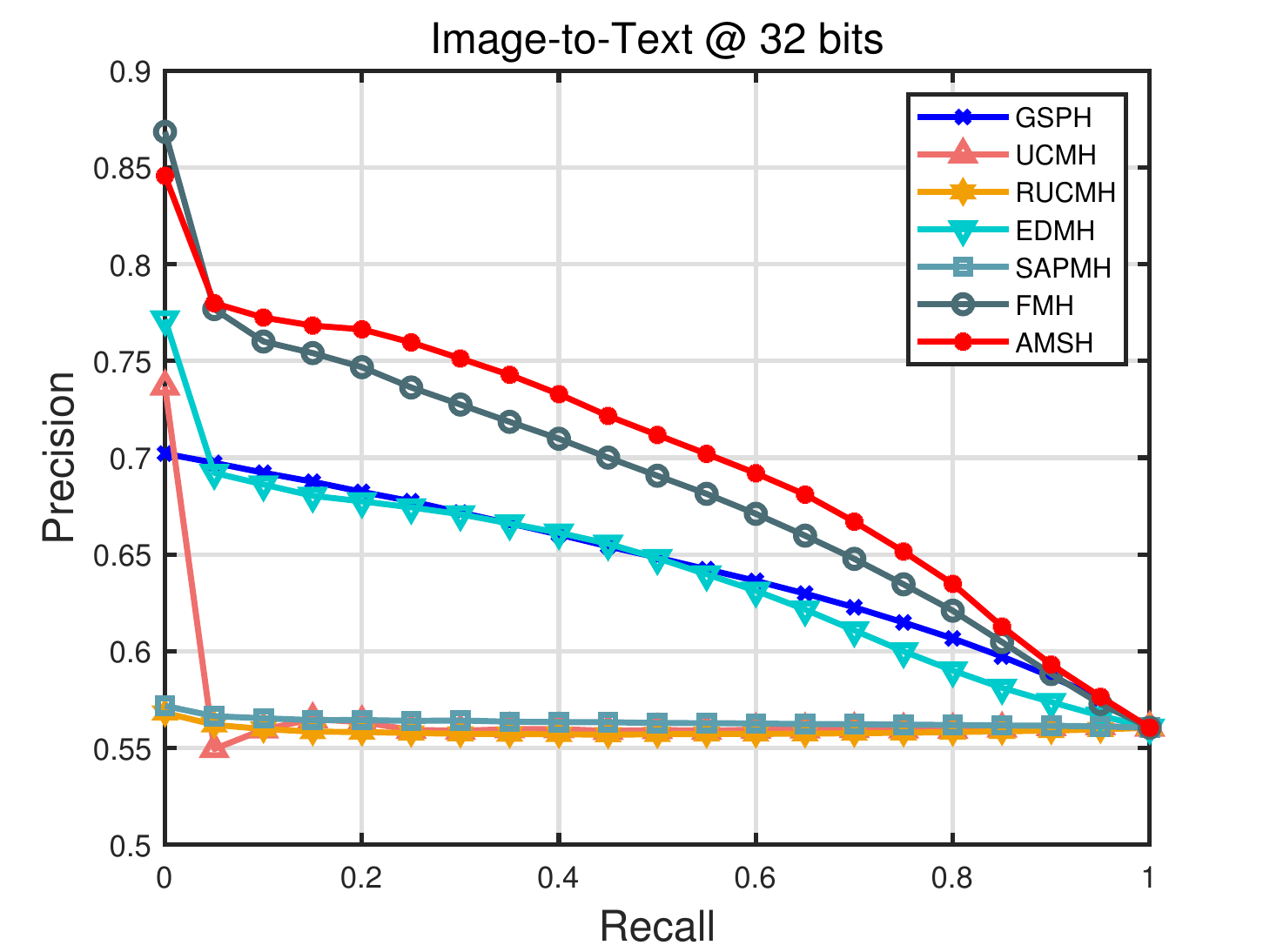}
		\includegraphics[width=4.2cm,height=3.3cm]{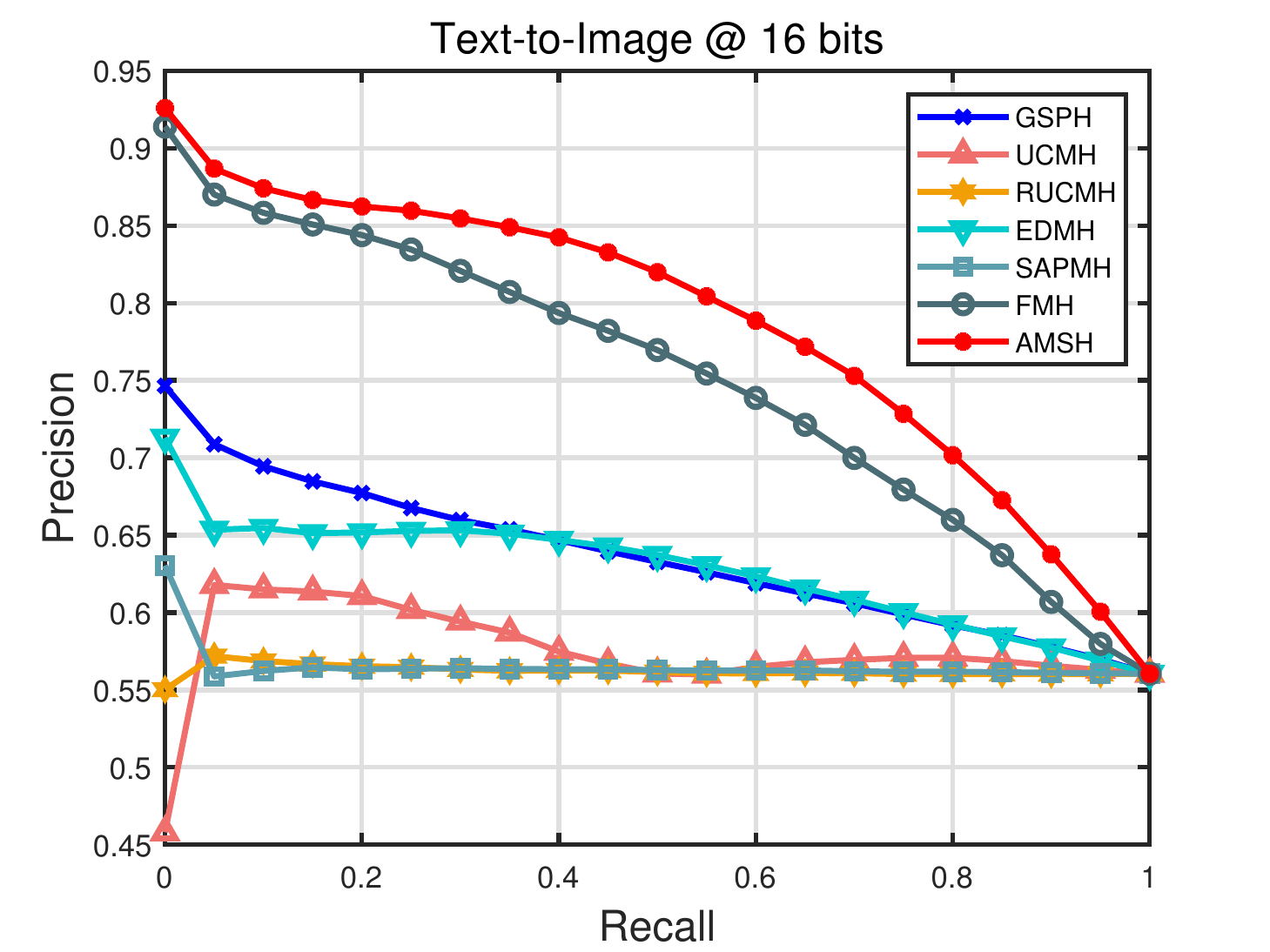}
		\includegraphics[width=4.2cm,height=3.3cm]{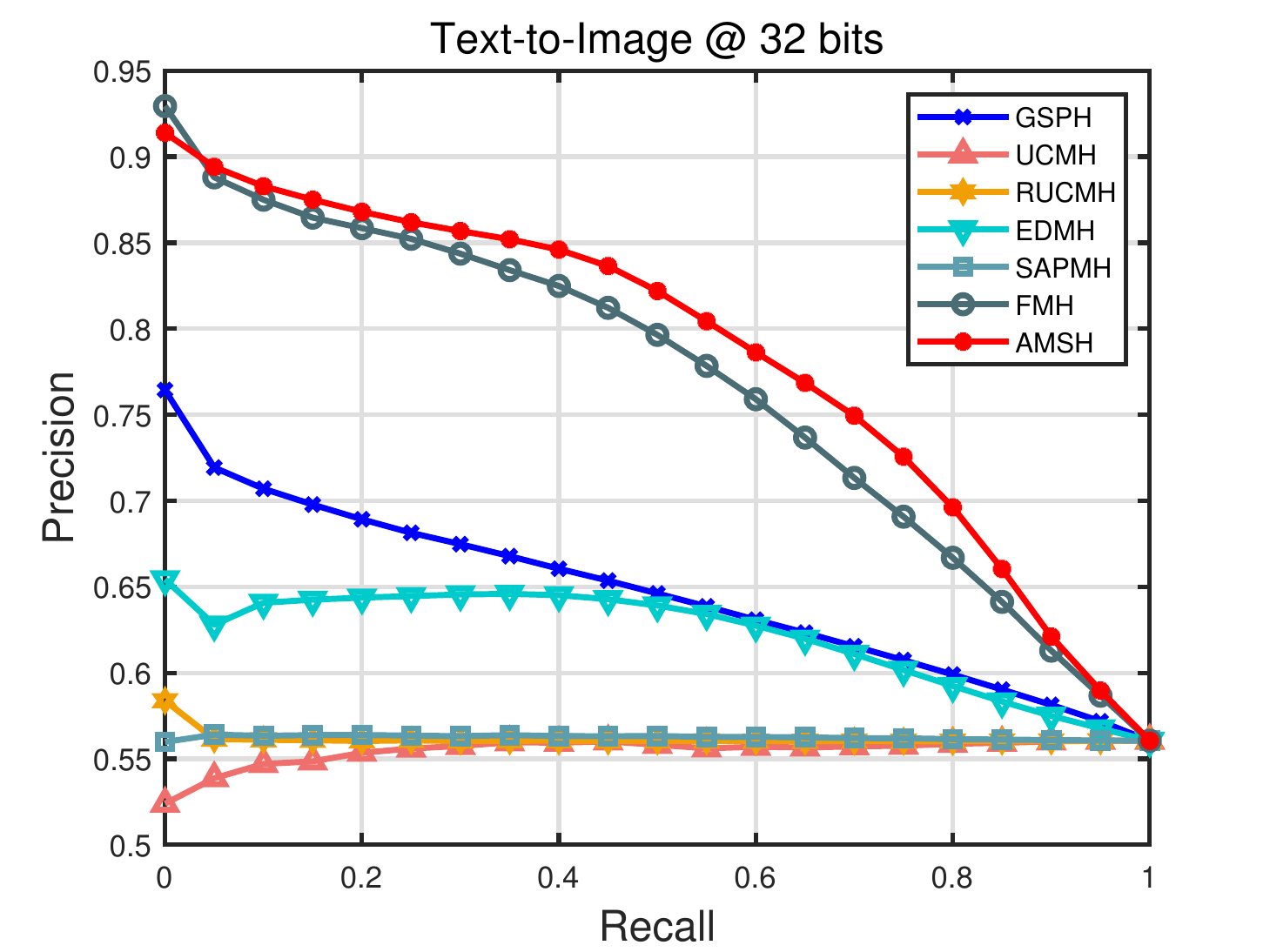}
		\includegraphics[width=4.2cm,height=3.3cm]{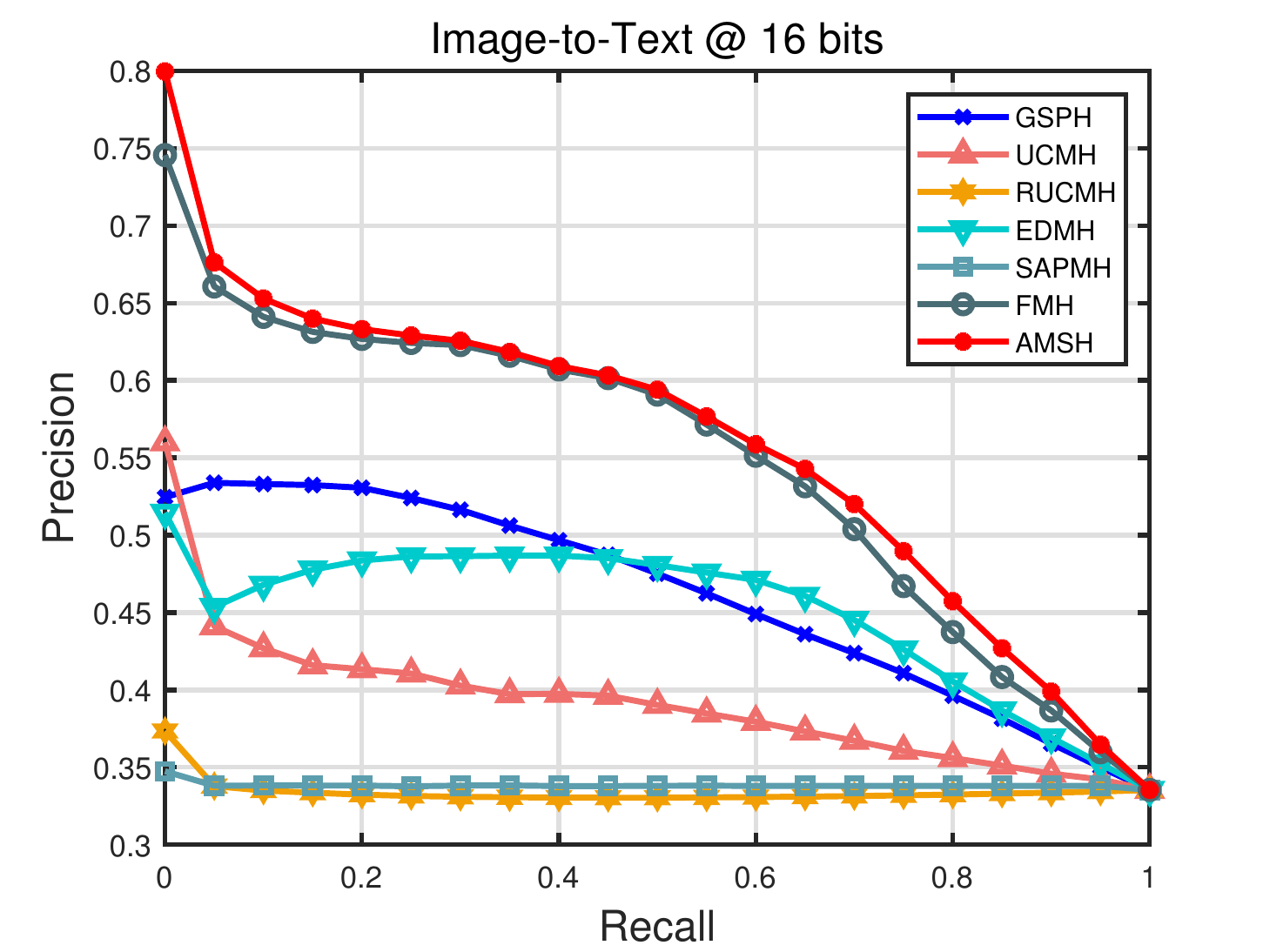}
		\includegraphics[width=4.2cm,height=3.3cm]{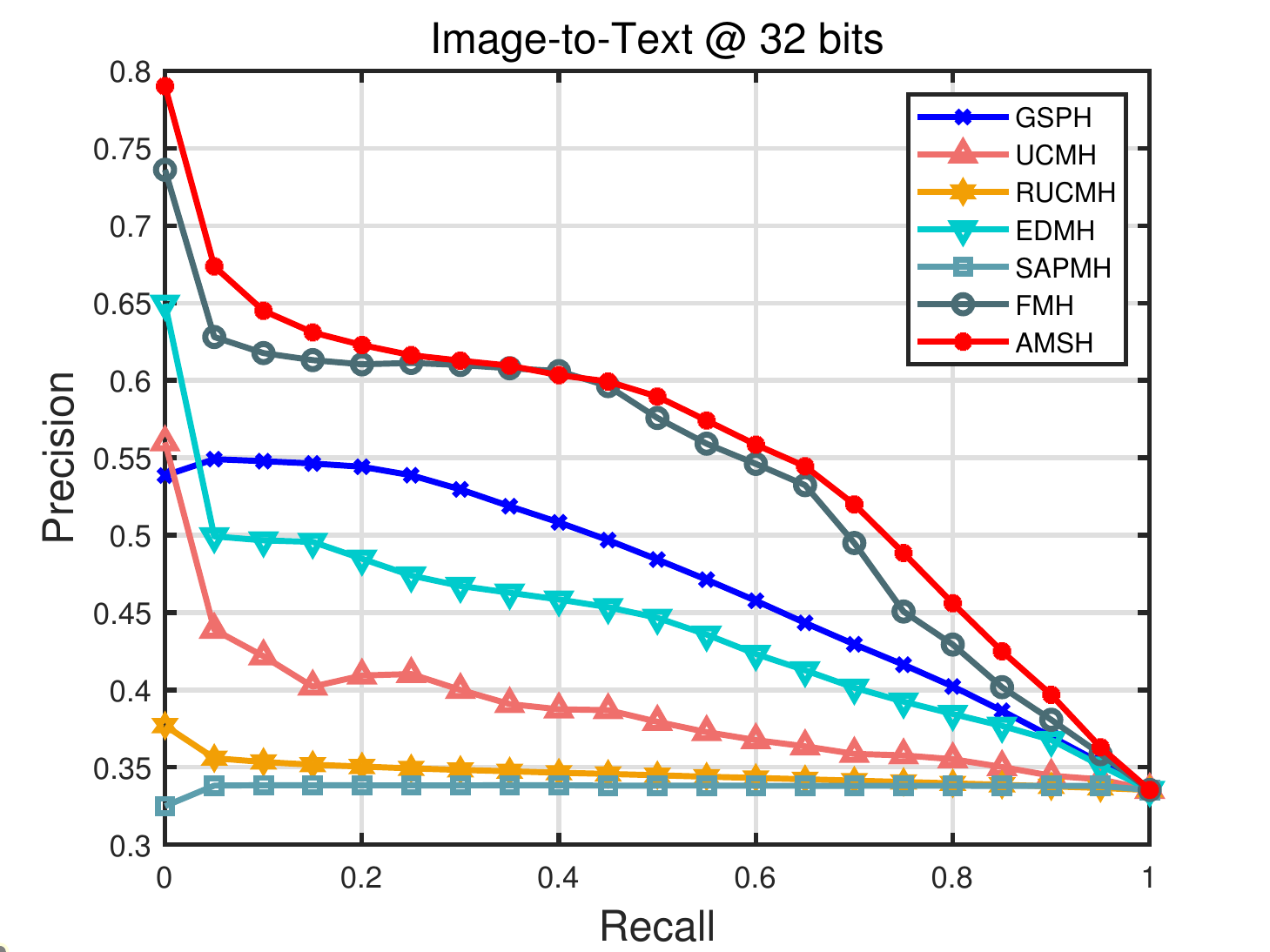}
		\includegraphics[width=4.2cm,height=3.3cm]{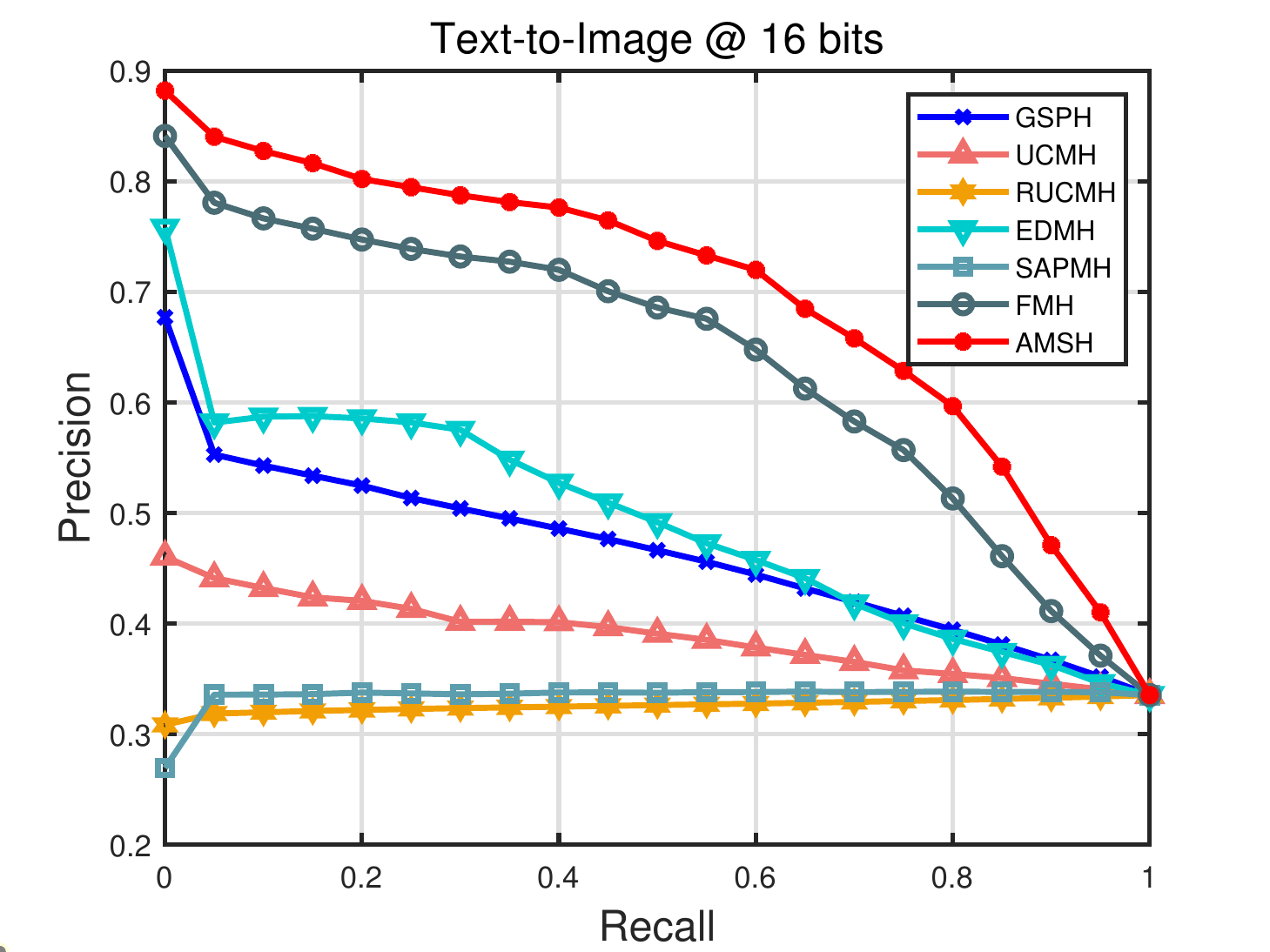}
		\includegraphics[width=4.2cm,height=3.3cm]{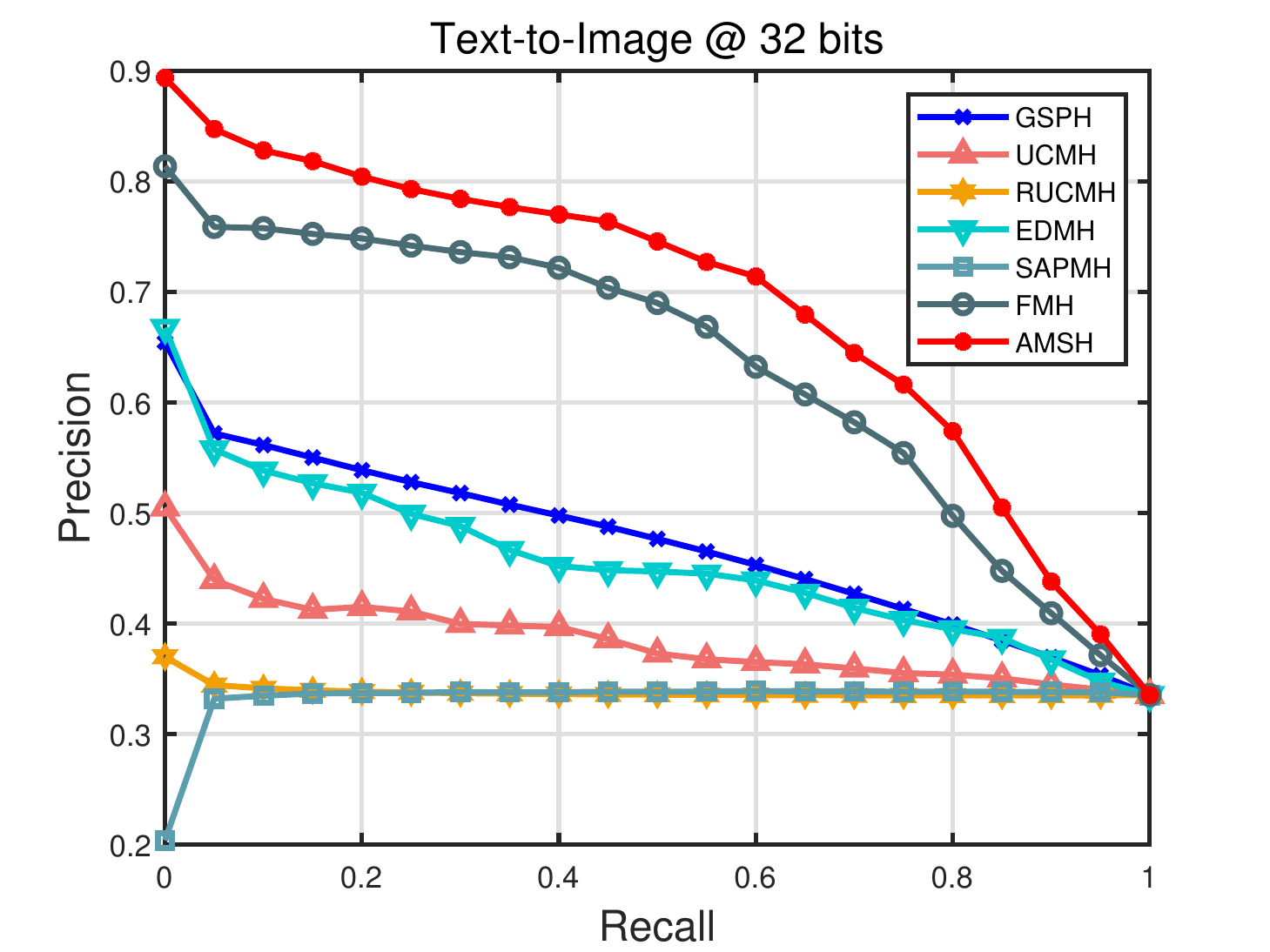}
		\includegraphics[width=4.2cm,height=3.3cm]{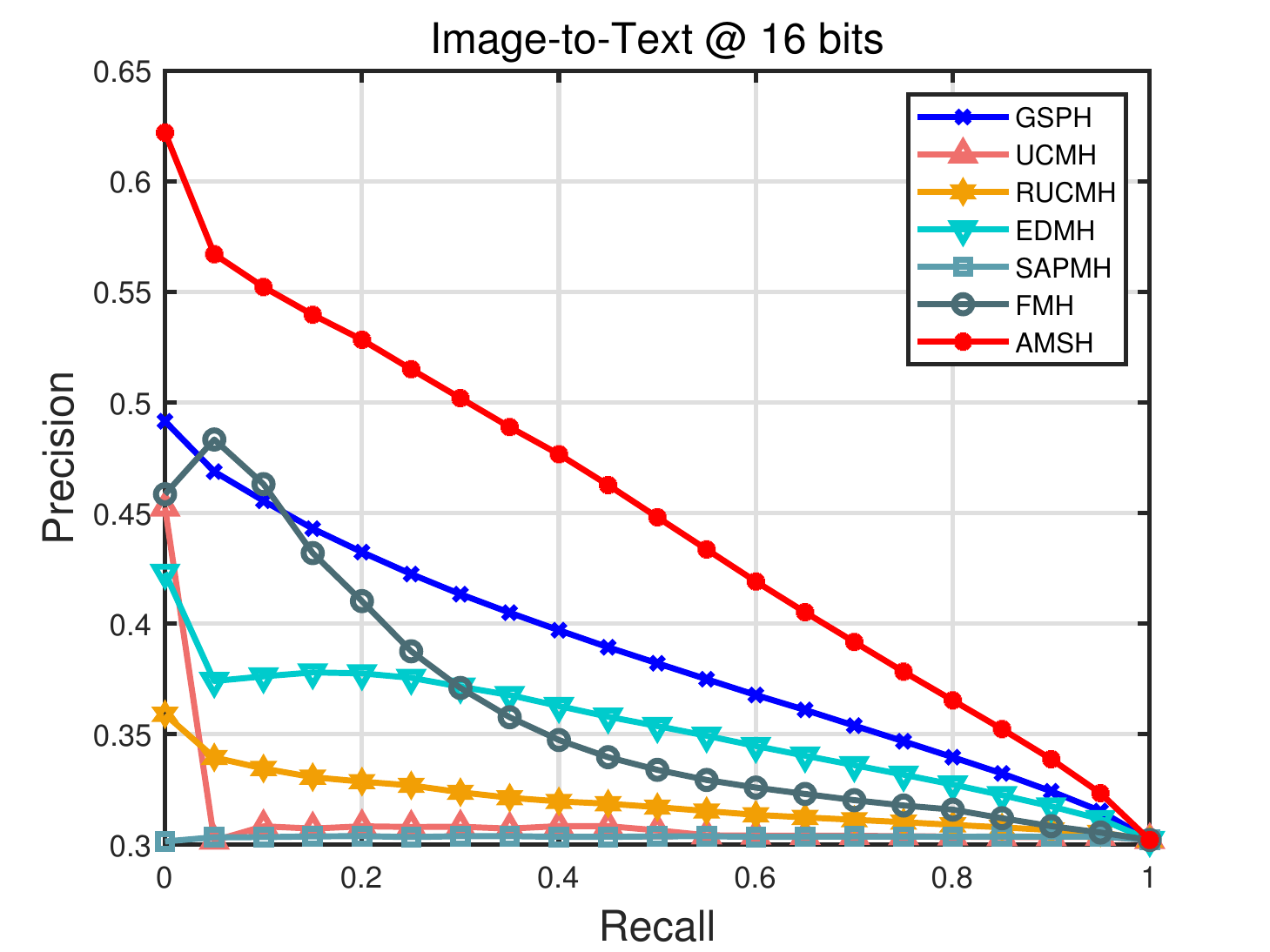}
		\includegraphics[width=4.2cm,height=3.3cm]{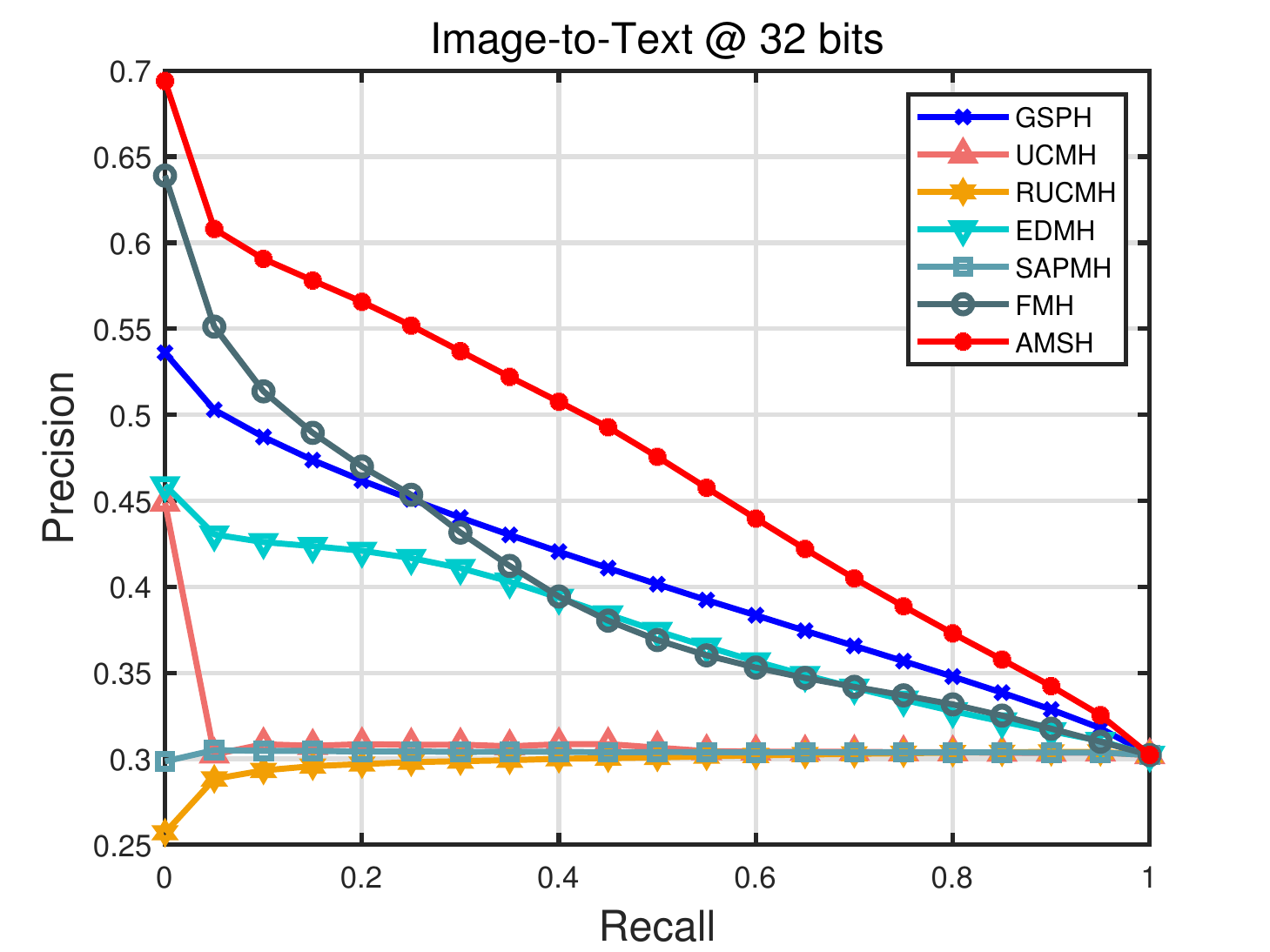}
		\includegraphics[width=4.2cm,height=3.3cm]{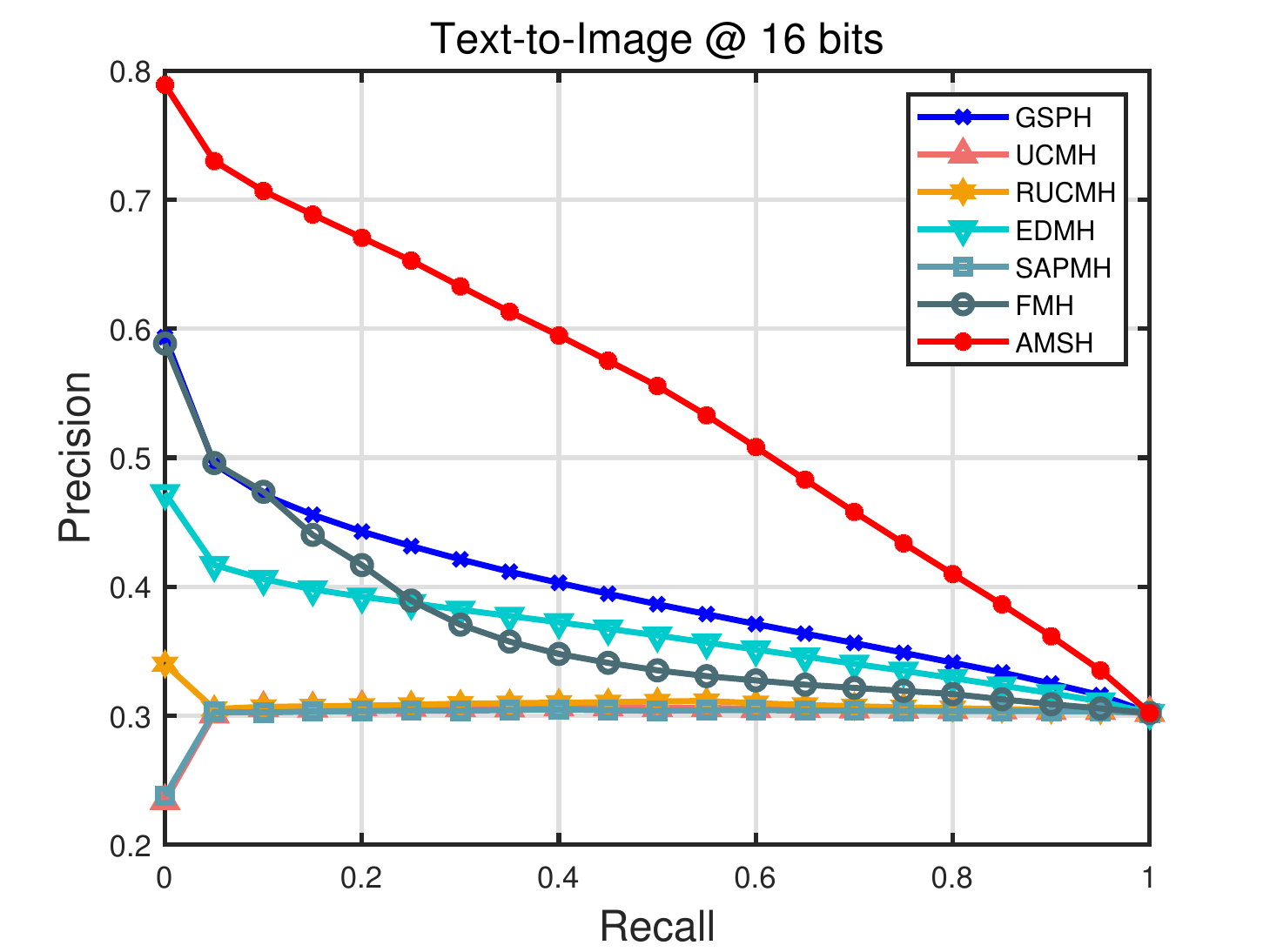}
		\includegraphics[width=4.2cm,height=3.3cm]{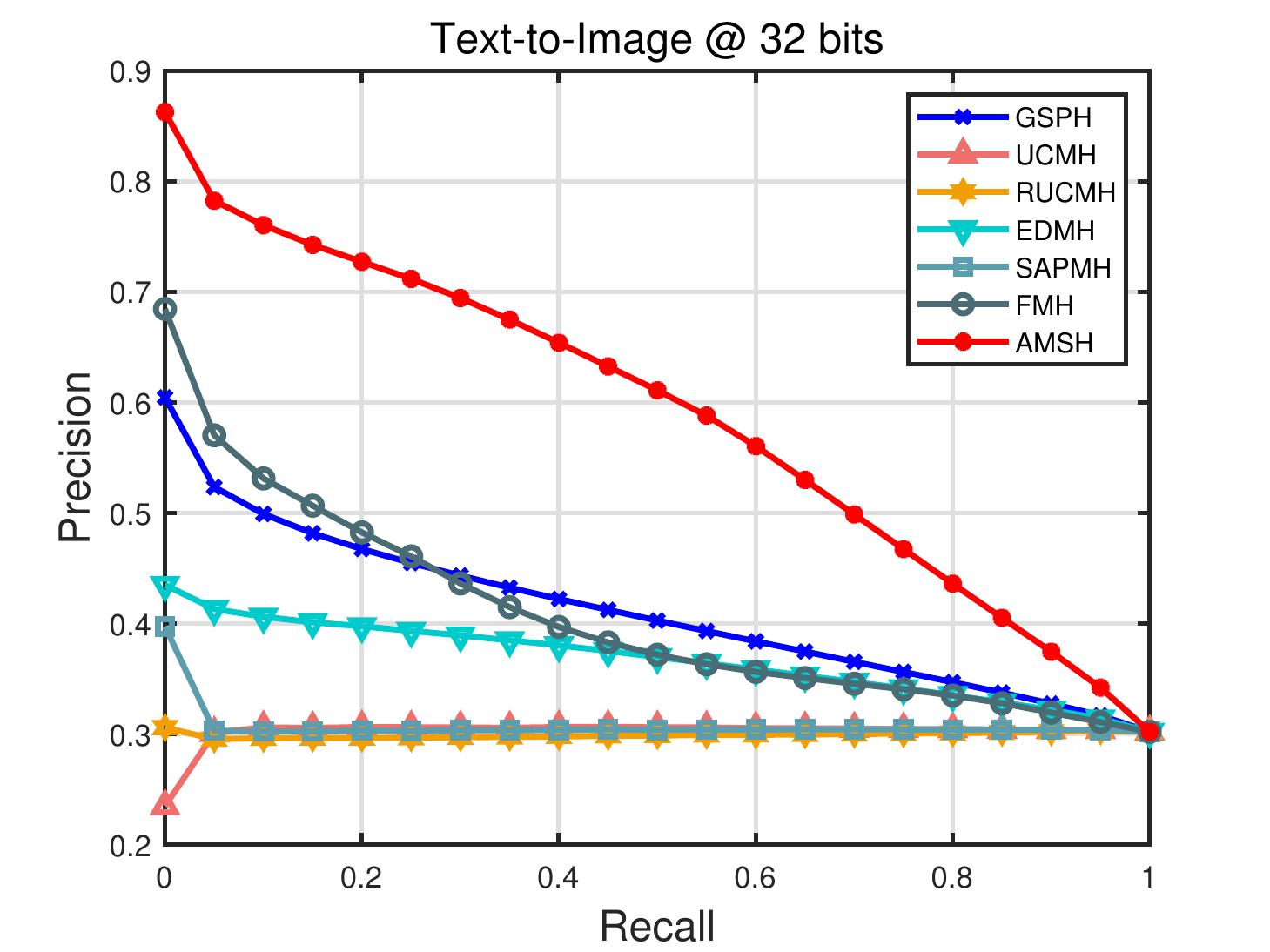}
		\caption{The PR curves of AMSH and baselines on MIRFlickr-25K (top), NUS-WIDE (middle) and IAPR TC-12 (bottom) with unpaired data.}
		\label{unpairedpr}
	\end{figure*}
	
	TABLE \ref{unpairedmap} shows the MAP results of all the methods, and the PR curves are illustrated in Fig. \ref{unpairedpr}. Some observations can be summarized as follows:
	\begin{itemize}
		\item AMSH achieves better performance than these UCMR methods on both Image-to-Text and Text-to-Image tasks with different code lengths, and the PR curves of AMSH are generally above those of other methods, demonstrating its effectiveness and superiority to handle unpaired multi-modal data.
		
		\item AMSH does not suffer from data correspondence, and the performance in the unpaired setting is almost the same as that in the paired setting, for AMSH considers both the inter-modal and intra-modal similarity, making it more robust under all circumstances. FMH and EDMH are both similarity-preserving methods. However, they fail to incorporate label discrimination with adaptive margins like AMSH. GSPH directly utilizes hash codes for similarity preservation without using latent representations, which may limit the representation capacity of hash codes. RUCMH bridges different modalities by linear reconstruction, which is insufficient to capture latent discriminative data structures. SAPMH obtains relatively low MAP results, which shows semi-paired hashing methods can not accurately solve unpaired data.
		
		\item Similar to the experimental results in the paired setting, most methods obtain better performance with longer code lengths. Besides, these methods generally achieve better results on Text-to-Image task than on Image-to-Text task. The results on MIRFlickr-25k are generally better than those on NUS-WIDE and IAPR TC-12.
	\end{itemize}

	\subsection{Parameter Analysis}
	In this subsection, we discuss the parameter sensitivity on MIRFlickr-25K, NUS-WIDE and IAPR TC-12. There are three parameters in AMSH, i.e., $\eta$, $\lambda$ and $\beta$, which adjust the weights of hash code learning, intra-modal similarity and inter-modal similarity, respectively. When analyzing each parameter, the others are fixed to default values specified in Section IV-B. 
	\begin{figure*}[t]
		\center
		\includegraphics[width=5.2cm,height=3.9cm]{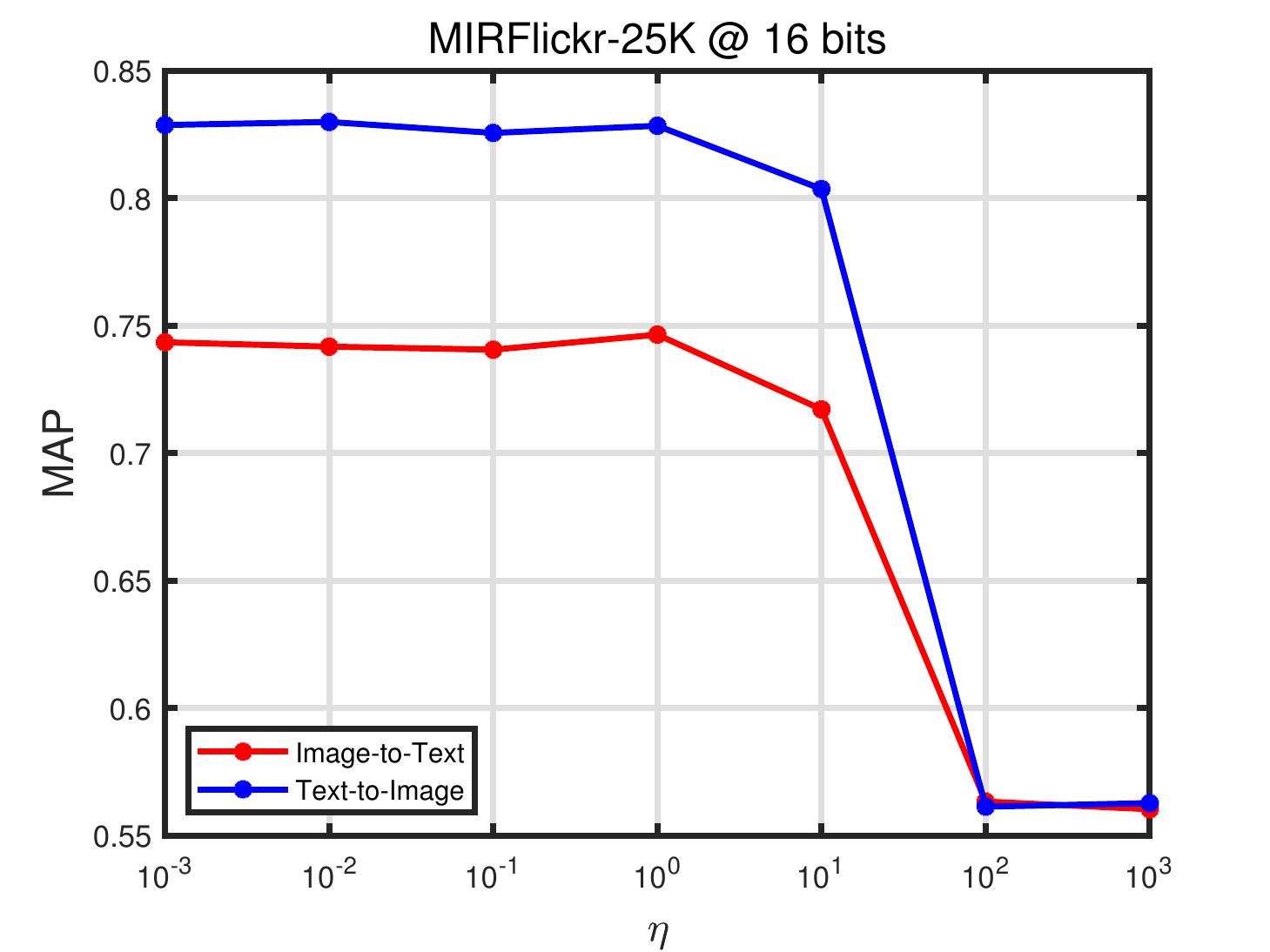}
		\includegraphics[width=5.2cm,height=3.9cm]{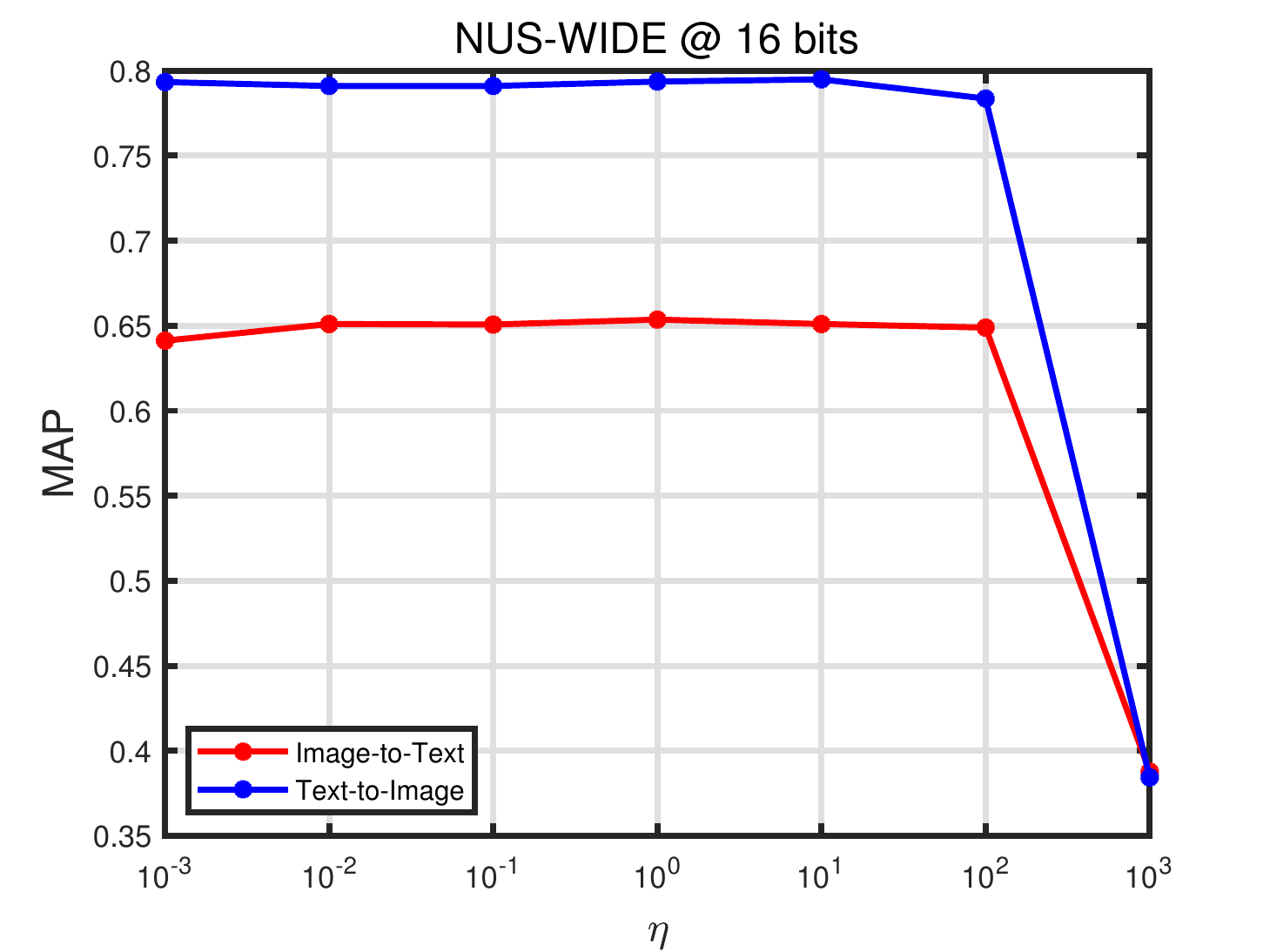}
		\includegraphics[width=5.2cm,height=3.9cm]{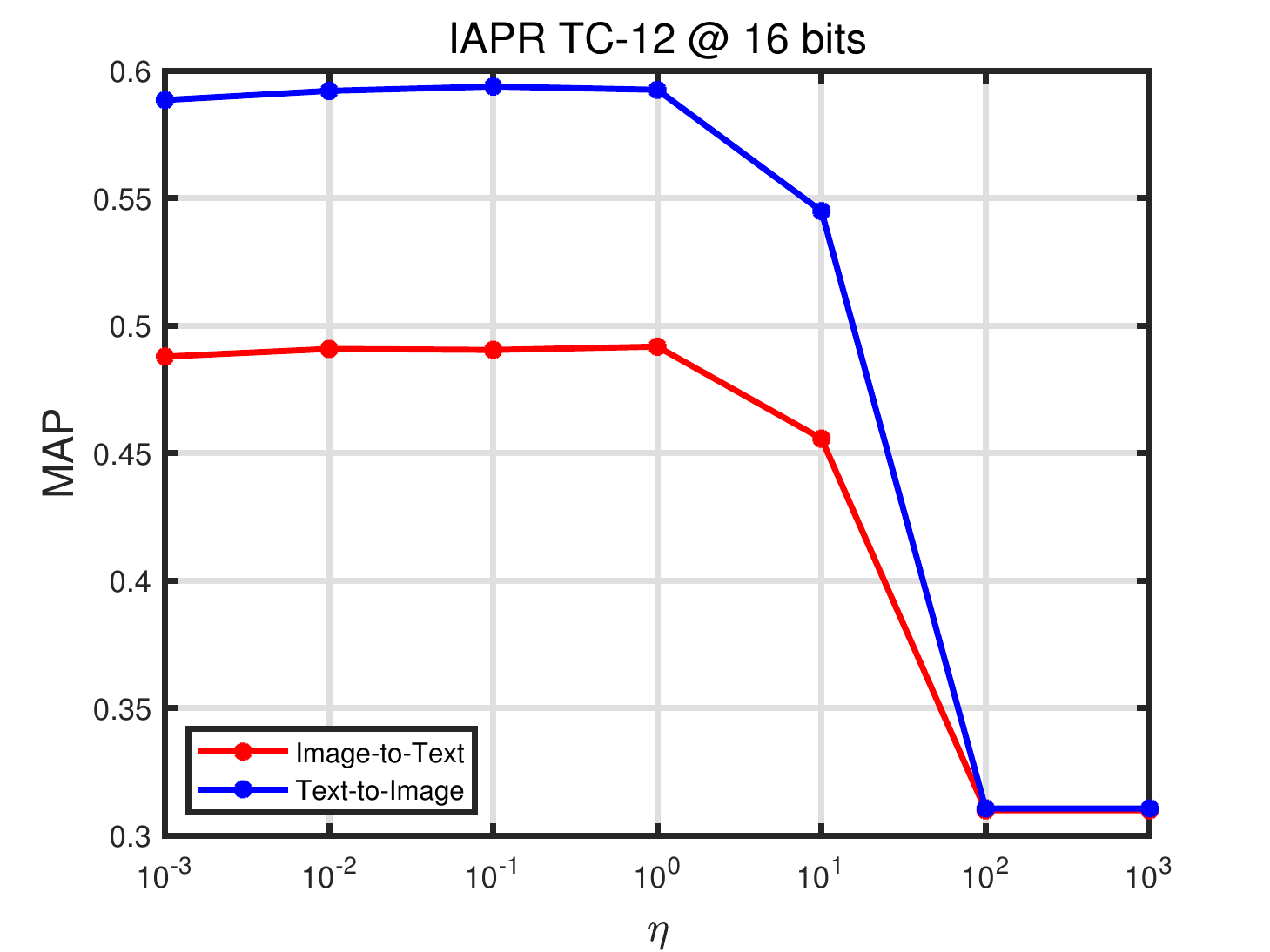}
		\includegraphics[width=5.2cm,height=3.9cm]{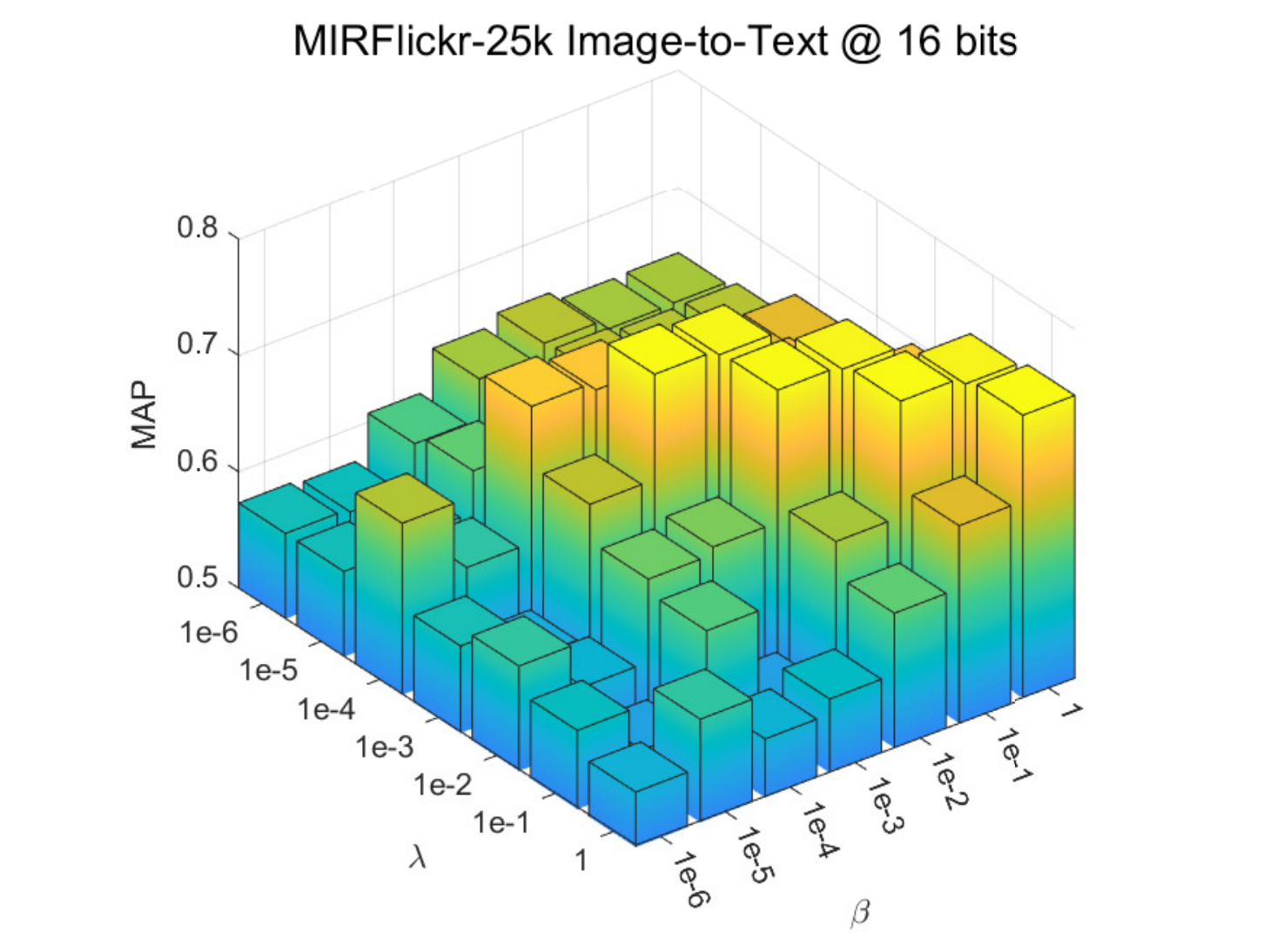}
		\includegraphics[width=5.2cm,height=3.9cm]{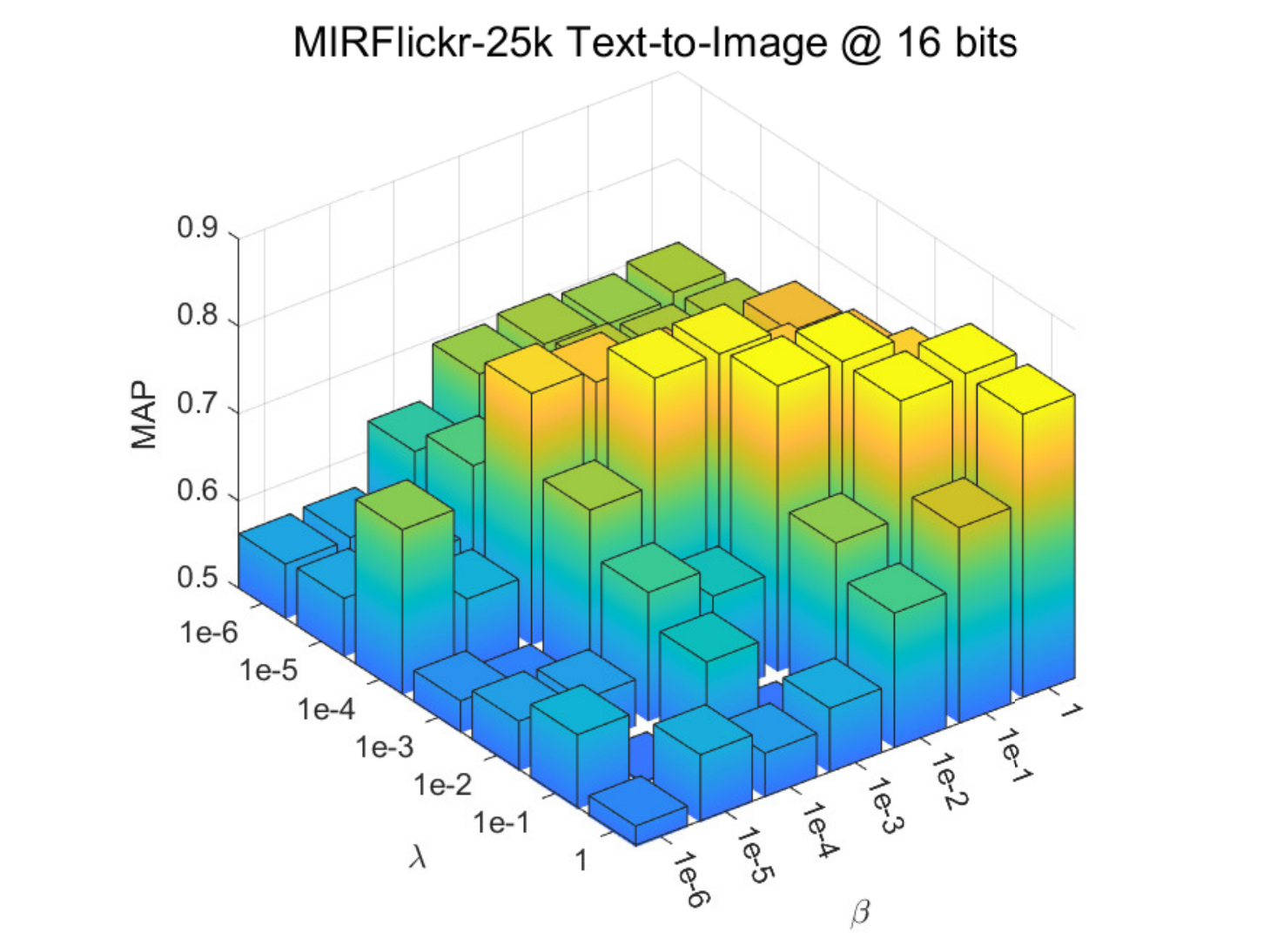}
		\includegraphics[width=5.2cm,height=3.9cm]{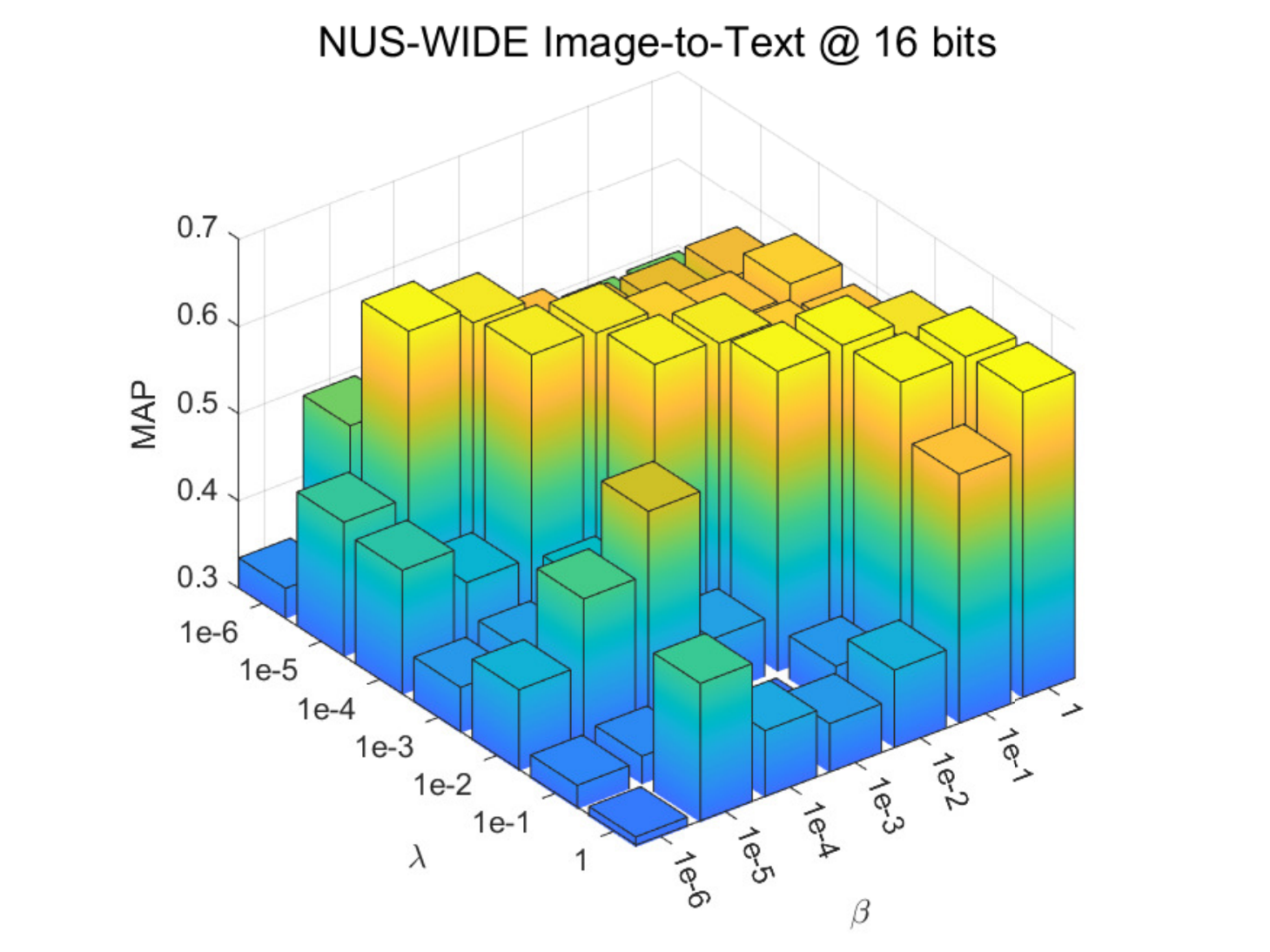}
		\includegraphics[width=5.2cm,height=3.9cm]{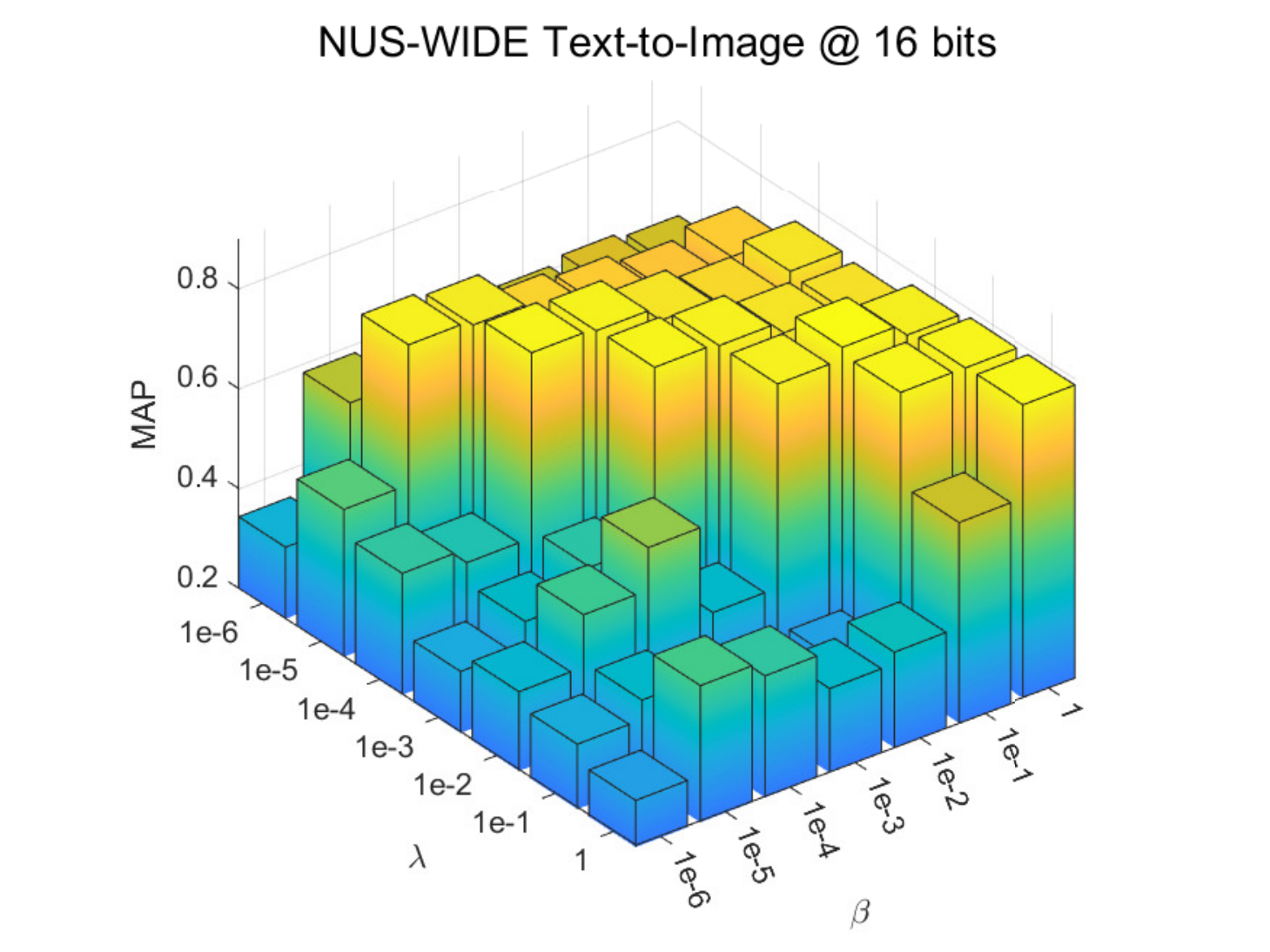}
		\includegraphics[width=5.2cm,height=3.9cm]{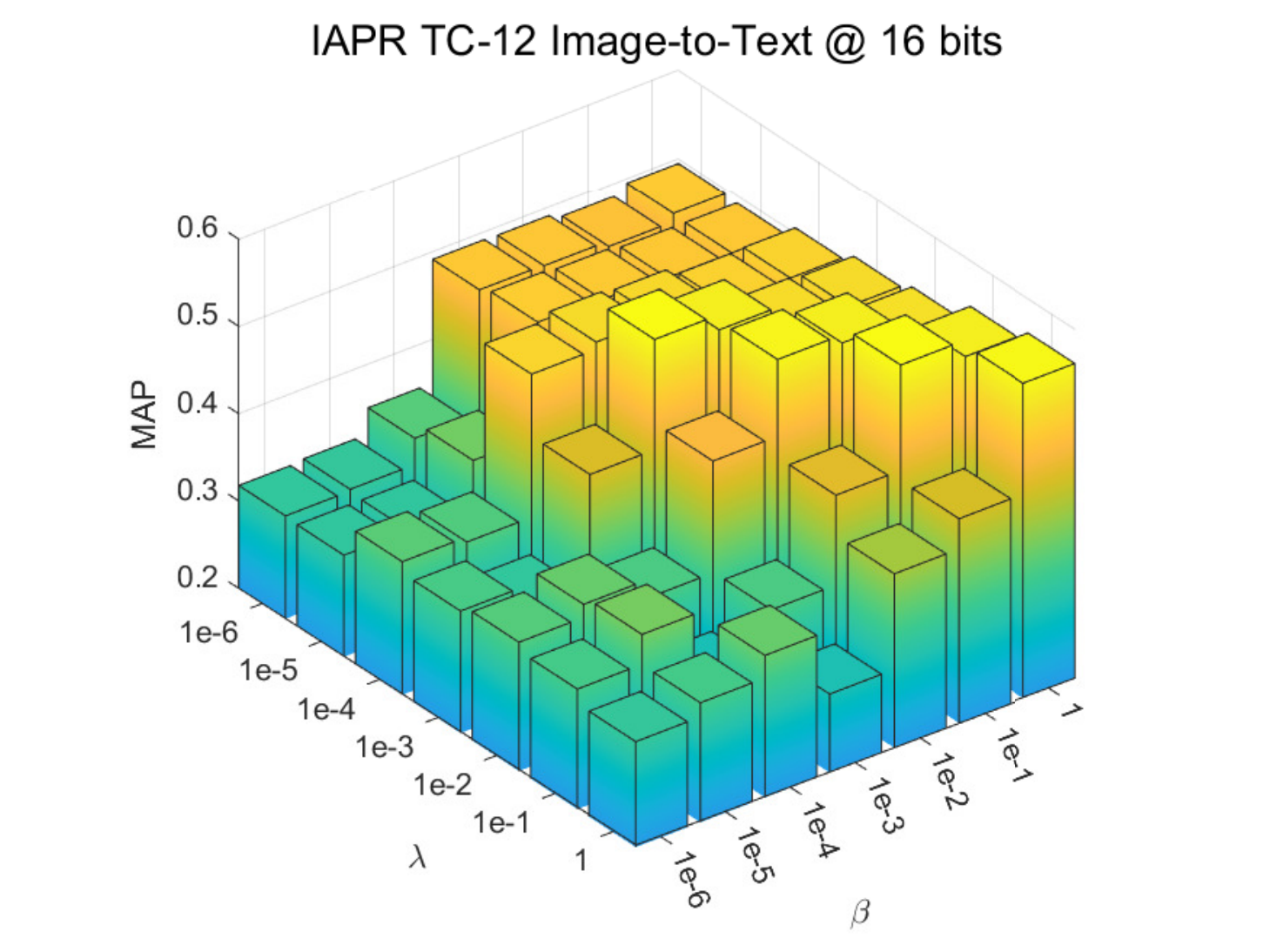}
		\includegraphics[width=5.2cm,height=3.9cm]{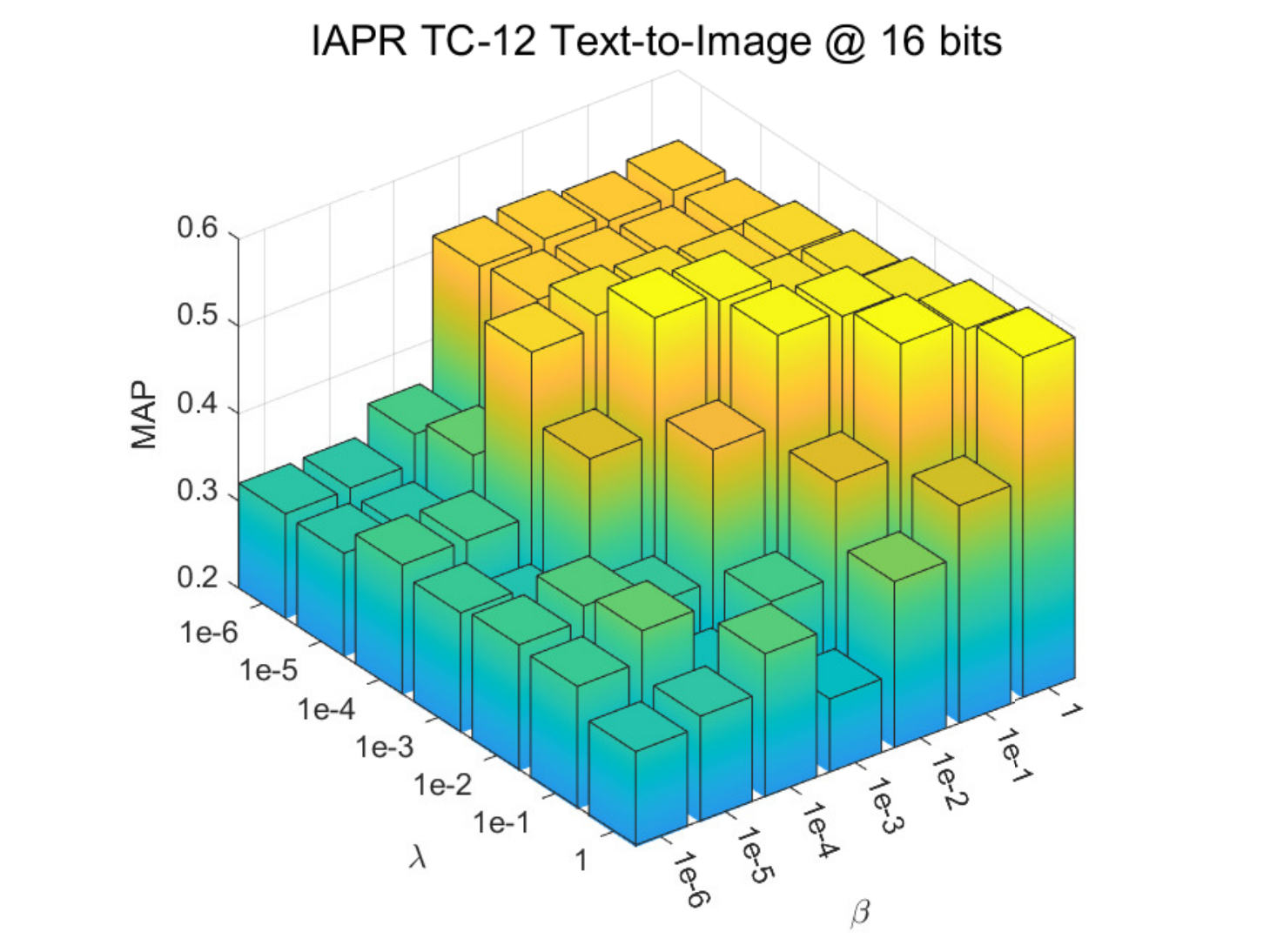}
		
		\caption{The MAP results of AMSH versus $\eta, \lambda$, and $\beta$ on three datasets.}
		\label{param}
	\end{figure*}
	
	\textbf{Parameter} \bm{$\eta$} The MAP results of Image-to-Text and Text-to-Image tasks are given in Fig. \ref{param}. $\eta$ controls the distance of binary codes and continuous representations. As $\eta$ ranges from $1e-3$ to $1e0$, the performance keeps steady and slightly improves, which shows the robustness and effectiveness of the second term in Eq. (\ref{overall}). However, when $\eta$ is set to $1e2$ and $1e3$, the MAP results drop drastically on two tasks. One possible reason is that larger $\eta$ causes $\mathbf{V}^{(i)}$ to be as close  to  $\mathbf{B}^{(i)}$ as possible, which limits the representation capacity of $\mathbf{V}^{(i)}$ to learn discriminative hash codes.
	
	\textbf{Parameter} \bm{$\lambda$} \textbf{and} \bm{$\beta$} The MAP results of AMSH versus $\lambda$ and $\beta$ tasks are given in Fig. \ref{param}. $\lambda$ influences the intra-modal similarity preservation. It can be observed that when $\beta$ is fixed, with the increase of $\lambda$, the performance is alleviated, which demonstrates the effectiveness of intra-modal similarity preservation. $\beta$ controls the inter-modal similarity preservation. With the increase of $\beta$, the performance of AMSH is improved. When the value of $\lambda$ is large, the MAP drastically drops, for which one possible reason is bigger value of $\lambda$ disturbs the inter-modal similarity preservation, thus compromising the learning of discriminative hash codes. It can also be observed that the sensitivity of AMSH to $\beta$ is less than that to $\lambda$. Even when $\beta$ is large, the MAP results only slightly drop, which demonstrates the effectiveness of inter-modal similarity preservation. From the parameter analysis, it can be summarized that the suitable parameter ranges for $\lambda$ and $\beta$ are from $1e-3$ to $1e0$. 
	
	\subsection{Training Time and Convergence}
	To solve AMSH, we develop an alternate optimization algorithm. TABLE \ref{Time} reports the training time of AMSH and all the baselines on MIRFlickr-25K, NUS-WIDE and IAPR TC-12, including the hash codes learning and hash functions learning. Generally speaking, BATCH reaches the fastest training speed for its efficient matrix factorization and semantic information preservation. Among all the methods, FSH is the slowest one, in which the fusion similarity embedding is time-consuming. For GSPH, the discrete optimization procedure of hash code learning is efficient. However, in the second step, it adopts kernel logistic regression, which is time-consuming. It can be observed that AMSH achieves a promising training speed, though it is not the fastest method. One important factor is that AMSH is a two-step method, where in the second step the adaptive margin matrix and the hash function are updated with an alternate optimization method, and each subproblem can be solved efficiently. Traditional two-step hash methods, i.e. SCRATCH \cite{SCRATCH} and BATCH \cite{BATCH}, obtain the hash functions with direct linear regression, which is fast, and yet compromises the performance, as discussed in Section III. AMSH can handle paired and unpaired scenarios while achieving better performance than these baselines, with an acceptable increase in training time.
	\begin{table*}[t]
		\centering
		\caption{The training time (in second) of AMSH and baselines on three datasets with different code lengths.}
		\begin{tabular}{p{1.8cm}p{0.7cm} p{0.7cm} p{0.7cm} p{0.7cm} p{0.15cm} p{0.8cm} p{0.8cm} p{0.8cm} p{0.8cm} p{0.15cm}p{0.9cm} p{0.9cm} p{0.9cm}p{0.9cm}}
			\toprule
			\multicolumn{1}{l}{\multirow{1}[4]{*}{Method}} & \multicolumn{4}{c}{MIRFlickr-25K}  &       & \multicolumn{4}{c}{NUS-WIDE} & &\multicolumn{4}{c}{IAPR TC-12} \\
			\cmidrule{2-5}\cmidrule{7-10}\cmidrule{12-15}
			& 16 b   & 32 b    & 64 b   & 128 b   & &  16 b    & 32 b   & 64 b   & 128 b & &  16 b   & 32 b   & 64 b    & 128 b \\
			\midrule
			CMFH \cite{CMFH} &2.12&2.44&2.65&3.16&&85.82 &90.30 &90.41& 95.47 &&43.5&43.92&46.12&46.85\\
			FSH \cite{FSH} &76.89&78.62&80.36&81.29&&6708.35&6710.78&6710.14&6714.35&&248.13&248.76&250.58&251.47\\
			CRE \cite{CRE} &56.86&61.03&58.91&63.07&&520.59&540.99&541.53&563.83&&75.32&73.11&73.98&77.24\\
			DCH \cite{DCH} &1.94&3.01&6.97&21.34&&20.81&31.60&75.48&245.37&&3.97&7.32&15.71&42.36\\
			SMFH \cite{SMFH} &4.71&4.94&13.54&16.94&&11.81&12.20&13.33&14.88&&31.67&33.84&38.87&50.51
			\\
			SCRATCH \cite{SCRATCH} &0.76&0.93&1.28&2.22&&24.16&26.78&33.34&46.49&&15.69&16.51&17.31&19.78\\
			SRLCH \cite{SRLCH} &0.93&1.26&2.11&3.63&&21.14&26.05&43.94&79.74&&7.41&9.02&13.41&22.31 \\
			BATCH \cite{BATCH} &0.61&0.80&1.30&2.09&&5.77&7.44&11.96&21.78&&0.78&1.03&1.61&2.75 \\
			GSPH \cite{GSPH} &75.25&93.82&184.92&360.08&&903.25&1125.84&2219.04&4329.6&&209.32&345.77&642.32&3828.65 \\
			EDMH \cite{EDMH}&1.13&1.54&2.26&3.96&&8.37&9.22&12.27&18.00&&6.27&6.50&7.05&8.05 \\
			AMSH& 2.39 &2.48& 3.07 &4.12 &&21.89 &24.05& 30.36 &42.84& & 3.71&4.04&4.98&7.20 \\
			\bottomrule
		\end{tabular}%
		\label{Time}%
	\end{table*}%
	\begin{figure}[t]
		\center
		\includegraphics[width=2.7cm,height=2.5cm]{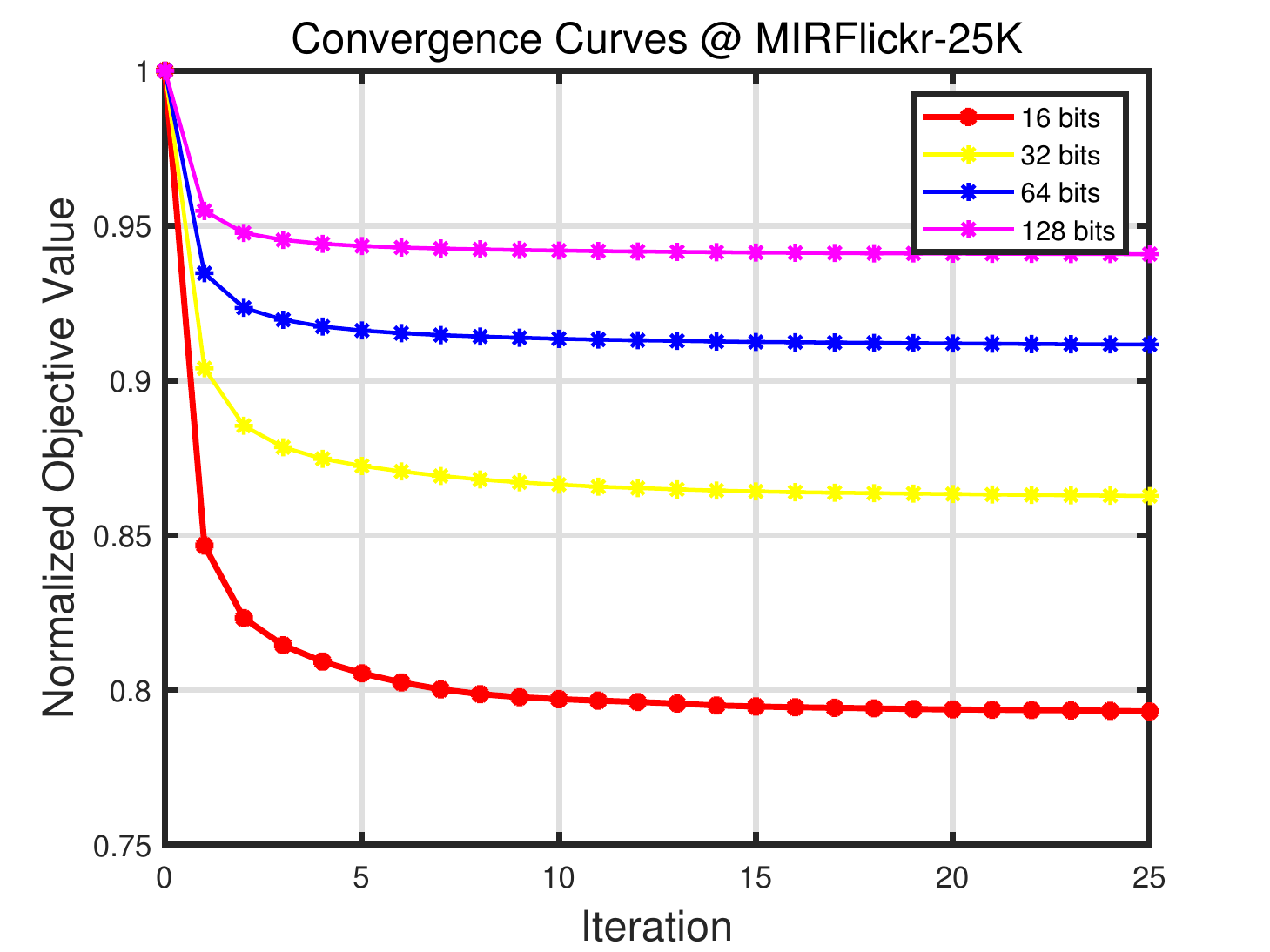}
		\includegraphics[width=2.7cm,height=2.5cm]{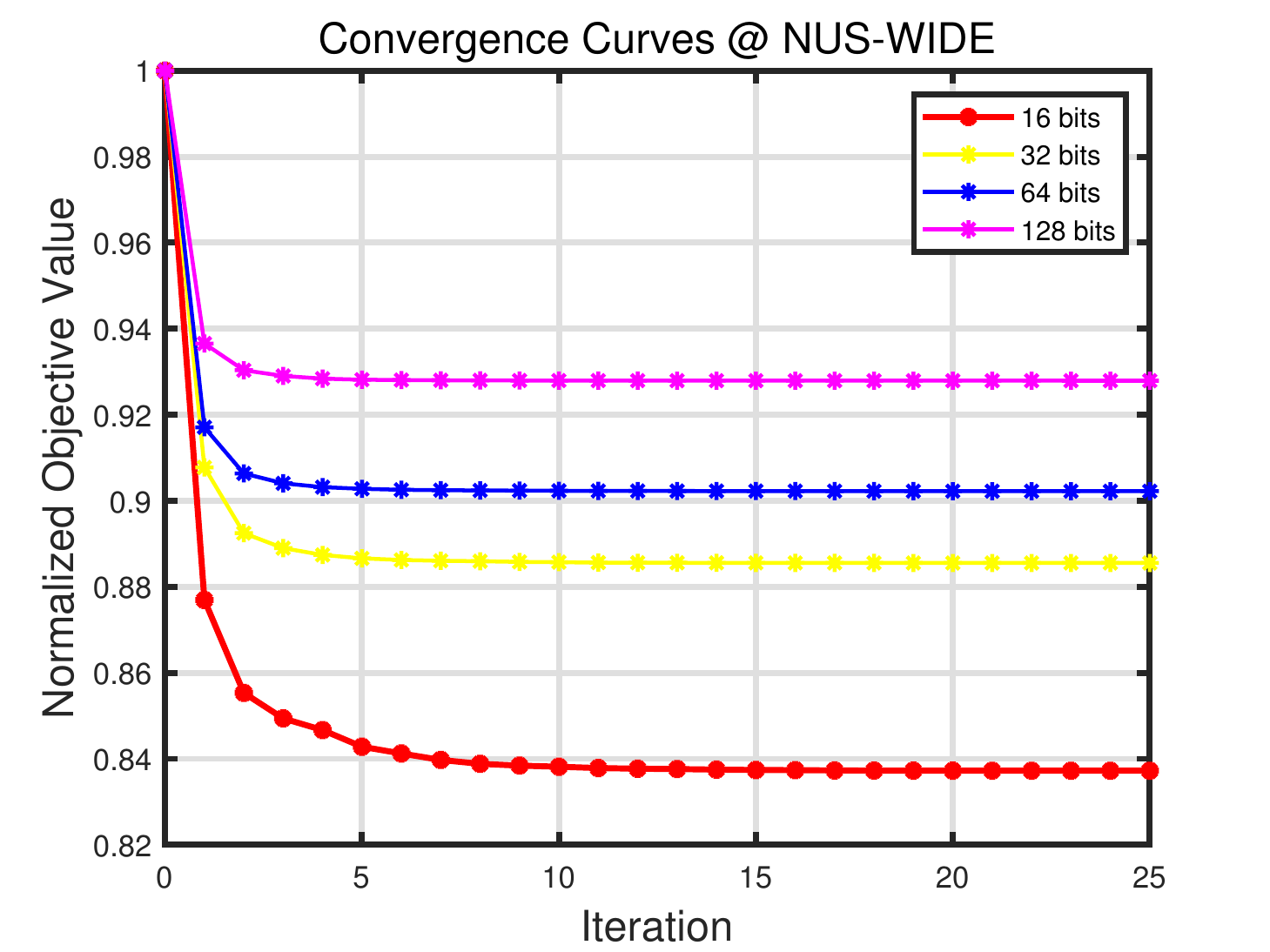}
		\includegraphics[width=2.7cm,height=2.5cm]{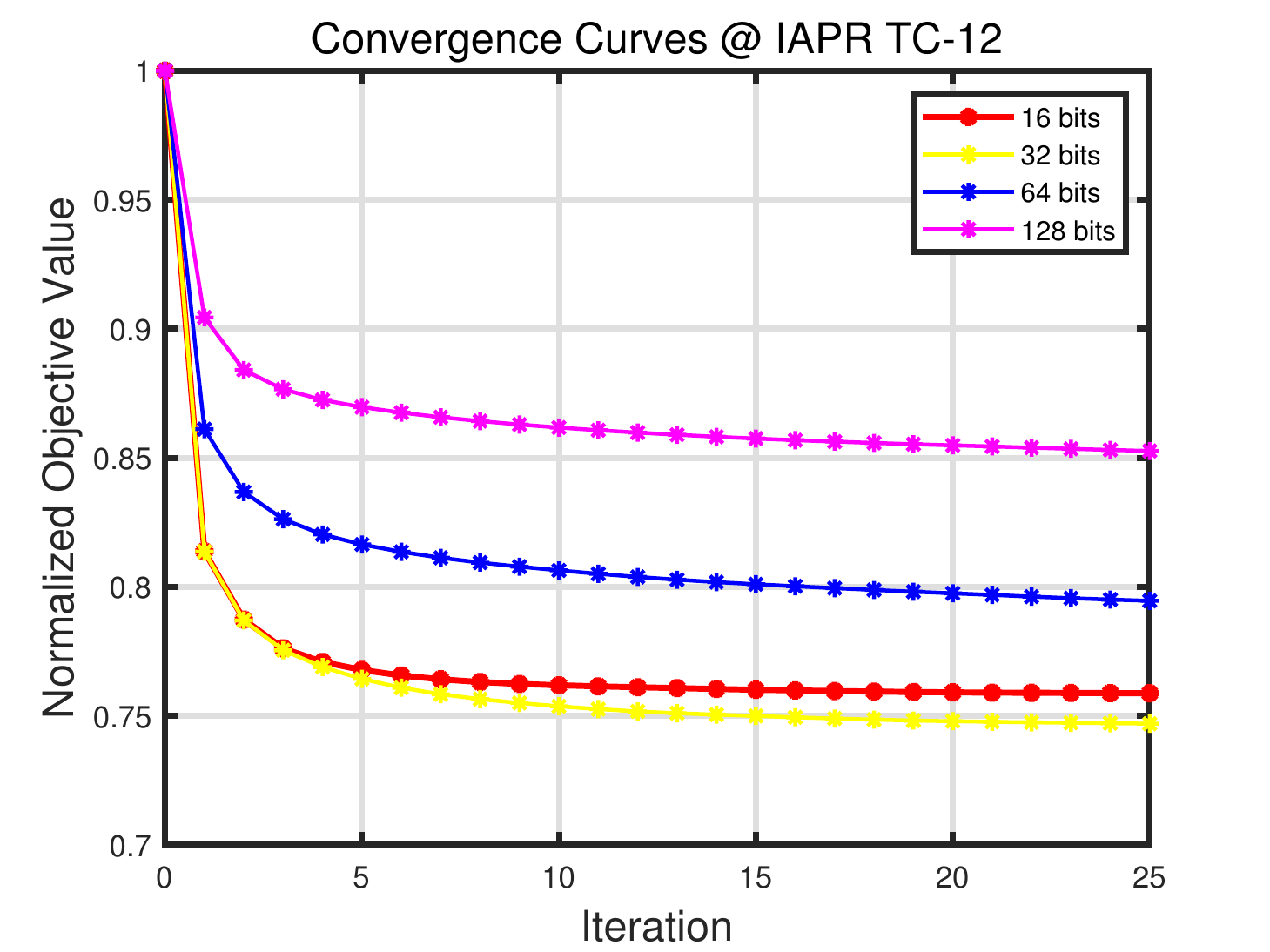}
		\caption{The convergence analysis of AMSH on MIRFlickr-25K (left), NUS-WIDE (middle) and IAPR TC-12 (right) with various code lengths.}
		\label{convplot}
	\end{figure}
	
	In Section III, we theoretically prove the convergence property of AMSH. In this subsection, we present the convergence curves with experiments conducted on MIRFlickr-25K, NUS-WIDE and {IAPR TC-12. For better observation, we normalize the objective function values by dividing each point with the largest value. As can be seen in Fig. \ref{convplot}, the objective function value monotonously decreases with each iteration, and after 15 iterations, it reaches a stable state. 
		
		\subsection{Ablation Study}
		For AMSH, adaptive marginalization and similarity preservation are adopted for hash codes and hash functions learning. In this subsection, we analyze their effects by ablation study. From AMSH, we derive four variants, i.e., AMSH-s, AMSH-t, AMSH-k and AMSH-m. AMSH-s discards the intra-modal similarity preservation, i.e., the third term in Eq. (\ref{overall}). AMSH-t drops the inter-modal similarity preservation term. AMSH-k adopts no kernel function and utilizes the original data. AMSH-m discards the adaptive margin matrix. 
		\begin{table*}[t]
			\centering
			\caption{The MAP results of AMSH and its four variants on three datasets with paired data.}
			\renewcommand{\arraystretch}{1.1}
			\begin{tabular}{p{1.8cm}<{\centering}p{1.3cm} p{0.6cm}<{\centering} p{0.6cm}<{\centering} p{0.6cm}<{\centering} p{0.7cm}<{\centering} p{0.1cm}<{\centering} p{0.6cm}<{\centering} p{0.6cm}<{\centering} p{0.6cm}<{\centering} p{0.7cm}<{\centering}p{0.1cm}<{\centering}
					p{0.7cm}<{\centering} p{0.7cm}<{\centering} p{0.7cm}<{\centering}
					p{0.7cm}<{\centering}}
				\toprule
				\multicolumn{1}{c}{\multirow{1}[4]{*}{Task}} & \multicolumn{1}{l}{\multirow{1}[4]{*}{Method}} & \multicolumn{4}{c}{MIRFlickr-25K}  &       & \multicolumn{4}{c}{NUS-WIDE} &       & \multicolumn{4}{c}{IAPR TC-12} \\
				\cmidrule{3-6}\cmidrule{8-11}\cmidrule{13-16}
				&   &  16 b   & 32 b   & 64 b   & 128 b   &      &  16 b    & 32 b    & 64 b    & 128 b & &  16 b    & 32 b   & 64 b    & 128 b \\
				\midrule
				\multirow{1}[14]{*}{$I\rightarrow T$}
				& AMSH-s&0.6425&0.6166&0.6012&0.5919&&0.5346&0.4499&0.5151&0.4720&&0.4334&0.4078&0.3878&0.3741\\
				&AMSH-t&0.6192&0.5767&0.6376&0.6625&&0.4704&0.3101&0.3744&0.4371&&0.3148&0.3444&0.3407&0.3030\\
				&AMSH-k&0.7115&0.7226&0.7240&0.7258&&0.6386&0.6391&0.6512&0.6552&&0.4642&0.4898&0.5085&0.5206\\
				&AMSH-m&0.7381&0.7490&0.7533&0.7580&&0.6432&0.6502&0.6534&0.6678&&0.4859&0.5128&0.5352&0.5498\\
				&AMSH&\textbf{0.7455}&\textbf{0.7516}&\textbf{0.7572}&\textbf{0.7591}&&\textbf{0.6544}&\textbf{0.6589}&\textbf{0.6642}&\textbf{0.6690}&&\textbf{0.4888}&\textbf{0.5163}&\textbf{0.5375}&\textbf{0.5507}\\
				\midrule
				\multirow{1}[14]{*}{$T\rightarrow I$}
				& AMSH-s&0.6915&0.6562&0.6327&0.6144&&0.6412&0.5266&0.5679&0.5043&&0.5146&0.4955&0.4753&0.4476\\
				&AMSH-t&0.6307&0.5666&0.6342&0.6902&&0.4387&0.2677&0.4528&0.4424&&0.3067&0.3571&0.3378&0.3224\\
				&AMSH-k&0.8208&0.8332&0.8372&0.8374&&0.7748&0.7844&0.8026&0.8015&&0.5701&0.6085&0.6397&0.6522\\
				&AMSH-m&0.8178&0.8288&0.8351&0.8382&&0.7715&0.7772&0.7876&0.8013&&0.5752&0.6167&0.6514&0.6709\\
				&AMSH&\textbf{0.8217}&\textbf{0.8359}&\textbf{0.8407}&\textbf{0.8477}&&\textbf{0.7784}&\textbf{0.7975}&\textbf{0.8040}&\textbf{0.8092}&&\textbf{0.5914}&\textbf{0.6309}&\textbf{0.6632}&\textbf{0.6787} \\
				\bottomrule
			\end{tabular}%
			\label{ablation}%
		\end{table*}%
		
		The performance of AMSH is compared with these four variants on three datasets in the paired scenario with different code lengths. The MAP results are reported in the TABLE \ref{ablation}. It can be observed that AMSH generally outperforms the four variants in all cases. AMSH-s and AMSH-t both achieve low MAP results, which demonstrates the importance of similarity-preserving regularization terms. AMSH-k  achieves promising results. However, it is still inferior to AMSH, which proves the effectiveness of the kernel methods to capture nonlinear data structures. The MAP results of AMSH-m are comparably high yet still lower than AMSH, which shows the adaptive marginalization contributes to the improvement of performance.
		\subsection{Comparison with Deep Hashing Methods}
		Over the recent years, due to the great capacity of neural networks to capture high-level semantic information, deep learning based methods have contributed to the improvement of the performance of cross modal retrieval. To further demonstrate the superiority of AMSH, we compare it with three state-of-the-art supervised deep hashing methods, including PRDH \cite{PRDH}, DCMH \cite{DCMH} and AADAH \cite{AADAH}, on MIRFlickr and NUS-WIDE. Following the setting of \cite{AADAH}, we adopt 4096-D feature vectors for the image modality, which are extracted by CNN network pretrained on ImageNet. For MIRFlickr, 18,015 image-text pairs are selected as the retrieval set, from which 10,000 pairs are set to be the training set, and 2,000 as the query set. NUS-WIDE consists of 195,834 image-text pairs, and 2,000 pairs are randomly selected as the query set, and the rest as the retrieval set. Furthermore, we randomly choose 10,500 instances from the retrieval set as the training set \cite{PRDH,DCMH,AADAH}. TABLE \ref{deepcompare} reports the MAP results of AMSH$_{cnn}$ and the baselines.
		\begin{table}[t]
			\centering
			\caption{MAP values of AMSH and its variants on three datasets with different code lengths.}
			\renewcommand{\arraystretch}{1.1}
			\begin{tabular}{p{0.4cm}<{\centering}p{0.7cm}  p{0.6cm}<{\centering} p{0.6cm}<{\centering} p{0.6cm}<{\centering} p{0.05cm}<{\centering} p{0.6cm}<{\centering} p{0.6cm}<{\centering} p{0.6cm}<{\centering} }
				\toprule
				\multicolumn{1}{c}{\multirow{1}[4]{*}{Task}} & \multicolumn{1}{l}{\multirow{1}[4]{*}{Method}} & \multicolumn{3}{c}{MIRFlickr-25K}  &       & \multicolumn{3}{c}{NUS-WIDE}\\
				\cmidrule{3-5}\cmidrule{7-9}
				&   &  16 b   & 32 b   & 64 b     &      &  16 b    & 32 b    & 64 b     \\
				\midrule
				\multirow{1}[10]{*}{$I\rightarrow T$}
				& DCMH&0.7410&0.7465&0.7485&&0.5903&0.6031&0.6093\\
				&PRDH&0.7499&0.7546&0.7612&&0.6107&0.6302&0.6276\\
				&AADAH&0.7563&0.7719&0.7720&&0.6403&0.6294&0.6520\\
				&AMSH$_{cnn}$&\textbf{0.7766}&\textbf{0.7797}&\textbf{0.7880}&&\textbf{0.6565}&\textbf{0.6642}&\textbf{0.6635}\\
				\midrule
				\multirow{1}[10]{*}{$T\rightarrow I$}
				&DCMH&0.7827&0.7900&0.7932&&0.6389&0.6511&0.6571\\
				&PRDH&0.7890&0.7955&0.7964&&0.6527&0.6916&0.6720\\
				&AADAH&\textbf{0.7922}&\textbf{0.8062}&\textbf{0.8074}&&\textbf{0.6789}&\textbf{0.6975}&\textbf{0.7039}\\
				&AMSH$_{cnn}$&0.7775&0.7839&0.7899&&0.6578&0.6639&0.6854\\
				\bottomrule
			\end{tabular}%
			\label{deepcompare}%
		\end{table}%
		
		It can be observed from TABLE \ref{deepcompare} that taking CNN features as inputs, AMSH$_{cnn}$ outperforms these baselines on Image-to-Text task in all cases, demonstrating the effectiveness of AMSH for semantic hash learning. AADAH achieves better performance on Text-to-Image task than AMSH. The main reason is that deep hashing methods design fully-connected layers to extract more discriminative features from the text modality, while we adopts original BOW features. In general, with CNN features as inputs, AMSH obtains comparable MAP results with deep learning based CMH methods, which proves the effectiveness of AMSH in the paired settings. Besides, AMSH is a flexible method that can handle the UCMR problem, while these deep hashing methods can only deal with fully paired data. In terms of efficiency, AMSH can be trained within 5 seconds on MIRFlickr, which is much faster than these deep learning methods. 
		\section{CONCLUSION}
		In this paper, we propose a novel Adaptive Marginalized Semantic Hashing method (AMSH) to address the CMR problem in both paired and unpaired cases. AMSH is a two-step method. In the first step, to enhance label discrimination, AMSH incorporates adaptive margin matrices to enlarge the distances between different classes, which alleviates the rigid zero-one linear regression. To handle unpaired data, AMSH generates semantic-aware modality-specific latent representations, and exploits labels to preserve the inter-modal and intra-modal semantic similarities into latent representations and hash codes. In the second step, AMSH utilizes adaptive margin matrices to enlarge the gaps between positive and negative bits, which increases the discrimination and robustness of hash functions. An discrete alternate optimization scheme is proposed to solve AMSH. Experiments on several datasets demonstrate that AMSH outperforms some state-of-the-art cross-modal hashing methods in terms of accuracy and computational costs.
		
		AMSH adopts linear regression to obtain hash codes and hash functions, which is not enough to preserve the intrinsic data correlations. AMSH can be further improved by integrating it with deep neural networks for better performance. Besides, the accurate label information of different modal data is not always available. Thus, in the future work, we will extend the proposed AMSH to handle unsupervised unpaired data.
		%
		%
		%
		%
		%
		%
		%
		%
		%
		%
		\bibliographystyle{IEEEtran}
		\bibliography{ref}
		%
	%
	
\end{document}